\documentclass[journal]{IEEEtran}
\usepackage{amsmath,amsfonts}
\usepackage{algorithm}
\usepackage{array}
\usepackage[caption=false,font=normalsize,labelfont=sf,textfont=sf]{subfig}
\usepackage{textcomp}
\usepackage{stfloats}
\usepackage{url}
\usepackage{verbatim}
\usepackage{graphicx}
\usepackage{cite}

\usepackage{algpseudocode}

\algnewcommand{\LineComment}[1]{\State \(\triangleright\) #1}
\usepackage{graphicx}
\usepackage{mathtools}
\usepackage{bm}
\usepackage{multicol,caption}
\usepackage{color}
\usepackage{amssymb}
\usepackage{bbm}
\usepackage{booktabs}
\usepackage[dvipsnames]{xcolor}
\usepackage{tikz}
\usepackage{multirow}
\usepackage[normalem]{ulem}
\usepackage{adjustbox}

\usepackage{mathtools}
\usepackage{graphicx}
\usepackage{bm}
\usepackage{pifont}
\usepackage{longtable}
\usepackage{adjustbox} 
\usepackage{makecell}
\usepackage{booktabs}
\usepackage{enumitem}
\usepackage{amsmath}
\usepackage{multicol,caption}
\usepackage{color}
\usepackage{amssymb}
\usepackage{bbm}
\usepackage{multirow}

\newcommand{\includegraphicsmaybe}[2][]{\IfFileExists{#2}{\includegraphics[#1]{#2}}{\includegraphics{img/dummy.pdf}}}
\usepackage{float}
\usepackage{amsmath}

\usepackage{cite}
\usepackage{CJKutf8}
\usepackage[utf8]{inputenc}
\allowdisplaybreaks[3]


\begin{document}
\title{New Adversarial Image Detection Based on Sentiment Analysis}

\author{
	Yulong Wang, \emph{Member}, \emph{IEEE},
	Tianxiang Li,
	Shenghong Li, \emph{Member}, \emph{IEEE}, \\
     Xin Yuan, \emph{Member}, \emph{IEEE}, and
     Wei Ni, \emph{Senior Member}, \emph{IEEE}
	\thanks{Y.~Wang and T.~Li are with the State Key Laboratory of Networking and Switching Technology, School of Computer Science (National Pilot Software Engineering School), Beijing University of Posts and Telecommunications, Beijing 100876, China (e-mail: \{wyl, tianxiangtx\}@bupt.edu.cn).}
	\thanks{S.~Li, X.~Yuan, and W.~Ni are with the Commonwealth Science and Industrial Research Organisation (CSIRO), Marsfield, Sydney, New South Wales, 2122, Australia (e-mail: \{shenghong.li,xin.yuan,wei.ni\}@data61.csiro.au)}
}

\markboth{}
{Shell \MakeLowercase{\textit{et al.}}: Bare Demo of IEEEtran.cls for IEEE Transactions on Magnetics Journals}

\IEEEtitleabstractindextext{%
\begin{abstract}
Deep Neural Networks (DNNs) are vulnerable to adversarial examples, while adversarial attack models, e.g., DeepFool, are on the rise and outrunning adversarial example detection techniques. This paper presents a new adversarial example detector that outperforms state-of-the-art detectors in identifying the latest adversarial attacks on image datasets. Specifically, we propose to use sentiment analysis for adversarial example detection, qualified by the progressively manifesting impact of an adversarial perturbation on the hidden-layer feature maps of a DNN under attack. Accordingly, we design a modularized embedding layer with the minimum learnable parameters to embed the hidden-layer feature maps into word vectors and assemble sentences ready for sentiment analysis. Extensive experiments demonstrate that the new detector consistently surpasses the state-of-the-art detection algorithms in detecting the latest attacks launched against ResNet and Inception neutral networks on the CIFAR-10, CIFAR-100 and SVHN datasets. The detector only has about 2 million parameters, and takes shorter than 4.6 milliseconds to detect an adversarial example generated by the latest attack models using a Tesla K80 GPU card.
\end{abstract}

}

\maketitle

\IEEEdisplaynontitleabstractindextext
\IEEEpeerreviewmaketitle

\section{Introduction}
Deep Neural Networks (DNNs) have demonstrated their excellent performance in image classification, voice recognition, and text categorization. However, recent studies indicate that adversarial instances can undermine DNNs. Specifically, intentionally perturbed inputs, also known as adversarial examples, can mislead DNNs to make highly confident erroneous predictions~\cite{9013065}. The perturbation required is typically imperceptible to human eyes, making the perturbation hard to detect~\cite{DBLP:conf/iclr/MadryMSTV18}. This undesirable property of DNNs has developed into a significant security concern in real-world applications, such as self-driving cars~\cite{8844593} and identity recognition~\cite{9252132}.

A recent and effective approach to detecting adversarial attacks takes the feature maps produced by the hidden layers of a DNN (e.g., a DNN-based image classifier) as input, and detects adversarial input examples by measuring the difference between benign and adversarial feature maps~\cite{DBLP:conf/iclr/Ma0WEWSSHB18,DBLP:journals/corr/abs-1803-04765,DBLP:conf/cvpr/CohenSG20,DBLP:conf/nips/LeeLLS18,DBLP:journals/corr/abs-2209-00005,LUO2022108383}. 
For instance, a detection method named Local Intrinsic Dimensionality (LID)~\cite{DBLP:conf/iclr/Ma0WEWSSHB18} uses the difference of dimension between the subspaces surrounding adversarial examples and clean examples. Another detection method, known as Deep $k$-Nearest Neighbors (D$k$NN)~\cite{DBLP:journals/corr/abs-1803-04765}, applies the $k$-nearest neighbors ($k$-NN) technique on feature maps to assess the difference between the feature maps of the input example's $k$-nearest neighbors and those of benign examples in the predicted class against a predefined threshold. Nearest Neighbor Influence Functions (NNIF)~\cite{DBLP:conf/cvpr/CohenSG20} is another popular adversarial example detector, which detects adversarial examples by assessing the correlation between the input example's $k$-nearest neighbors and the most influential benign examples identified during training. A Mahalanobis distance-based algorithm developed in~\cite{DBLP:conf/nips/LeeLLS18} fits the feature maps to a class-conditional Gaussian distribution and then detects adversarial examples by measuring the Mahalanobis distances of the feature maps. {\color{black} Besides the hidden-layer feature maps of a DNN, Be Your Own Neighborhood (BEYOND)~\cite{DBLP:journals/corr/abs-2209-00005} utilizes the output of the DNN to detect adversarial examples. It uses the  hidden-layer representations provided by self-supervised learning (SSL) and the DNN's predicted label to examine the relation between adversarial examples and their augmented versions. Moreover, Positive–Negative Detector (PNDetector)~\cite{LUO2022108383} trains a positive-negative classifier against both benign examples (positive representations) and their negative representations that complement the benign examples in each pixel to identify adversarial examples.} 


The above existing adversarial example detectors~\cite{DBLP:conf/iclr/Ma0WEWSSHB18,DBLP:journals/corr/abs-1803-04765,DBLP:conf/cvpr/CohenSG20,DBLP:conf/nips/LeeLLS18,LUO2022108383,DBLP:journals/corr/abs-2209-00005} depend primarily on machine learning techniques or hand-crafted measures. Despite performing reasonably well against some mild types of attacks (e.g., FGSM~\cite{DBLP:journals/corr/GoodfellowSS14} and Jacobian-based salience map attack (JSMA)~\cite{DBLP:conf/eurosp/PapernotMJFCS16}), the existing adversarial example detectors are less effective in detecting mighty attacks, such as DeepFool~\cite{DBLP:conf/cvpr/Moosavi-Dezfooli16} and Elastic-Net Attacks on DNNs (EAD)~\cite{DBLP:conf/aaai/ChenSZYH18}.

In this paper, we propose a new and effective adversarial example detector for DNN-based image classification. The new detector is a shallow neural network with only a few layers and a small number of parameters, and outperforms the state-of-the-art detectors in identifying the latest attacks, including DeepFool and EAD, on widely used image datasets. 

{\color{black}
The key idea is that we propose to detect adversarial examples by extracting the progressively and increasingly manifesting impact of adversarial perturbations on the hidden-layer feature maps of the DNN (as opposed to the feature maps only). In light of the progressive manifest of sentiment in a sentence, we propose to embed the hidden-layer feature maps into word vectors (i.e., a sentence) and detect adversarial examples using sentiment analysis.} 

Another important aspect is a new and efficient embedding layer that embeds the differently-sized, three-dimensional (3D), hidden-layer feature maps to word vectors with consistent lengths and assembles sentences ready for sentiment analysis.
Specifically, a modular design is taken to create a trainable module to match the dimensions between the feature maps of successively selected hidden layers. Then, each feature map can be embedded into a word vector via a cascade of modules, thereby minimizing the number of trainable modules and learnable parameters.

{\color{black}
The main contributions of this paper are as follows. 
\begin{itemize}
    \item New sentiment analysis-based interpretation of adversarial example detection and meticulous selection of TextCNN for sentiment analysis, through rigorous experimental comparisons with other candidate neural network structures; 
    \item  New modular design of an embedding layer, which reshapes and embeds the differently-sized, 3D hidden-layer feature maps of a DNN to word vectors with equal length (and assembles sentences for sentiment analysis) using the minimum number of trainable parameters; 
    \item Extensive experiments that corroborate the superior effectiveness and generalization ability of the proposed adversarial example detector under the latest adversarial example attacks compared to the state-of-the-art adversarial example detectors.
\end{itemize}
}
The experiments demonstrate that the new adversarial example detector consistently outperforms the state-of-the-art detection algorithms, such as LID~\cite{DBLP:conf/iclr/Ma0WEWSSHB18}, D$k$NN~\cite{DBLP:journals/corr/abs-1803-04765}, NNIF~\cite{DBLP:conf/cvpr/CohenSG20}, BEYOND~\cite{DBLP:journals/corr/abs-2209-00005}, PNDetector~\cite{LUO2022108383} and Mahalanobis algorithm~\cite{DBLP:conf/nips/LeeLLS18}, in identifying the latest attacks, including AutoAttack~\cite{DBLP:conf/icml/Croce020a}, DeepFool~\cite{DBLP:conf/cvpr/Moosavi-Dezfooli16} and EAD~\cite{DBLP:conf/aaai/ChenSZYH18}, on the CIFAR-10, CIFAR-100, and SVHN datasets. We use Bhattacharyya distance~\cite{7121017}, hidden layer visualization, and ablation study to shed insight on the gain of the new detector.

The remainder of this paper is organized as follows.  Section~\ref{sec: related_work} reviews the state-of-the-art attacks and detectors. Section~\ref{sec: our_approach} provides the system and threat models. In Section~\ref{sec: new detector}, we elaborate on the design of the new sentiment analysis-based adversarial example detector. The new detector is experimentally examined against cutting-edge attack models and compared with state-of-art detection algorithms in Section~\ref{sec: experiment}, followed by concluding remarks in Section~\ref{sec: conclusion}.

\section{Related Work}
\label{sec: related_work}
In this section, we briefly review the latest attacks on DNNs, and the state-of-the-art adversarial example detectors. These attack models and detectors are employed in our comparison studies with the proposed adversarial example detector, as will be presented in Section~\ref{sec: experiment}.

\subsection{Adversarial Example Attack Algorithms}
\label{sec: attack_method}
Recently, several attack algorithms for maliciously perturbing images have been proposed for off-the-shelf DNNs~\cite{DBLP:journals/corr/GoodfellowSS14, DBLP:conf/iclr/MadryMSTV18, DBLP:conf/cvpr/Moosavi-Dezfooli16, DBLP:conf/sp/Carlini017, DBLP:conf/aaai/ChenSZYH18, DBLP:conf/eurosp/PapernotMJFCS16}.  Most attack algorithms exploit a DNN's gradients to obtain a small perturbation. The corresponding attack algorithms are classified as targeted attacks or untargeted attacks depending on whether the adversarial examples are misclassified to a specific target class or simply misclassified to a different class from their source classes.

\textbf{FGSM}~\cite{DBLP:journals/corr/GoodfellowSS14} perturbs an image by changing its pixel values towards the direction of increasing the DNN-based image classifier's classification loss. FGSM generates an adversarial example using
\begin{align}
	\mathbf{x} + \epsilon\, \text{sign} (\nabla_{\mathbf{x}} \mathcal{L}( \mathbf{x}, y)), \nonumber
\end{align}
where $\epsilon \in \mathbb{R}^+$ is the perturbation magnitude, $y$ indicates the ground-truth class, $\text{sign}(\cdot)$ takes the sign of a real value, and $\nabla_{\mathbf{x}} \mathcal{L}(\mathbf{x}, y)$ is the gradient of the loss function $\mathcal{L}(\mathbf{x}, y)$ in regards to the input image $\mathbf{x}$. FGSM runs fast since it only perturbs the input once, but it needs a relatively large perturbation magnitude $\epsilon$ for a high attack success rate.

\textbf{Projected Gradient Descent (PGD)}~\cite{DBLP:conf/iclr/MadryMSTV18} improves FGSM by generating an adversarial example iteratively: 
\begin{equation}
    \label{eq: pgd}
	\mathbf{x}^{i+1} = \textstyle\prod_{\mathbf{x} + \mathcal{S}_{\epsilon}} \left( \mathbf{x}^{i} + \alpha \, \text{sign}(\nabla_{\mathbf{x}} \mathcal{L}(\mathbf{x}, y)) \right), 
\end{equation}
where $i$ is the index to an iteration, $\alpha \leq \epsilon$ is the perturbation step size, $\mathcal{S}_{\epsilon} \subseteq \mathbb{R}^d$ is the set of allowed perturbations under the maximum perturbation magnitude~$\epsilon$, and the projector $\prod_{\mathbf{x} + \mathcal{S}_{\epsilon}}(\cdot)$ maps its input to the closest element to the input in the set $\mathbf{x} + \mathcal{S}_{\epsilon}$. PGD conducts a fine-grained perturbation on images and can achieve a higher attack success rate than FGSM under the maximum perturbation magnitude, at the cost of a longer running time.

\textbf{DeepFool}~\cite{DBLP:conf/cvpr/Moosavi-Dezfooli16} perturbs an image towards the region that is the nearest to the image but belongs to a different class. DeepFool generates an adversarial example by iteratively updating its input with
\begin{align}
	\mathbf{x}_{i+1} \leftarrow \mathbf{x}_{i} + \frac{|f_{\hat{c}}^{'}|}{|| \mathbf{w}^{'}_{\hat{c}} ||^2_2} | \mathbf{w}^{'}_{\hat{c}} | \odot \text{sign}(\mathbf{w}^{'}_{\hat{c}}), \nonumber
\end{align}
until the adversarial example is misclassified or the maximum iteration number is reached. $\mathbf{x}_0$ is the benign example without any perturbation. $f_{\hat{c}}^{'}$ is the difference between the output of the softmax function of the closest different class $\hat{c}$ and that of the predicted class of the benign example $\mathbf{x}_0$. $\mathbf{w}^{'}_{\hat{c}}$ is the difference between the gradients of the softmax function of class $f_{\hat{c}}$ and that of the softmax function of the predicted class of the benign example $\mathbf{x}_0$. $\odot$ is the point-wise product. A softmax function of class $\hat{c}$ takes an input image and outputs a percentage indicating the confidence that the input belonging to class $\hat{c}$. Because DeepFool tends to perturb an image to just cross the classification boundary of the image's original class, DeepFool can generate adversarial examples with considerably small perturbations.

\textbf{Carlini and Wagner's (C\&W) algorithm}~\cite{DBLP:conf/sp/Carlini017} solves the following optimization problem to obtain the perturbation applied to an image: 
\begin{align}
	\min_{\delta}~~&||\mathbf{\delta}||_2 + a \, \mathcal{L}(\mathbf{x} + \mathbf{\delta}, t) \nonumber \\
	\text{s.t.} ~~&\mathbf{x} + \mathbf{\delta} \in [0,1]^P, \nonumber
\end{align}
where $\mathbf{\delta}$ is the perturbation on the image $\mathbf{x}$; $a$ is a constant specified in prior by running a variant of binary search; $\mathcal{L}(\cdot,\cdot)$ is one of seven loss functions specified in C\&W, such that $\mathbf{x} + \mathbf{\delta}$ is misclassified to the target class $t$ only if $\mathcal{L}(\mathbf{x} + \mathbf{\delta},t) \leq 0$; and $P$ is the dimension of the input image $\mathbf{x}$ and the perturbation $\mathbf{\delta}$. C\&W can deliver a high attack success rate but requires a perturbation with a relatively large magnitude.

\textbf{EAD}~\cite{DBLP:conf/aaai/ChenSZYH18} is inspired by C\&W and crafts an adversarial example by solving the optimization problem:
\begin{align}
	\min_{\mathbf{\tilde{x}}}~~& a \, \mathcal{L}(\tilde{\mathbf{x}},t) + b||\mathbf{\tilde{x}} - \mathbf{x}||_1 + ||\mathbf{\tilde{x}} - \mathbf{x}||_2^2 \nonumber \\
	\text{s.t.} ~~ &\mathbf{\mathbf{\tilde{x}}} \in [0,1]^P, \nonumber
\end{align}
where $a\ge 0$ and $b \ge 0$ are the regularization coefficients of the loss function $\mathcal{L}(\cdot,\cdot)$ and the $\ell_1$-norm penalty, respectively. EAD can reach the same attack success rate as C\&W, with smaller perturbations.

\textbf{JSMA}~\cite{DBLP:conf/eurosp/PapernotMJFCS16} extends saliency maps~\cite{8297086} to produce adversarial saliency maps. These maps reveal the input features that an adversary can most effectively perturb, to achieve the anticipated misclassification outcome. JSMA determines the perturbation to each pixel by using a modified saliency map:
\begin{align}
	S(\mathbf{x},t)[i,j] = 
	\left\{ 
	\begin{array}{ll}
		0, ~~ \text{if}~ J_{it}(\mathbf{x})<0~\text{or}~\sum_{j \neq t} J_{ij}(\mathbf{x})>0,\\
		J_{it}(\mathbf{x})|\sum_{j \neq t} J_{ij}(\mathbf{x})|, ~~ \text{otherwise,}
	\end{array}
	\right. \nonumber
\end{align}
where $i$ and $j$ are the indexes of elements in the saliency map $S$; $J_{ij}(\mathbf{x}) = \frac{\partial f_j(\mathbf{x})}{\partial \mathbf{x}_i}$ is the $(i,j)$-th entry of the Jacobian matrix of the image classifier $f$; and $f_j$ is the softmax function of the $j$-th class.

{\color{black}
\textbf{AutoAttack (Auto)}~\cite{DBLP:conf/icml/Croce020a} is a suite of parameter-free attacks. It contains two white-box attacks, i.e., Auto-PGD (APGD) with the cross entropy loss function and with the Difference of Logits Ratio (DLR) loss function, and two other attacks, i.e., Fast Adaptive Boundary (FAB) Attack ~\cite{DBLP:conf/icml/Croce020} and Square Attack~\cite{DBLP:conf/eccv/AndriushchenkoC20}. APGD aims to produce
adversarial examples inside an $\ell_p$-ball, with the DLR loss function defined as
\begin{align}
    \text{DLR}(\mathbf{x},y) = -\frac{z_y - \text{max}_{i \neq y}z_i}{z_{\pi_1} - z_{\pi_3}}, \nonumber
\end{align}
where $z_i$ is the logit of example class $i$ after taking $\mathbf{x}$ as input; $y$ is the ground-truth label of $\mathbf{x}$; $\pi$ is the permutation ordering the components of $z$ in decreasing order.
FAB~\cite{DBLP:conf/icml/Croce020} is a white-box attack that does not need to restart for every threshold $t_\epsilon$ if one wants to evaluate the success rate of attacks with perturbations constrained to within $\{\epsilon \in \mathbb{R}\, \vert\, \|{\epsilon}\|_p \leq t_\epsilon \}$. $\epsilon$ is the perturbation magnitude. $\mathbb{R}$ stands for the set of real values. Square Attack~\cite{DBLP:conf/eccv/AndriushchenkoC20} produces norm-bounded perturbations to launch score-based black-box attacks. It needs no knowledge of the gradient of the DNN under attack.} 

\subsection{State-of-The-Art Adversarial Example Detectors}
\label{sec: detectors}
To prevent adversarial example attacks, several detectors have been developed, including D$k$NN~\cite{DBLP:journals/corr/abs-1803-04765}, LID~\cite{DBLP:conf/iclr/Ma0WEWSSHB18}, Mahalanobis' algorithm~\cite{DBLP:conf/nips/LeeLLS18}, and NNIF~\cite{DBLP:conf/cvpr/CohenSG20}.
\begin{itemize}
    \item \textbf{D$k$NN}~\cite{DBLP:journals/corr/abs-1803-04765} combines the $k$-NN algorithm with the input representation in a DNN's hidden layers (i.e., feature map). D$k$NN identifies an adversarial example when the group of the representations of the example's $k$-nearest neighbors in the hidden layers differs from that of examples of the predicted class.
    
    \item \textbf{LID}~\cite{DBLP:conf/iclr/Ma0WEWSSHB18} is under the assumption that the dimensions of the subspaces surrounding adversarial (perturbed) examples and benign (unperturbed) examples differ. LID estimates the dimension and accordingly detects adversarial examples.
    
    \item \textbf{Mahalanobis}' algorithm~\cite{DBLP:conf/nips/LeeLLS18} assumes pre-trained input features can be fitted by a class-conditional Gaussian distribution. The Mahalanobis distance to the closest class-conditional distribution reveals adversarial examples.
    
    \item \textbf{NNIF} algorithm~\cite{DBLP:conf/cvpr/CohenSG20} assumes that the $k$-NN training samples (i.e., the nearest neighbors in the feature map space) and the most influential training samples (identified using an influence function) correlate for benign examples, but do not correlate for adversarial examples. The correlation is measured to detect if an attack is underway. 

    {\color{black}
    \item \textbf{BEYOND}~\cite{DBLP:journals/corr/abs-2209-00005} assumes that benign perturbations, i.e. random noises, with bounded budgets cause minor variations on the feature space, and then detect anomalous behaviors by distinguishing an adversarial example's relation with its augmented version, or neighbors, from representation similarity and label consistency.

    \item \textbf{PNDetector}~\cite{LUO2022108383} assumes the misclassification space is randomly distributed in the ideal feature space of a pre-trained classifier. PNDetector is a positive-negative classifier trained by original examples (positive representations) and their negative representations that share the same structural and semantic features.
    }
    
\end{itemize}
According to~\cite{DBLP:conf/cvpr/CohenSG20}, LID~\cite{DBLP:conf/iclr/Ma0WEWSSHB18}, Mahalanobis~\cite{DBLP:conf/nips/LeeLLS18}, and NNIF~\cite{DBLP:conf/cvpr/CohenSG20} yield their respective best detection performance when using all hidden layers of a DNN, while D$k$NN~\cite{DBLP:journals/corr/abs-1803-04765} achieves its best detection by only using the penultimate layer of the DNN.

\section{System Model}
\label{sec: our_approach}

\subsection{System Architecture}
\begin{figure}[t]
	\centering
	\includegraphics[width=3.5in]{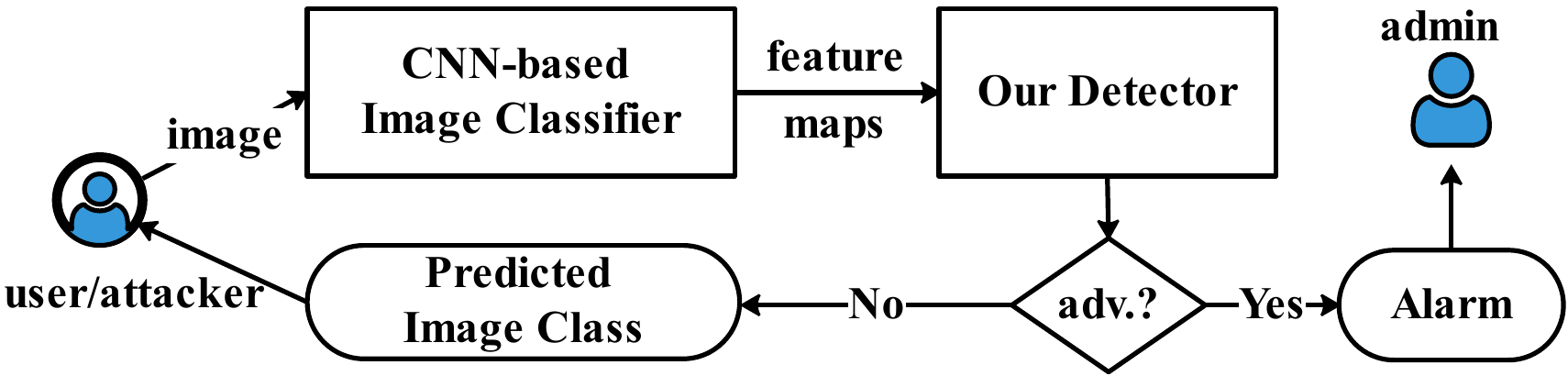}
	\caption{The proposed detector's location when used in a DNN-based computer vision application. }
	\label{fig: overall_view}
\end{figure}

The proposed adversarial example detector runs in parallel with a DNN-based image classifier in computer vision applications to protect the image classifier, as illustrated in Fig.~\ref{fig: overall_view}. When the DNN-based image classifier classifies an input image, the feature maps produced by several hidden layers of the image classifier are copied and sent to the detector for adversarial example detection. 

If adversarial perturbations are detected on the input image, the proposed adversarial example detector generates a notification to alert the administrator of the computer vision application. The image classifier's prediction is stopped from making any decision, such as granting access based on face recognition~\cite{8936536}. 
If the detector does not detect any hostile alteration in the input image, the DNN-based computer vision application continues functioning as usual.

\subsection{Threat Model}
We adopt the threat model described in~\cite{9013065}, where an adversarial attacker attempts to mislead the DNN-based image classifier by feeding the classifier adversarial examples. The attacker can repeatedly perturb the pixels of an image until the DNN-based image classifier misclassifies the image to a different class from the correct one.

Assume that the attacker has complete knowledge of the DNN-based image classifier (or, in other words, the classifier is a white box to the attacker). Accordingly, the attacker can generate adversarial examples that can be misclassified by the classifier. This is due to the fact that the neural network architectures of the best-performing image classifiers (e.g., the ResNet models~\cite{8804390}) are often common knowledge. Even if the classifiers' parameters are unknown to the attacker, the attacker can learn a good surrogate of the classifier by sending queries to the classifier and collecting responses~\cite{9252132}. 

We consider the typical situation where the attacker has no knowledge of the adversarial example detector. In other words, the detector is a black box to the attacker. This is because, in most cases, the detection results are generally inaccessible to the attacker, and hence the attacker can hardly learn a surrogate of the detector.
We also consider a relatively rare situation where the attacker somehow gets access to the adversarial example detector and its gradient (e.g., due to a compromised server or a rogue employee). In this case, a white-box attack~\cite{9462227} to both the DNN-based image classifier and the adversarial example detector is evaluated.

\section{New Sentiment Analysis-based Adversarial Example Detector}
\label{sec: new detector}
We propose to interpret a series of feature maps (of an input image) produced by the different hidden layers of the DNN-based image classifier under an adversarial example attack. We detect the presence of adversarial perturbations on the images by embedding the hidden-layer feature maps into a sentence and analyzing the sentiment of the sentence. The presence or absence of adversarial perturbation is translated to the positive or negative sentiment of the sentence, respectively. A sentiment analysis model developed originally for natural language processing (NLP), such as TextCNN~\cite{DBLP:conf/emnlp/Kim14} and Long Short-Term Memory (LSTM)~\cite{8788581}, can be applied to detect the perturbations.

The rationale behind our interpretation of hidden-layer feature maps to a sentence for sentiment analysis is that the feature maps account for the subtle transition from a perturbed image of one class to a recognizable sample of another class. The feature map of a perturbed image (i.e., an adversarial example) can be closer to the target class than the correct class of the unperturbed (i.e., a benign example) at the penultimate layer of an image classifier. 
On the other hand, the perturbed image is typically indistinguishable from the unperturbed at the input layer of the image classifier, due to the typical imperceptibility of perturbations~\cite{DBLP:conf/iclr/MadryMSTV18}. 
By using sentiment analysis, the progressively manifesting impact of perturbation on the hidden-layer feature maps can be exploited to detect adversarial examples.

\begin{figure}[t]
	\centering
	\includegraphics[width=3.5in]{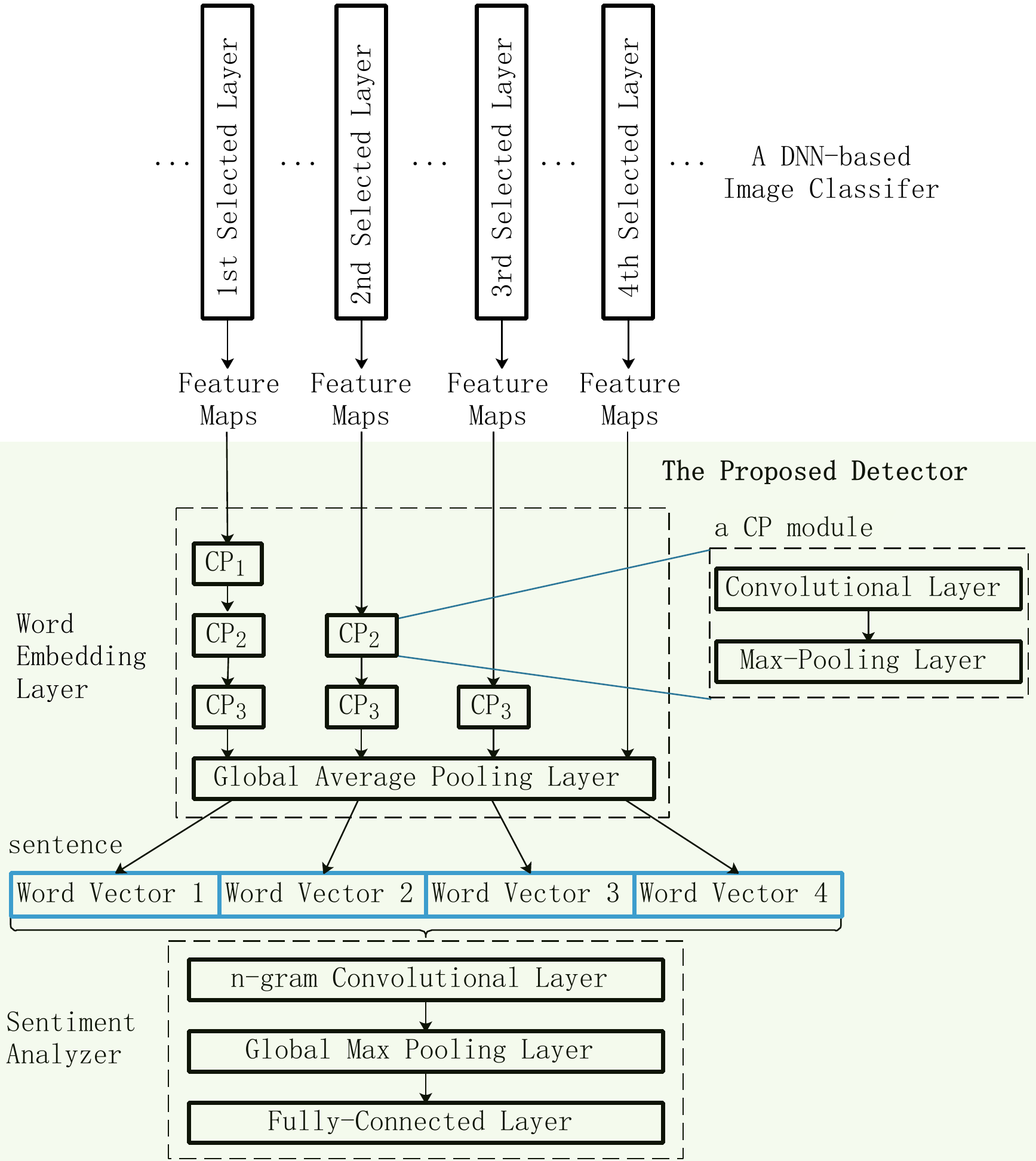}
	\caption{The architecture of the proposed detector. }
	\label{fig: arch}
\end{figure}

The proposed detector is made up of two components: a Word Embedding Layer $\mathcal{E}$ and a Sentiment Analyzer $\mathcal{A}$. As illustrated in Fig.~\ref{fig: arch}, the input of the proposed detector (i.e., a series of feature maps) is first mapped by the Word Embedding Layer $\mathcal{E}$ to a sentence, and then analyzed by the Sentiment Analyzer $\mathcal{A}$.

\subsection{Word Embedding Layer}
\label{sec: sen_gen}
The Word Embedding Layer translates the hidden-layer feature maps (i.e., the outputs of the hidden layers) of the DNN potentially under attack as a sentence for follow-on sentiment analysis. In sentiment analysis, a fixed-length vector, referred to as ``word vector'', is used to represent a word in a sentence~\cite{8241844}. A string of word vectors is used as the input to classify the sentence between positive and negative sentiments or, in other words, benign and adversarial examples. In NLP, word vectors are typically obtained by embedding words in a vector space through unsupervised learning based on a large text dataset. For instance, Mikolov \textit{et al.}~\cite{DBLP:conf/nips/MikolovSCCD13} trained a set of word vectors on 100 billion words of Google News. However, there is no trained word vector for the hidden-layer feature maps of images. The feature maps produced at the different hidden layers of the DNN can have different sizes. A new approach is needed to embed the hidden-layer feature maps into word vectors to be used in sentiment analysis.

We design a new Convolution-Pooling (CP) module, which resizes the feature map produced by a selected hidden layer of the DNN to have the same dimension as the feature map of the next selected hidden layer, as shown in Fig.~\ref{fig: arch}. Suppose that $L$ hidden layers are selected from the DNN under attack for adversarial example detection. There are $(L-1)$ CP modules in the word embedding layer.

A 3-tuple $(c_i,w_i,h_i),\, i=1,\cdots,L$ is used to describe the dimension of the feature map produced by the $i$-th selected hidden layer of the DNN, where $c_i$, $w_i$, and $h_i$ denote the number of channels, and the width and the height per channel, respectively. For the $i$-th CP module, denoted by CP$_i$ ($i=1,\cdots,L-1$), the input and the output dimensions are $(c_i,w_i,h_i)$ and $(c_{i+1},w_{i+1},h_{i+1})$, respectively. In other words, CP$_i$ converts a $(c_i,w_i,h_i)$-dimensional feature map to a $(c_{i+1},w_{i+1},h_{i+1})$-dimensional feature map. 

Each CP module, i.e., CP$_i$ ($i=1,\cdots,L-1$), comprises a convolutional layer and a max-pooling layer. The convolutional layer of CP$_i$ has $c_{i+1}$ convolutional kernels to convert the $c_i$ feature maps, one per channel, produced by the $i$-th selected hidden layer of the DNN to $c_{i+1}$ feature maps, one per channel. Then, the max-pooling layer of CP$_i$ converts the width $w_i$ and height $h_i$ of each of the $c_{i+1}$ feature maps to $w_{i+1}$ and $h_{i+1}$, respectively. The convolutional layer and the pooling layer of each CP module are constructed with kernels of appropriate dimensions accordingly. 

By concatenating $\text{CP}_i,\cdots,\text{CP}_{L-1}$, the feature map of the $i$-th selected hidden layer of the classifier is resized to be consistent with the feature map of the last ($L$-th) selected hidden layer, i.e., $(c_L,w_L,h_L)$, as shown in Fig.~\ref{fig: arch}. Likewise, the sizes of all $L$ selected feature maps are unified to $(c_L,w_L,h_L)$. The feature maps with the unified dimension of $c_L \times w_L \times h_L$ are passed into a global average pooling layer, as shown in Fig.~\ref{fig: arch}. The global average pooling layer flattens the feature maps by replacing each of the feature maps with the average value of its elements, and translates them to word vectors of the same dimension of $1 \times c_L$. A sentence is constructed by concatenating the word vectors, and then output to the Sentiment Analyzer. 

This modular design allows the CP modules to be reused for resizing different feature maps while keeping the number of CP modules to the minimum of only $(L-1)$, hence minimizing the number of learnable parameters in the Word Embedding Layer. As a result, the Word Embedding Layer is fast to train and computationally efficient.  

\subsection{Sentiment Analyzer}
\label{sec: sentiment analyzer}
The Sentiment Analyzer is responsible for classifying the input sentences (i.e., strings of word vectors) into positive sentiments (i.e., perturbed images) or negative sentiments (i.e., unperturbed images). The Sentiment Analyzer is a shallow neural network, which typically contains a convolutional layer, a global max-pooling layer, and a fully-connected layer. We choose TextCNN as the Sentiment Analyzer, due to its simple architecture and good sentence classification accuracy~\cite{DBLP:conf/emnlp/Kim14}. 

Different from a traditional neural network with equal-sized 2D convolutional kernels in each hidden layer, the Sentiment Analyzer utilizes one-dimensional (1D) $n$-gram convolutional kernels in its convolutional layer~\cite{DBLP:conf/emnlp/Kim14}. Each of the convolutional kernels can take $n$ word vectors as the input. $n$ ranges from one to the number of word vectors in a  sentence generated by the Word Embedding Layer. There are multiple $n$-gram convolutional kernels with randomly initialized parameters to extract features from the $n$-length segments of a sentence.

\begin{figure}[t]
	\centering
	\includegraphics[width=3.3in]{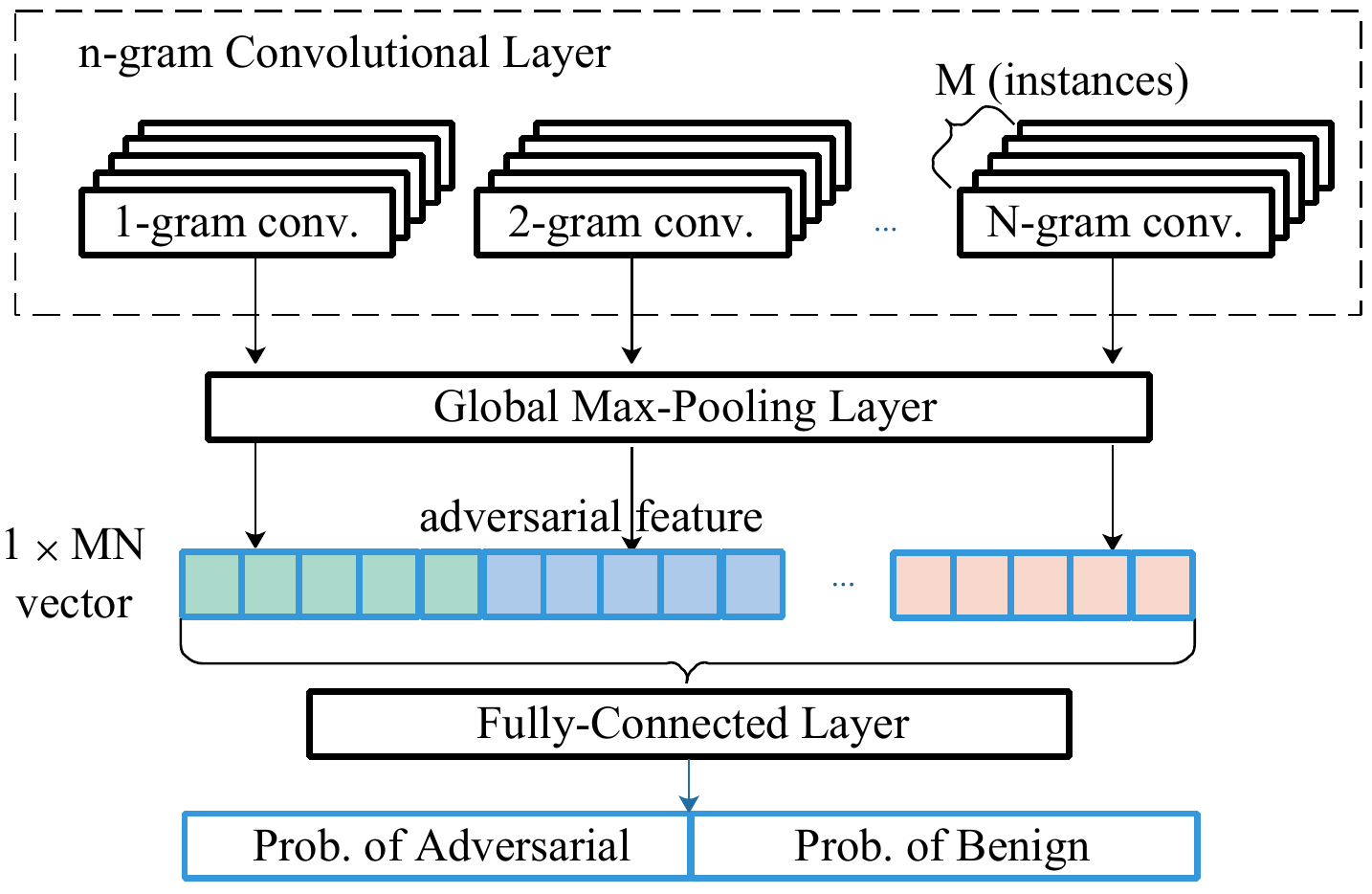}
	\caption{The architecture of the sentiment analyzer. There are $M$ instances of $n$-gram convolutional kernels.$n = 1,\cdots,N$.}
	\label{fig:sentiment_analyzer}
\end{figure}

The global max-pooling layer in the Sentiment Analyzer has two functions. First, it reduces the dimension of the feature maps output by the convolutional layer in the Sentiment Analyzer, hence helping counteract overfitting. Second, the global max-pooling layer can change the shape of hidden-layer  feature maps so that the feature maps are flattened to a vector and fit the input size of the following fully-connected layer. Moreover, the fully-connected layer generates a $1\times 2$ row vector that provides the likelihood of the input example being benign or adversarial. To avoid overfitting, the fully-connected layer has a dropout parameter of 0.5. The architecture of the Sentiment Analyzer is illustrated in Fig.~\ref{fig:sentiment_analyzer}.

\subsection{Algorithm Summary}
\begin{algorithm}
	\caption{Proposed sentiment analysis-based adversarial example detector}
	\label{alg: sentence_generator}
	\textbf{input}: $\{F_l\}$ ($l = 1,\dots, L$), the set of feature maps output by $L$ selected hidden layers of the image classifier.
	
	\textbf{output}: the probability of the input example being adversarial.
	\begin{algorithmic}[1]
 	
 	\LineComment{Operation of the Word Embedding Layer $\mathcal{E}$}
 	\State Initialize $\{W_l\}$ by setting $W_l = F_l$  
	\For{$l=$ 1 to $L$}
	    \For{$q=l$ to $L-1$} \Comment{unify dimension}
	        \State $W_l \leftarrow \text{CP}_q(W_l)$, \Comment{alter $W_l$'s dimension}
	    \EndFor
	   \State Feed $W_l$ to the Global Average Pooling layer to obtain a $1\times c_L$ word vector and save it in $W_l$.
	\EndFor 
	\State Concatenate $W_l$ ($l=1,\dots, L$) to construct a $1 \times Lc_L$-dimensional sentence.
    \LineComment{ Operation of the Sentiment Analyzer $A$}
	\State Let $\mathcal{T} \leftarrow \emptyset$ be a set of hidden-layer feature maps.
	\For {$n=1$ to $N$}
	    \For {$i=1$ to $M$}
	        \State Generate a sentiment hidden-layer map by applying the $i$-th $n$-gram convolutional kernel to the sentence;
	        \State Add the sentiment hidden-layer map to $\mathcal{T}$.
	    \EndFor
	\EndFor
	\State Feed all hidden-layer feature maps in $\mathcal{T}$ to the Global Max Pooling Layer and concatenate the outcomes in a vector;
	\State Feed the vector to the fully-connected layer that outputs a $1\times 2$ vector containing the probabilities of the input example being benign and adversarial.
    \end{algorithmic}
\end{algorithm}

Algorithm~\ref{alg: sentence_generator} summarizes the adversarial example detection using the proposed detector, where $L$ hidden layers are selected from the DNN-based image classifier to output feature maps to the detector. The time complexity of the algorithm is~$\mathcal{O}(L^2)$. The space complexity of Algorithm~\ref{alg: sentence_generator} is measured by the number of learnable parameters of the proposed detector. Because the pooling layers do not have any learnable parameters, we only consider the learnable parameters in the convolutional layers. The space complexity of the Word Embedding Layer is $\mathcal{O}(L^2c_Lw_Lh_L)$. The space complexity of the Sentiment Analyzer is $\mathcal{O}(MN^2c_L)$, where $N$ is the number of types of $n$-gram convolutional kernels and $M$ is the number of instances of a type of $n$-gram convolutional kernel. As a result, the space complexity of Algorithm~\ref{alg: sentence_generator} is $\mathcal{O}(L^2c_Lw_Lh_L+MN^2c_L)$.

The learnable parameters of the proposed adversarial example detector, namely, the model weights and bias, can be optimized using supervised learning based on unperturbed and perturbed examples. Existing attack algorithms described in Section~\ref{sec: attack_method} can be used to perturb images and generate adversarial examples for training.

\section{Experiment Results}
\label{sec: experiment}
In this section, we evaluate the performance of the proposed detector, and then present a visualized explanation of the detection mechanism. Our code is available at {https://github.com/wangfrombupt/adversarial\_detector}.

\subsection{Experiment Setup}
Our experiment setup is consistent with~\cite{DBLP:conf/cvpr/CohenSG20} in terms of image classifiers, datasets, attack models, benchmark detectors, and performance indicators. Since~\cite{DBLP:conf/cvpr/CohenSG20} presented the latest study on adversarial example detection in the literature, and developed the state-of-the-art detector, namely, NNIF, which is also used as a benchmark in our experiments. 

\textbf{Image classifier:} By default, the image classifier (a DNN) under attack is a Deep Residual Network~\cite{8804390} with 34 hidden layers, referred to as ResNet-34. The feature extraction layers of ResNet-34 are divided into five successive hidden blocks, i.e., Batch Normalization 1 ($\text{BN}_1$), Residual Block 1 ($\text{Res}_1$), $\text{Res}_2$, $\text{Res}_3$, and $\text{Res}_4$. A convolutional layer is followed by a batch normalization layer in $\text{BN}_1$. The rest of the hidden blocks are residual blocks, which are basic building blocks in a deep residual network model. 

{\color{black}
We also adopt Inception to build another image classifier based on the third version of the Inception Network (referred to as Inception-V3). The feature extraction layers of Inception-V3 are divided into seven hidden blocks: Stem block, Inception-A block (Inception-A), Reduction-A block (Reduction-A), Inception-B, Reduction-B, Inception-C, and global Avg-pool block. The Stem block is divided into seven successive hidden layers, including five convolution layers and two pool layers. An Inception Block consists of three parallel sub-blocks of convolution layers and pooling layers, whose outputs are later concatenated. The rest of the hidden blocks are Reduction blocks, which are made up of three parallel sub-blocks (two convolution layers and one pooling layer). The feature maps from the following hidden blocks of Inception-V3 are used as inputs to the proposed detector: Stem, Inception-A, Inception-C, Reduction-B, and Avg-pool.} 

\textbf{Datasets:} Three popular image datasets are considered: CIFAR-10, CIFAR-100, and SVHN. Each of the three image datasets is divided into three subsets: A training set of 49,000 images, a validation set of 1,000 images, and a testing set of 10,000 images.

\textbf{Attack models:} {\color{black}Seven latest adversarial attack strategies are considered: 
AutoAttack~\cite{DBLP:conf/icml/Croce020a},
FGSM~\cite{DBLP:journals/corr/GoodfellowSS14}, JSMA~\cite{DBLP:conf/eurosp/PapernotMJFCS16}, DeepFool~\cite{DBLP:conf/cvpr/Moosavi-Dezfooli16}, C\&W~\cite{DBLP:conf/sp/Carlini017}, PGD~\cite{DBLP:conf/iclr/MadryMSTV18}, and EAD~\cite{DBLP:conf/aaai/ChenSZYH18}; see Section~II-A. The neural network tool used in support of the defense algorithms is PyTorch, except for the case when PNDetector is taken as the defense technique. This is because the PNDetector is based on TensorFlow~\cite{DBLP:journals/corr/AbadiABBCCCDDDG16}. In this case, Cleverhans~\cite{Papernot2016CleverHans}, which supports Tensorflow, is used as the toolbox to support the attack strategies to produce adversarial samples against PNDetector. The {\color{black}other} parameter configurations of the attacks are summarized in Table~\ref{tab:attack_cfg}. }

\begin{table}
	\caption{\color{black}The hyper-parameter setting of the adversarial attacks considered. We employ the Adversarial Robustness Toolbox (ART)~\cite{https://doi.org/10.48550/arxiv.1807.01069}, Foolbox~\cite{rauber2017foolboxnative}, and TorchAttack~\cite{DBLP:journals/corr/abs-2010-01950} to launch the attacks.}
	\label{tab:attack_cfg}
	 \renewcommand{\arraystretch}{1.2}
      \renewcommand\tabcolsep{1pt}
	\centering
        \begin{tabular}{l|l|l}
		\hline
		Attack Algo. &Hyper-parameters & Tool\\
		\hline          
  	AutoAttack (8/255)&eps = 8/255 & TorchAttacks\\ 
        AutoAttack (0.02)&eps = 0.02 & TorchAttacks\\ 
		\hline
        FGSM (0.1)&eps = 0.1 & ART\\
        FGSM (8/255)&eps = 8/255 & ART\\
  	\hline
  	DeepFool& max\_iter = 50, overshoot=0.02& ART\\
		\hline
        JSMA (1.0, 0.1)& theta=1.0, gamma=0.1 &ART\\
        JSMA (0.8, 0.3)& theta=0.8, gamma=0.3 &ART\\
        \hline	
        PGD (0.02)&eps = 0.02, eps\_step = 0.002, & ART\\
        & max\_iter=10 &\\
        PGD (8/255)&eps = 8/255, eps\_step = 0.002, & ART\\
        & max\_iter=10 &\\
		\hline 
		C\&W&binary\_search\_steps = 5, steps = 1000, &Foolbox\\
        &stepsize = 0.01, confidence = 0.8, &\\
        &initial\_const = 0.1&\\
		\hline
        EAD & binary\_search\_steps = 9, steps = 1000, & Foolbox\\
        &confidence = 0.8, initial\_const = 0.1, & \\
        &regularization = 0.01, &\\
        &initial\_stepsize = 0.01, &\\
        &decision\_rule = `L1' &\\
        \hline
        \end{tabular}
\end{table}

{\color{black}Table~\ref{tab:examples} illustrates the adversarial examples generated by the latest attacks. All of the attacks can mislead ResNet-34 into misclassifying inputs, often with high confidence. The residuals in the third column of the table show that these attack algorithms cause minor perturbation to the benign images, and hence may evade human inspection.}
\begin{table}
     \caption{\color{black}Examples and noises on CIFAR-10 dataset under the latest attack algorithms. The DNN-based image classifier under attack is ResNet-34. The parameters of the attack algorithms are provided in Table~\ref{tab:attack_cfg}.}
     \label{tab:examples}
    \centering
    \begin{tabular}{m{5.2em}m{4em}m{4em}m{5em}m{5.5em}}
       \hline 
    Algorithm & Image& Residual & Prediction & Confidence~(\%)\\ 
    \hline 
        no attack&\adjustbox{center,valign=m,margin=1}{\includegraphics[height=0.5in]{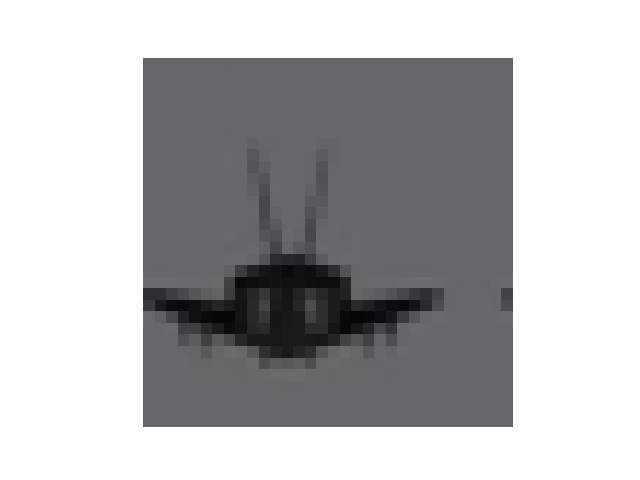}}&&airplane&90.59\\    
         \hline
        PGD (0.02)&\adjustbox{center,valign=m,margin=1}{\includegraphics[height=0.5in]{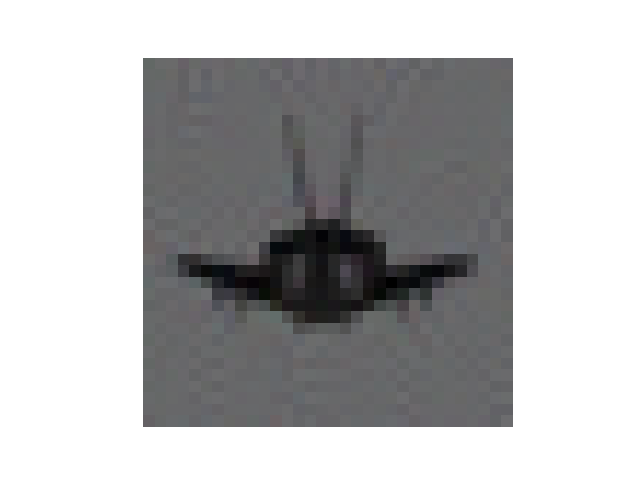}}&
        \adjustbox{center,valign=m,margin=1}{\includegraphics[height=0.5in]{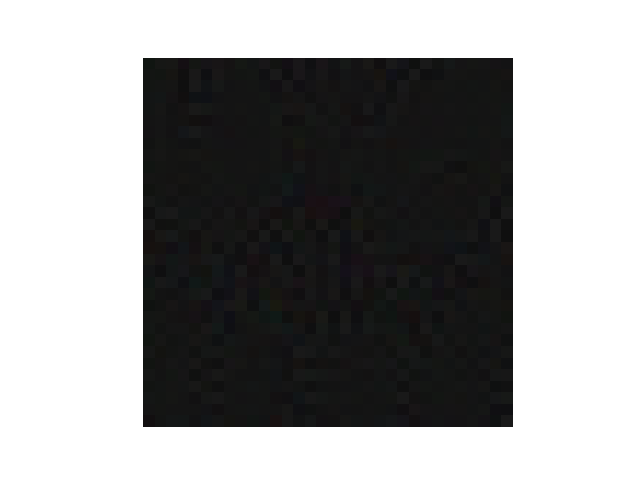}}&dog&67.56\\
        PGD (8/255)&\adjustbox{center,valign=m,margin=1}{\includegraphics[height=0.5in]{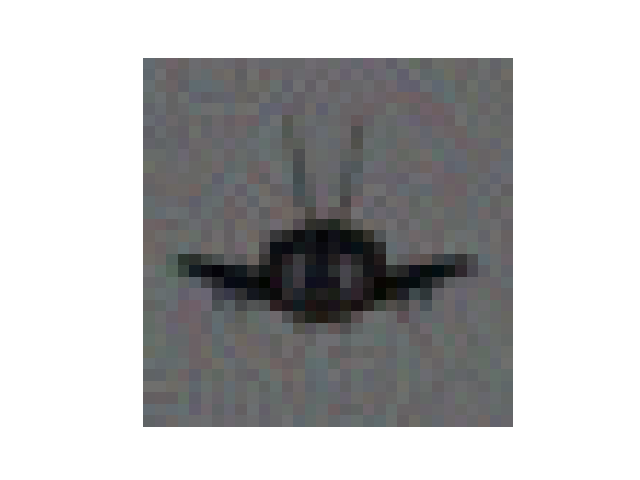}}&
        \adjustbox{center,valign=m,margin=1}{\includegraphics[height=0.5in]{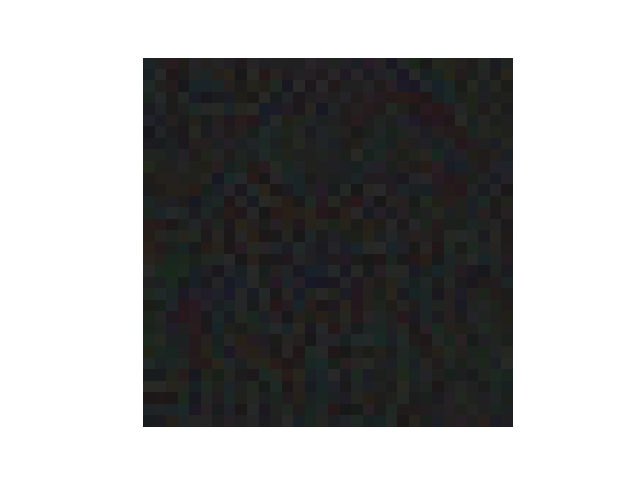}}&dog&91.15\\
        \hline
        Auto (8/255)&\adjustbox{center,valign=m,margin=1}{\includegraphics[height=0.5in]{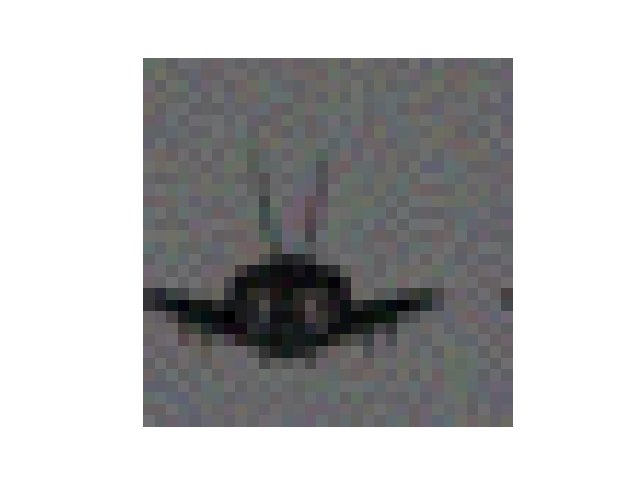}}&
        \adjustbox{center,valign=m,margin=1}{\includegraphics[height=0.5in]{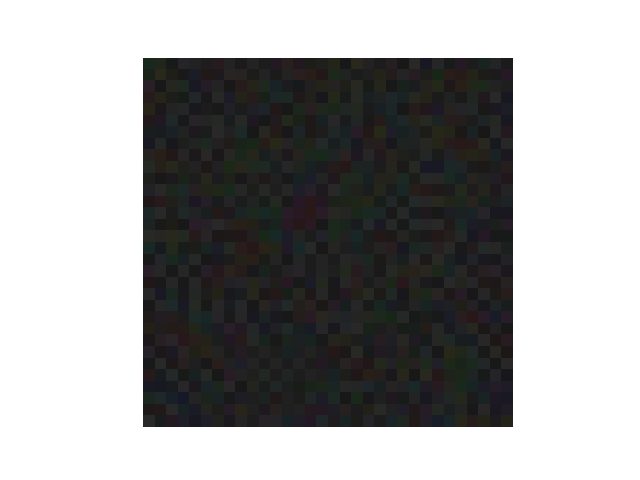}}&cat&91.13\\
        Auto (0.02)&\adjustbox{center,valign=m,margin=1}{\includegraphics[height=0.5in]{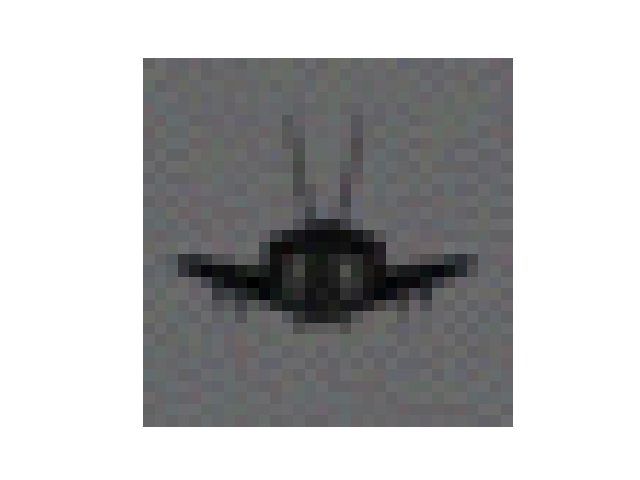}}&
        \adjustbox{center,valign=m,margin=1}{\includegraphics[height=0.5in]{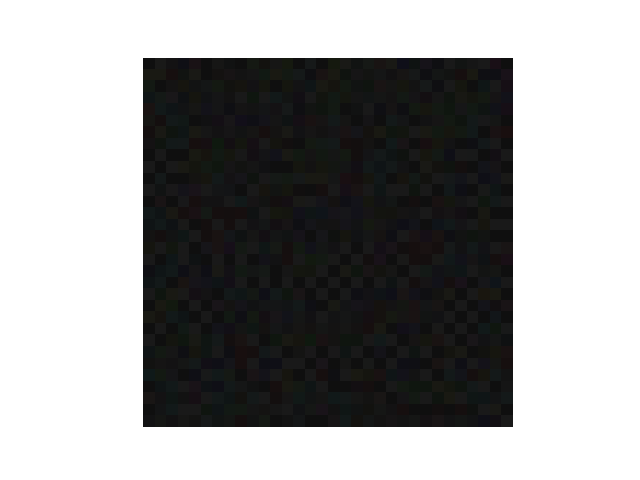}}&cat&91.57\\
        \hline
        EAD&\adjustbox{center,valign=m,margin=1}{\includegraphics[height=0.5in]{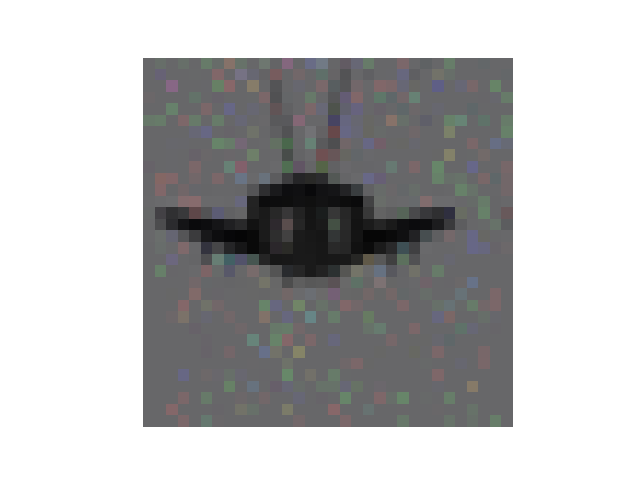}}&\adjustbox{center,valign=m,margin=1}{\includegraphics[height=0.5in]{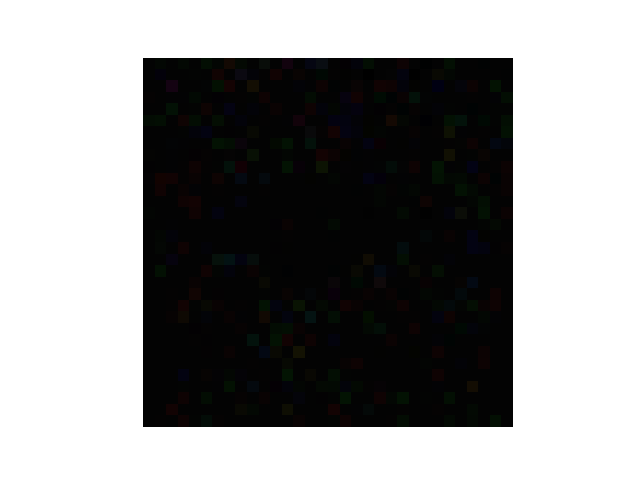}}&cat&77.44\\
        \hline
        FGSM (8/255)&\adjustbox{center,valign=m,margin=1}{\includegraphics[height=0.5in]{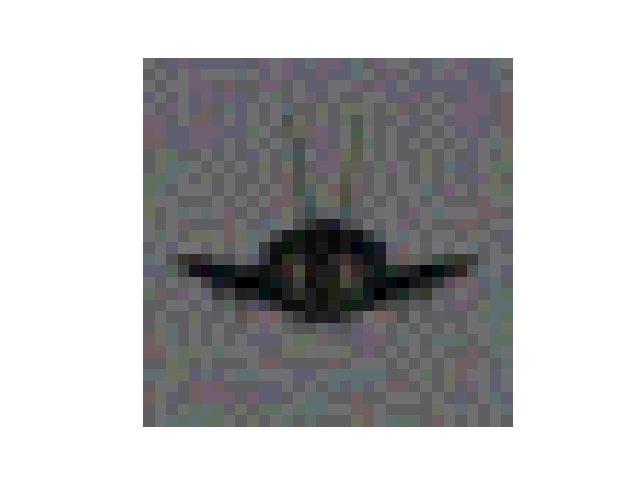}}&
        \adjustbox{center,valign=m,margin=1}{\includegraphics[height=0.5in]{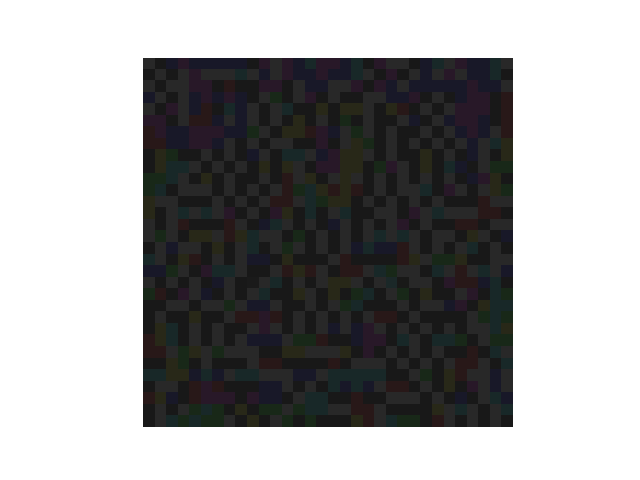}}&cat&91.87\\
        FGSM (0.1)&\adjustbox{center,valign=m,margin=1}{\includegraphics[height=0.5in]{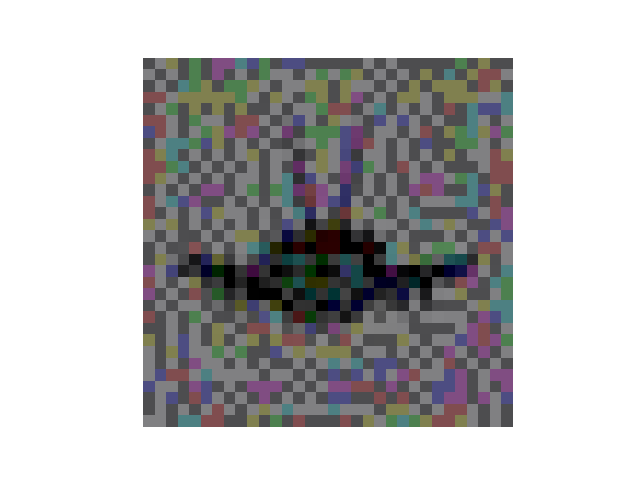}}&
        \adjustbox{center,valign=m,margin=1}{\includegraphics[height=0.5in]{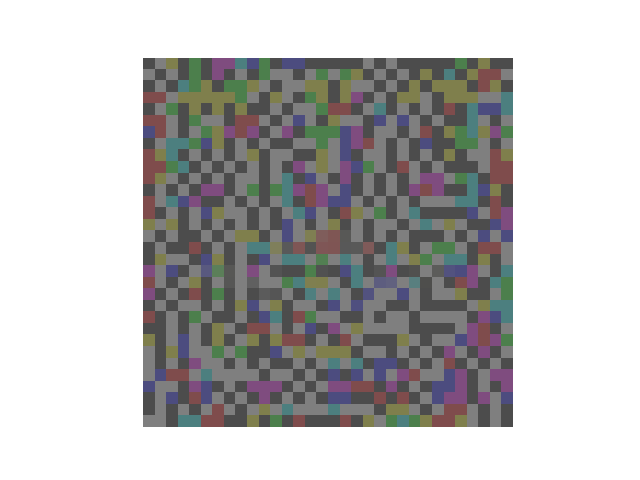}}&bird&86.29\\
        \hline
        C\&W&\adjustbox{center,valign=m,margin=1}{\includegraphics[height=0.5in]{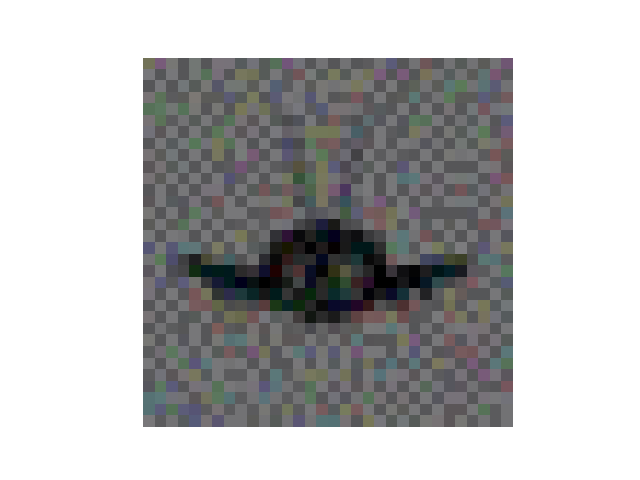}}&
        \adjustbox{center,valign=m,margin=1}{\includegraphics[height=0.5in]{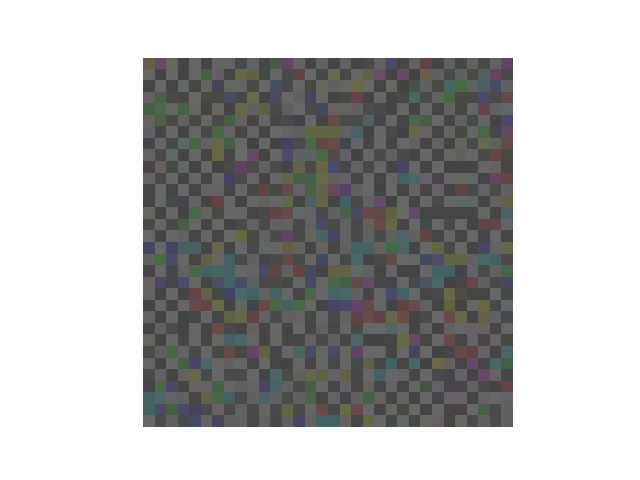}}&cat&52.86\\
        \hline
        DeepFool&\adjustbox{center,valign=m,margin=1}{\includegraphics[height=0.5in]{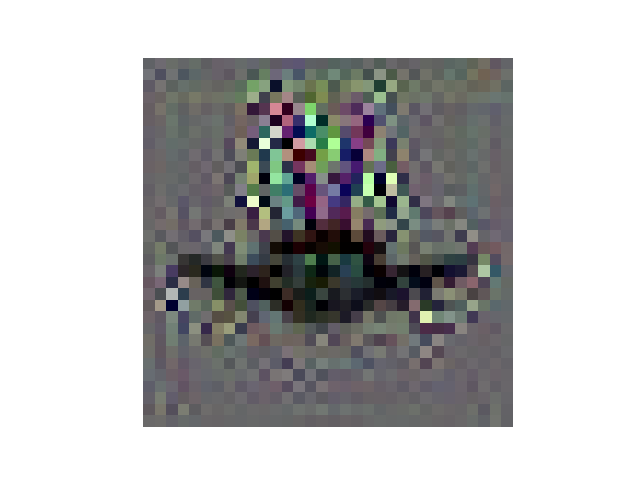}}&
        \adjustbox{center,valign=m,margin=1}{\includegraphics[height=0.5in]{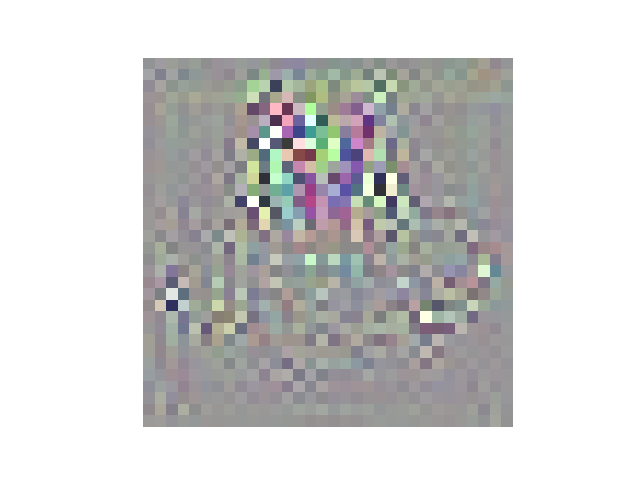}}&bird&90.81\\
        \hline
        JSMA (1.0,0.1)&\adjustbox{center,valign=m,margin=1}{\includegraphics[height=0.5in]{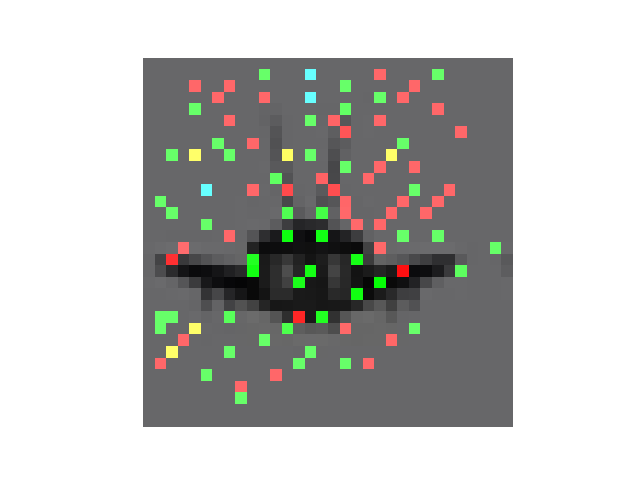}}&\adjustbox{center,valign=m,margin=1}{\includegraphics[height=0.5in]{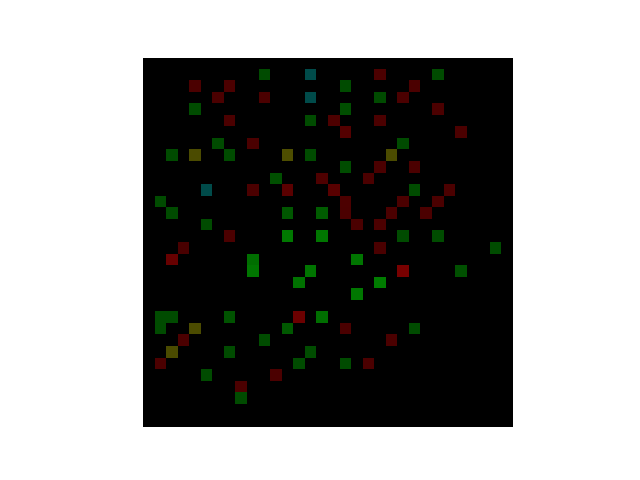}}&deer&44.31\\
        JSMA (0.8,0.3)&\adjustbox{center,valign=m,margin=1}{\includegraphics[height=0.5in]{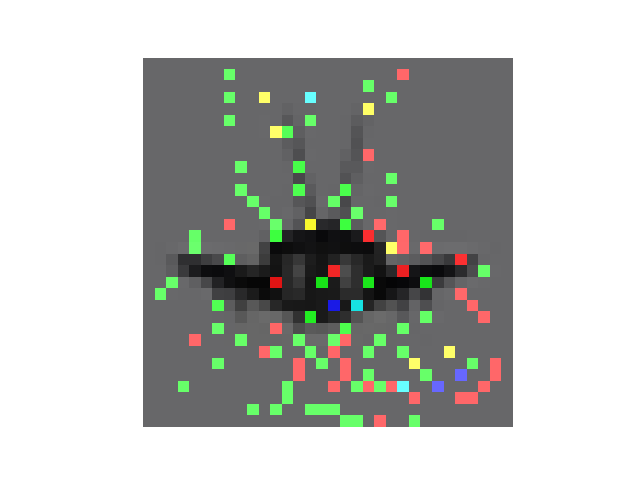}}&\adjustbox{center,valign=m,margin=1}{\includegraphics[height=0.5in]{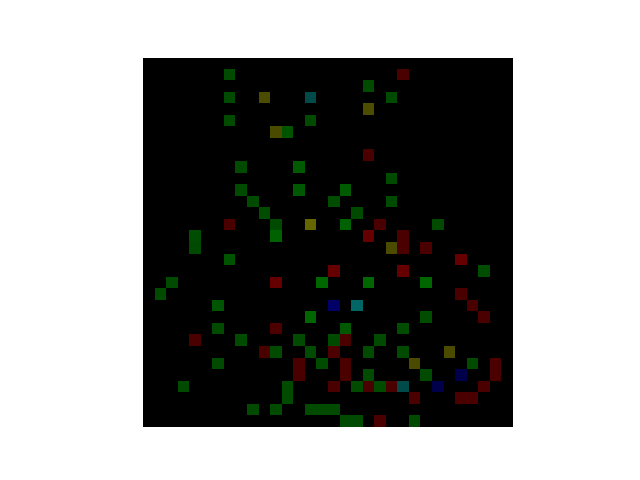}}&cat&36.27\\
        \hline
    \end{tabular}
    \label{tab: examples_and_noises}
\end{table}

\textbf{Performance indicator:} The area under receiver operating characteristic (ROC) curve, or ``AUC'', serves as a performance metric to evaluate the adversarial example detectors. The AUC of a model with 100\% incorrect predictions is 0. The AUC of a model with 100\% correct predictions is 1 (or 100\%). AUC is a useful tool. It assesses the accuracy of the model's predictions, regardless of classification threshold~\cite{DBLP:journals/tip/JingSYYN17}.

\textbf{State-of-the-art detectors:} The proposed detector is compared with the state-of-the-art adversarial example detectors, namely, LID~\cite{DBLP:conf/iclr/Ma0WEWSSHB18}, D$k$NN~\cite{DBLP:journals/corr/abs-1803-04765}, NNIF~\cite{DBLP:conf/cvpr/CohenSG20}, Mahalanobis~\cite{DBLP:conf/nips/LeeLLS18}, {\color{black}BEYOND~\cite{DBLP:journals/corr/abs-2209-00005}, and PNDetector~\cite{LUO2022108383},} as summarized in Section~\ref{sec: detectors}. The setups of the benchmark detectors are optimized under each of the considered attack models and datasets. Specifically, we optimize the number of neighbors, denoted by $k$, for BEYOND, D$k$NN and LID; the noise magnitude, denoted by $\gamma$, for the Mahalanobis algorithm; the number of high-influence samples, denoted by $H$, for the NNIF; and the false positive ratio (FPR) for the PNDetector. 
Based on the AUC values of the detection ROC curve, all hyper-parameters are validated with the validation set using nested cross entropy validation (except that the original hyper-parameters of NNIF in~\cite{DBLP:conf/cvpr/CohenSG20} are used because of significant time to re-train the NNIF, as also pointed out in~\cite{DBLP:conf/cvpr/CohenSG20}). The hyper-parameters of the benchmark detectors are summarized in Table~\ref{tab:detector_cfg}.

\begin{table}
	\caption{\color{black}The optimally chosen parameter values of the benchmark detectors. $k$ is the number of neighborhoods. $\gamma$ is the noise magnitude. FPR stands for false positive rate. }
	\label{tab:detector_cfg}    
	\scriptsize
 	 \renewcommand{\arraystretch}{1.2}
      \renewcommand\tabcolsep{0.8pt}
	\centering
    \begin{tabular}{l | l|c c c c c c c c c c c}
		\hline
		& &\multicolumn{8}{c}{ }\\
		&&\rotatebox[origin=c]{-90}{FGSM (0.1)} &\rotatebox[origin=c]{-90}{FGSM (8/255)} &\rotatebox[origin=c]{-90}{JSMA (1,0.1)} &\rotatebox[origin=c]{-90}{JSMA (0.8,0.3)} &\rotatebox[origin=c]{-90}{DeepFool}&\rotatebox[origin=c]{-90}{C\&W}&\rotatebox[origin=c]{-90}{PGD (0.02)} &\rotatebox[origin=c]{-90}{PGD (8/255)} &\rotatebox[origin=c]{-90}{EAD}& \rotatebox[origin=c]{-90}{Auto (8/255)} & \rotatebox[origin=c]{-90}{Auto (0.02)}\\
		\hline   
		\multirow{5}{*}{\rotatebox[origin=c]{90}{CIFAR-10}}&D$k$NN($k$)& 75 & 75 & 75 & 75 & 100 & 75 & 75 & 75 & 75 & 75 & 75\\
		&LID($k$)& 50 & 50 & 20 & 20 & 80 &50& 30 &  30 & 75 & 50& 50\\
		&Mahalanobis($\gamma$)& 0.0 & 0.0 & 0.01 & 0.01 & 0.0 & 0.001 & 0.01 & 0.01 & 0.01 & 0.0& 0.01\\
		&NNIF($H$)& 50 & 50 & 200 & 200 & 100 & 200 & 450 & 450 & 500 & --& --\\
        &PNDetector(FPR)& 0.2 & 0.2 & 0.2 & 0.2 & 0.2 & 0.2 & 0.2 & 0.2& 0.2 & 0.2& 0.2\\
        &BEYOND($k$)& 50 & 50 & 50 & 50 & 50 & 50 & 50 & 50 & 50 & 50 & 50\\
		\hline
		\multirow{5}{*}{\rotatebox[origin=c]{90}{SVHN}}&D$k$NN($k$)& 75 & 75 & 100 & 100 &  125 & 75 & 75 & 75 & 75 & 75& 75\\
		&LID($k$)& 60 & 60 & 30 & 30 & 30 &	40 & 40 & 40 & 30 & 40& 30\\
		&Mahalanobis($\gamma$)& 0.0014 & 0.0014 & 0.0 & 0.0 & 0.0 & 0.001 & 0.0 &	0.0 & 0.0 & 0.0& 0.0\\
		&NNIF($H$)& 300 & 300 & 50 & 50 & 300 & 50 & 100 & 100 & 100 & -- & --\\
         &PNDetector(FPR)& 0.05 & 0.05 & 0.05 & 0.05 & 0.05 & 0.05 & 0.05 & 0.05 & 0.05 & 0.05&0.05 \\
         &BEYOND($k$)& 55 & 55 & 55 & 55 & 55 & 55 & 55 & 55 & 55 & 55 & 55\\
		\hline
		\multirow{5}{*}{\rotatebox[origin=c]{90}{CIFAR-100}}&D$k$NN($k$)& 100 & 100 & 150 & 150 &	100  & 75 & 125 & 125 & 200& 75& 75\\
		&LID($k$)& 80 & 80 & 60 & 60 & 50 & 70 & 10 & 10 & 50 & 50& 90\\
		&Mahalanobis($\gamma$)& 0.01 & 0.01 &	0.005 & 0.005 &	0.0 & 0.01 & 0.0001 & 0.0001 & 0.0 & 0.005& 0.0014\\
		&NNIF($H$)& 30 & 30 & 30 & 30 & 40 & 40 & 50& 50 & 30 &-- &-- \\
        &PNDetector(FPR)&0.2 & 0.2 & 0.2 & 0.2 & 0.2 & 0.2 & 0.2 & 0.2& 0.2 & 0.2& 0.2\\
        &BEYOND($k$)& 50 & 50 & 50 & 50 & 50 & 50 & 50 & 50 & 50 & 50 & 50\\
		\hline
	\end{tabular}
\end{table}

\textbf{Setting of the proposed detector:} We select five hidden layers from the ResNet-34 model as inputs to the word embedding layer of our detector. Each one is the last layer of a hidden block in the ResNet-34 model (i.e., $\text{BN}_1$, $\text{Res}_1$, $\text{Res}_2$, $\text{Res}_3$, and $\text{Res}_4$). {\color{black}For the Inception-V3 model, we choose the last layers of its five hidden blocks (i.e., Stem, Inception-A, Inception-C, Reduction-B, and Avg-pool) as inputs to the word embedding layer of the proposed detector.} The size of the convolutional kernel used in the CP module is $3\times3$. We use 1-, 2-, 3- and 4-gram convolutional kernels in the sentiment analyzer of the proposed detector. Each of the convolutional kernels has 100 instances with randomly initialized parameters. The proposed detector is trained for 10 epochs to minimize the cross-entropy loss, using the Adam optimizer 
with a learning rate of 0.0001.

\subsection{Detection Performance}
We examine the performance of the proposed detector and the baseline detection algorithms in defending the considered latest attacks. As shown in Table~\ref{tab:auc}, the new detector consistently outperforms all the existing detectors on all the considered datasets and image classifiers. The table also shows that the proposed detector is effective in defending against the DeepFool and EAD attacks, both of which are particularly destructive on the CIFAR-100 dataset and invalidate all the existing detectors. Particularly, none of the existing detectors can provide an AUC (i.e., the detection rate) of over 91\%. {\color{black}In contrast, our new detector is able to achieve an AUC of over 94\% towards both the DeepFool and EAD attacks on the CIFAR-100 dataset when ResNet-34 model is deployed as the image classifier, and achieve an AUC of over 89\% towards these attacks when the Inception-V3 model is deployed as the image classifier. On the other hand, the proposed detector is computationally efficient. As shown in Table~\ref{tab:runtime}, it takes the detector shorter than $4.6$ milliseconds to detect an adversarial example, which is acceptable for many practical applications.}
 \begin{table*}[t]
	\caption{\color{black}The AUC scores (\%) of the considered adversarial example detection algorithms under the different attacks on different datasets. The adopted backbones of the image classifier are ResNet-34 and Inception-V3, respectively. }
	\label{tab:auc}
	\footnotesize
 	 \renewcommand{\arraystretch}{0.8}
      \renewcommand\tabcolsep{2.5pt}
	\centering
    \begin{tabular}{l|l | l|c c c c c c c c c c c}
		\hline
		Backdone&Dataset&Detector&\multicolumn{11}{c}{Attacking Algorithms}\\
  &&&\rotatebox[origin=c]{-90}{FGSM (0.1)} &\rotatebox[origin=c]{-90}{FGSM (8/255)} &\rotatebox[origin=c]{-90}{JSMA (1,0.1)} &\rotatebox[origin=c]{-90}{JSMA (0.8,0.3)} &\rotatebox[origin=c]{-90}{DeepFool}&\rotatebox[origin=c]{-90}{C\&W}&\rotatebox[origin=c]{-90}{PGD (0.02)} &\rotatebox[origin=c]{-90}{PGD (8/255)} &\rotatebox[origin=c]{-90}{EAD}& \rotatebox[origin=c]{-90}{Auto (8/255)} & \rotatebox[origin=c]{-90}{Auto (0.02)}\\
		\cline{1-14}   
		&\multirow{5}{*}{\rotatebox[origin=c]{90}{CIFAR-10}}&D$k$NN& 80.35&68.17&76.87&80.64&76.91&86.12&84.93&86.80&76.61&83.24&84.57\\
		&&LID& 99.99 & 97.51 &97.87&98.62&96.08&99.92&72.91& 86.51&94.06&92.74&88.89\\
		&&Mahalanobis& \textbf{100}&98.30 &88.58&90.44&93.01&99.79&88.47&97.84&79.53&98.63&97.28\\
		&&NNIF& 99.96 & -- & 99.50& -- & 99.32& 99.50& 98.31& -- & 95.09& --&--\\
        &&PNDetector& 86.24& 82.24& 97.79& 97.63& 96.89& 97.84& 48.37& 41.20& 82.85& 33.79&38.72\\
        &&BEYOND & 94.02 & 90.55 & 93.89 & 91.78 & 95.08 & 95.13 & 94.53 & 95.09 & 90.90 &95.57 & 94.79\\
		&&Proposed& \textbf{100}&\textbf{100}& \textbf{100}&99.99&\textbf{99.43}&\textbf{100}&\textbf{99.75}& \textbf{100}&\textbf{99.47}&\textbf{100}&\textbf{99.99}\\
		\cline{2-14}  
		&\multirow{5}{*}{\rotatebox[origin=c]{90}{SVHN}}&D$k$NN& 90.49 & 73.33&82.50&83.40& 87.34& 91.95& 	87.65& 	91.04&77.61 &79.07&91.26\\
		&&LID& 99.97 & 95.68 &	96.49 &96.52&	95.95 &	99.68 &	77.79 &89.81	&91.68 &92.80&88.83\\
		&&Mahalanobis&96.66 &88.50 & 95.96 &95.62&91.11 &90.84 &84.07 &	88.24&91.37 &92.21&92.59\\
		ResNet&&NNIF& \textbf{100}& -- & 99.76& --& 99.06& 99.59& 96.18&-- &97.40& --&--\\
         &&PNDetector& 96.43& 85.63& 97.88& 98.35& 98.76& 99.32& 81.66& 68.11&88.11& 51.68&64.91\\
         &&BEYOND & 91.18 & 87.93 & 90.78 & 92.03 & 96.01 & 95.72 & 90.94 & 91.00 & 89.65 & 92.20 & 91.24 \\
		&&Proposed& \textbf{100}& \textbf{99.90}& \textbf{100}& \textbf{100}& \textbf{99.53}& \textbf{100}& \textbf{99.56}& \textbf{100}&\textbf{99.73}&\textbf{100}&\textbf{99.97}\\
		\cline{2-14}
		&\multirow{5}{*}{\rotatebox[origin=c]{90}{CIFAR-100}}&D$k$NN& 77.67 &70.06 &74.48 & 78.03&	75.90 &	74.55 &	79.17 &	78.61&77.77 &67.30&68.93\\
		&&LID& 99.95 & 95.34 & 87.34 & 90.17& 	61.05& 	99.34 & 	73.90& 	83.66 &54.01 &95.02&92.69\\
		&&Mahalanobis& 99.83 &94.42 &	95.97& 96.76&	65.99 &	97.95 &	76.57 &	90.02 &60.82 &89.29& 87.81\\
		&&NNIF& 99.96& -- &97.50 & --& 77.17& 96.51& 96.60&-- &74.86& --&--\\
        &&PNDetector& 74.31& 68.24& 79.25& 80.63&90.54& 90.48& \textbf{22.96}&\textbf{13.28}& 79.64&70.08&67.46 \\
        &&BEYOND & 91.39 & 84.03 & 87.35 & 91.05 & 89.39 & 92.93 & 90.32 & 91.14 & 85.10 & 92.07 & 91.21  \\
		&&Proposed& \textbf{100}& \textbf{100}&\textbf{99.99}& \textbf{100}& \textbf{94.52}& \textbf{100}& \textbf{99.50}& \textbf{100}&\textbf{94.39}&\textbf{100}&\textbf{100}\\
		\hline
		&\multirow{5}{*}{\rotatebox[origin=c]{90}{CIFAR-10}}&D$k$NN& 75.16&63.29 &76.94 &77.61 &68.76 &77.84 &77.34 &79.39 &73.82&78.07&77.24\\
		&&LID& 99.99& 98.35& 97.64& 98.13&94.97 &97.09 &76.76 &91.04 &94.36&96.62&93.27\\
		&&Mahalanobis& \textbf{100}&99.92 &99.24 &99.40 &97.06 &98.15 &86.99 &98.85 &96.58&99.71&98.70\\
        &&PNDetector&85.55 &82.17 &92.55 &95.25 &93.71 &96.37 &72.68 &46.98 &95.62&42.11&58.73\\
        &&BEYOND & 96.50 & 93.81 & 95.04 & 94.76 & 94.79 & 93.06 & 93.75 & 94.46 & 93.95 & 94.58 & 93.96\\
		&&Proposed& \textbf{100}& \textbf{100}& \textbf{99.98}& \textbf{99.98}& \textbf{99.92}& \textbf{99.36}& \textbf{98.63}& \textbf{99.95}& \textbf{99.29}&\textbf{100}& \textbf{99.96}\\
		\cline{2-14}
		&\multirow{5}{*}{\rotatebox[origin=c]{90}{SVHN}}&D$k$NN&87.72 &75.79 &76.69 &75.81 &81.13 &84.91 &84.45 &87.62 &82.18&85.89&87.30\\
		&&LID& 99.91&89.29 &96.45 &96.68 &92.62 &98.10 &79.65 &84.89 &92.35&89.96&88.62\\
		Inception&&Mahalanobis&99.90 &93.89 &98.68 &98.74 &96.45 &99.01 &84.00 &90.90 &96.93&96.06&94.13\\
        &&PNDetector&97.27 &92.11 &98.34 &98.30 &97.89 &99.07 &74.46 &62.93 &78.54&45.69&52.39\\
        &&BEYOND & 93.91 & 92.28 & 92.59 & 92.62 & 92.95 & 94.81 & 94.56 & 94.45 & 94.79 & 92.10 & 91.46\\
		&&Proposed& \textbf{100}& \textbf{99.79}& \textbf{99.95}& \textbf{99.95}& \textbf{99.32}& \textbf{99.28}& \textbf{95.89}& \textbf{99.81}& \textbf{99.12}& \textbf{99.93}& \textbf{99.74}\\
		\cline{2-14}
		&\multirow{5}{*}{\rotatebox[origin=c]{90}{CIFAR-\textbf{100}}}&D$k$NN&81.11 &74.71 &77.45 &77.99 &78.87 &90.40 &81.70 &79.65 &78.13&68.74&66.00\\
		&&LID&99.89 &92.87 &88.71 &88.42 &69.39 &82.83 &73.12 &81.40 &62.64&98.44&97.15\\
		&&Mahalanobis&99.99 &99.68 &99.31 &99.47 &87.46 &95.70 &83.99 &96.65 &85.15&99.63&98.60\\
        &&PNDetector&69.93 &70.01 &80.05 &83.63 &89.41 &89.86 &\textbf{20.08} &\textbf{12.33} &85.56&70.86&65.18\\
        &&BEYOND & 90.76 & 88.23 & 94.20 & 94.88 & 88.94 & 94.09 & 94.45 & 98.14 & 87.62 & 94.36 & 94.90\\
		&&Proposed& \textbf{100}& \textbf{100}&\textbf{99.97} &\textbf{99.99} & \textbf{89.52}&\textbf{96.60} & \textbf{96.77}& \textbf{99.89}&\textbf{89.73}&\textbf{100}&\textbf{99.98}\\
		\hline
	\end{tabular}
\end{table*}

\begin{table}[t]
	\caption{\color{black}The running time (in milliseconds) of our detector to detect an adversarial example on a Tesla K80 GPU card.}
	\label{tab:runtime}
     \renewcommand{\arraystretch}{1}
    \renewcommand\tabcolsep{1pt}
    \scriptsize
    \centering
    \begin{tabular}{c|c|c c c c c c c c c c c |c}
    \hline
     Model&Dataset &	\rotatebox[origin=c]{-90}{FGSM (0.1)} &\rotatebox[origin=c]{-90}{FGSM (8/255)} &\rotatebox[origin=c]{-90}{JSMA (1,0.1)} &\rotatebox[origin=c]{-90}{JSMA (0.8,0.3)} &\rotatebox[origin=c]{-90}{DeepFool}&\rotatebox[origin=c]{-90}{C\&W}&\rotatebox[origin=c]{-90}{PGD (0.02)} &\rotatebox[origin=c]{-90}{PGD (8/255)} &\rotatebox[origin=c]{-90}{EAD}& \rotatebox[origin=c]{-90}{Auto (8/255)} & \rotatebox[origin=c]{-90}{Auto (0.02)}&\rotatebox[origin=c]{-90}{Average}\\
     \hline
         &CIFAR-10&4.74 &	4.32 &	4.91 &	4.18 &	4.67 &	4.78 &	4.45& 	4.35 &	4.86& 	4.24& 	4.28 &	4.53 \\
     ResNet-34&SVHN&4.32 &	4.27& 	4.90 &	4.31& 	4.56 &	4.51 &	4.77 &	4.16 &	4.46 &	4.29 &	4.27 &	4.44\\
         &CIFAR-100&4.75 &	4.27 &	4.73& 	4.26& 	4.31 &	4.70 &	4.45& 	4.23& 	4.65& 	4.25&	4.23 &	4.44 \\
    \hline
&CIFAR-10  &	2.45	&3.25	&3.13&3.45	&3.03	&3.28&	2.41&	2.33&	2.16&	1.89&	1.91&	2.66\\
Inception-V3&SVHN&	2.97	&2.66	&2.54	&2.45	&2.18&	2.73	&1.93	&2.95&	2.17&	2.53&	2.46&	2.51\\
&CIFAR-100  &	2.45&	3.90	&2.57&	3.05&	3.46&	3.62&	3.52&	2.49&	2.38&	2.17&	1.85&	2.86\\

\hline
    
    \end{tabular}
\end{table}

\subsection{Visualization of the New Detector}
\label{sec: visual}
We use t-distributed stochastic neighbor embedding (t-SNE)~\cite{9064929} to visualize each word vector in a sentence generated as described in Section~\ref{sec: sen_gen}. Here, t-SNE is a statistical tool that can visualize high-dimensional data by assigning a position in a low-dimensional map to each data point. Related objects are portrayed as close points, and dissimilar objects are represented by distant points. 

Consider the DeepFool attack for its popularity and hard-to-detect property~\cite{DBLP:conf/cvpr/CohenSG20}. The visualization results based on DeepFool are shown in Table~\ref{tbl:table_of_sne_all_adv}, where red and blue points correspond to adversarial and benign examples, respectively. Here, ResNet-34 serves as the image classifier. the feature maps output by the last layer of the hidden blocks, i.e., $\text{BN}_1$, $\text{Res}_1$, $\text{Res}_2$, $\text{Res}_3$, and $\text{Res}_4$, are considered. 

\begin{table*}[t]
	\caption{The t-SNE visualization of the word distribution generated from each selected hidden block of ResNet-34 under the DeepFool attack. Each point represents a word, corresponding to either an adversarial (red) or a benign example (blue).}
	\label{tbl:table_of_sne_all_adv}
	\centering
	\begin{tabular}{m{2cm}<{\centering}m{2.5cm}<{\centering}m{2.5cm}<{\centering}m{2.5cm}<{\centering}m{2.5cm}<{\centering}m{2.5cm}<{\centering}}
		\toprule
		Dataset & $\text{BN}_1$ & $\text{Res}_1$ & $\text{Res}_2$ & $\text{Res}_3$ & $\text{Res}_4$ \\
		\midrule
		CIFAR-10 & 
		\includegraphics[height=1in]{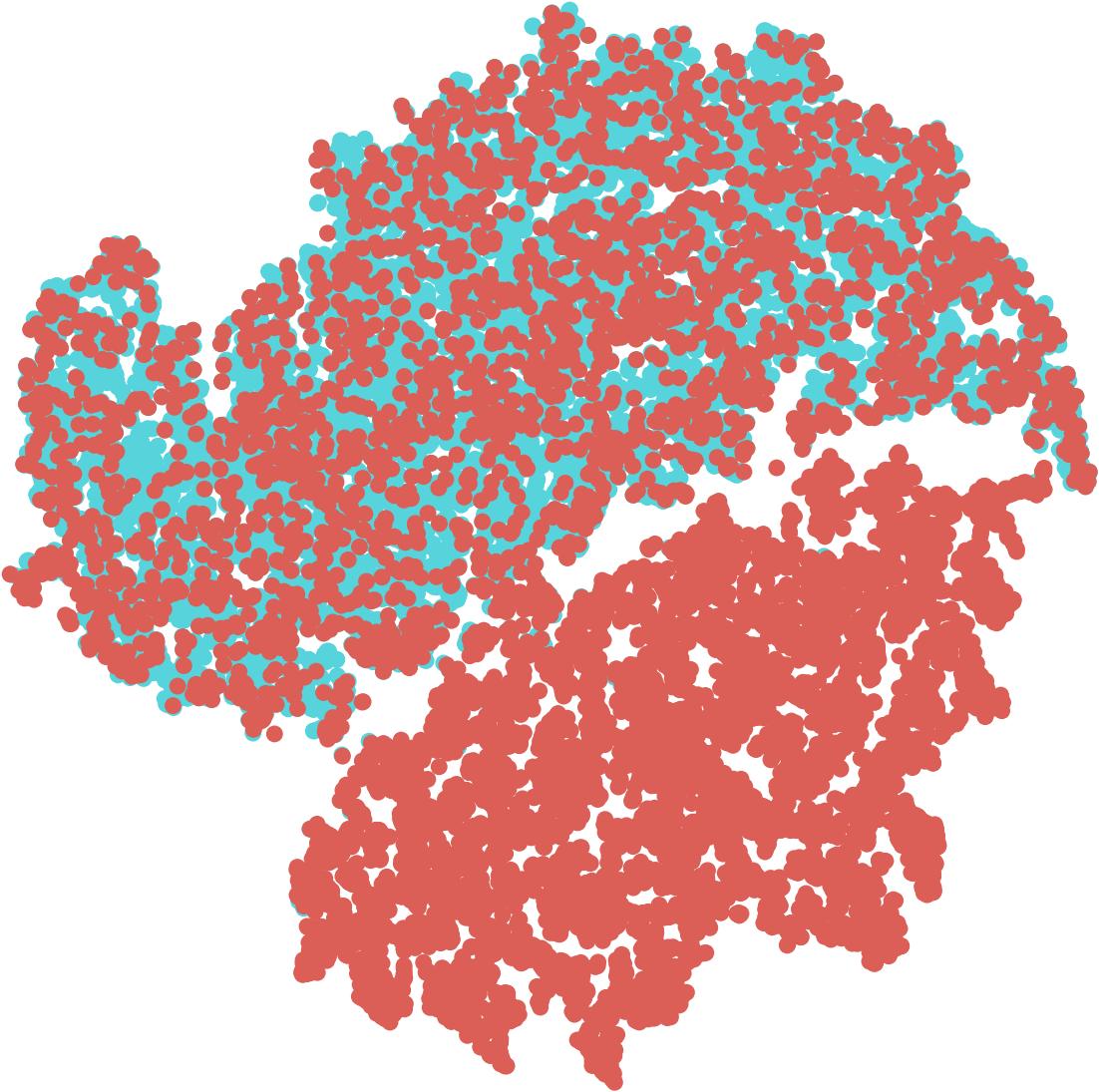} & \includegraphics[height=1in]{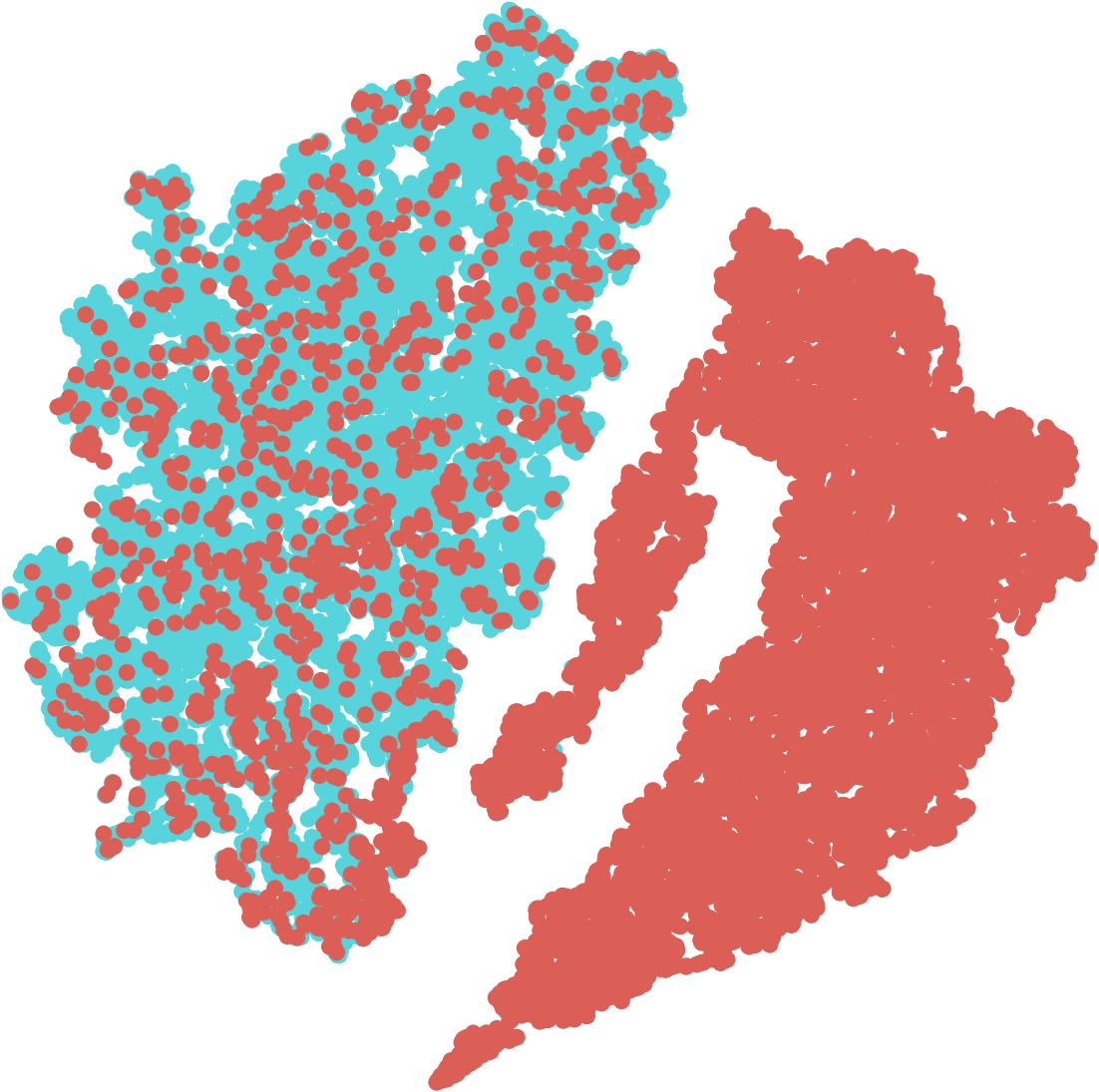} & \includegraphics[height=1in]{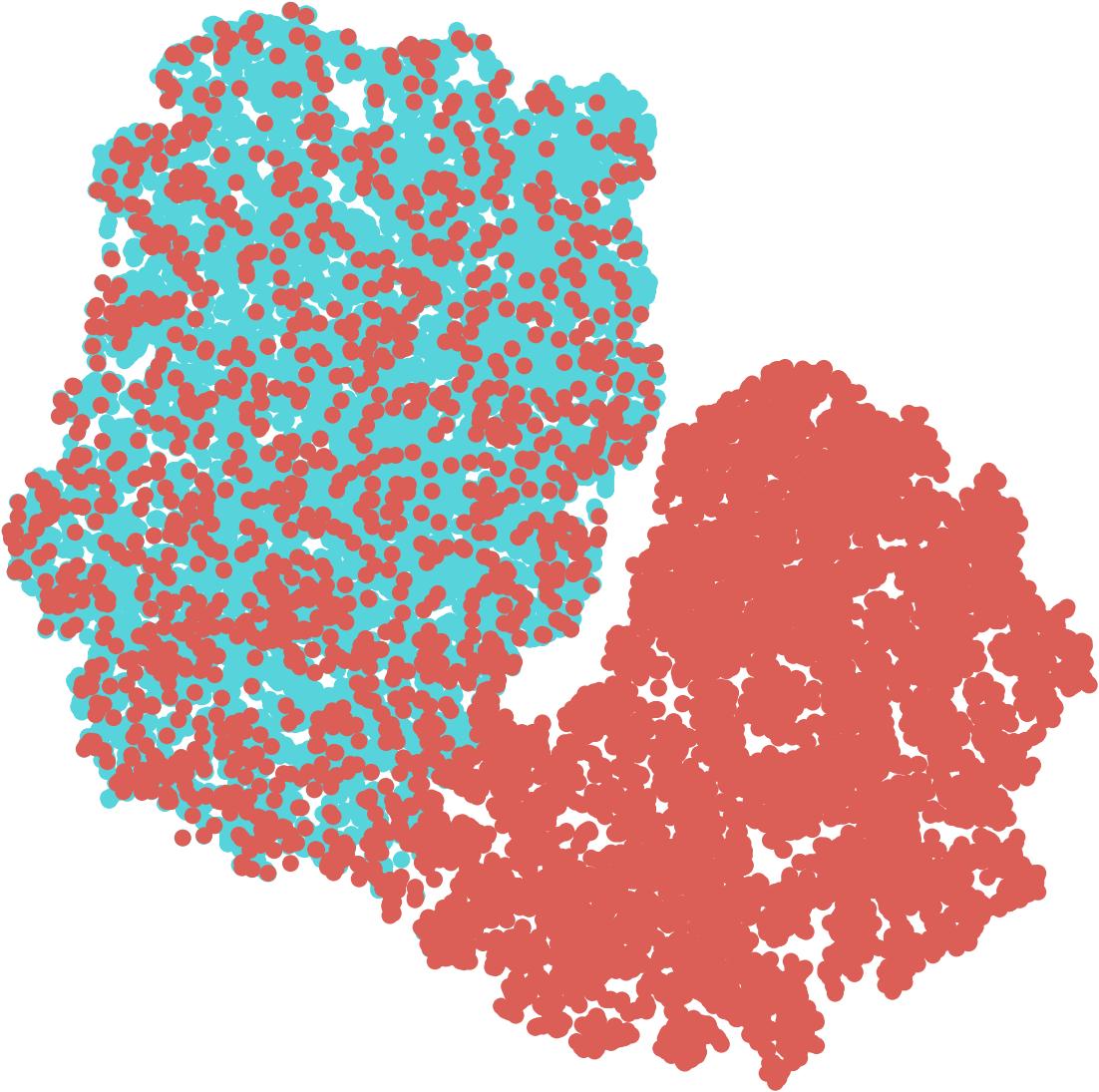} & \includegraphics[height=1in]{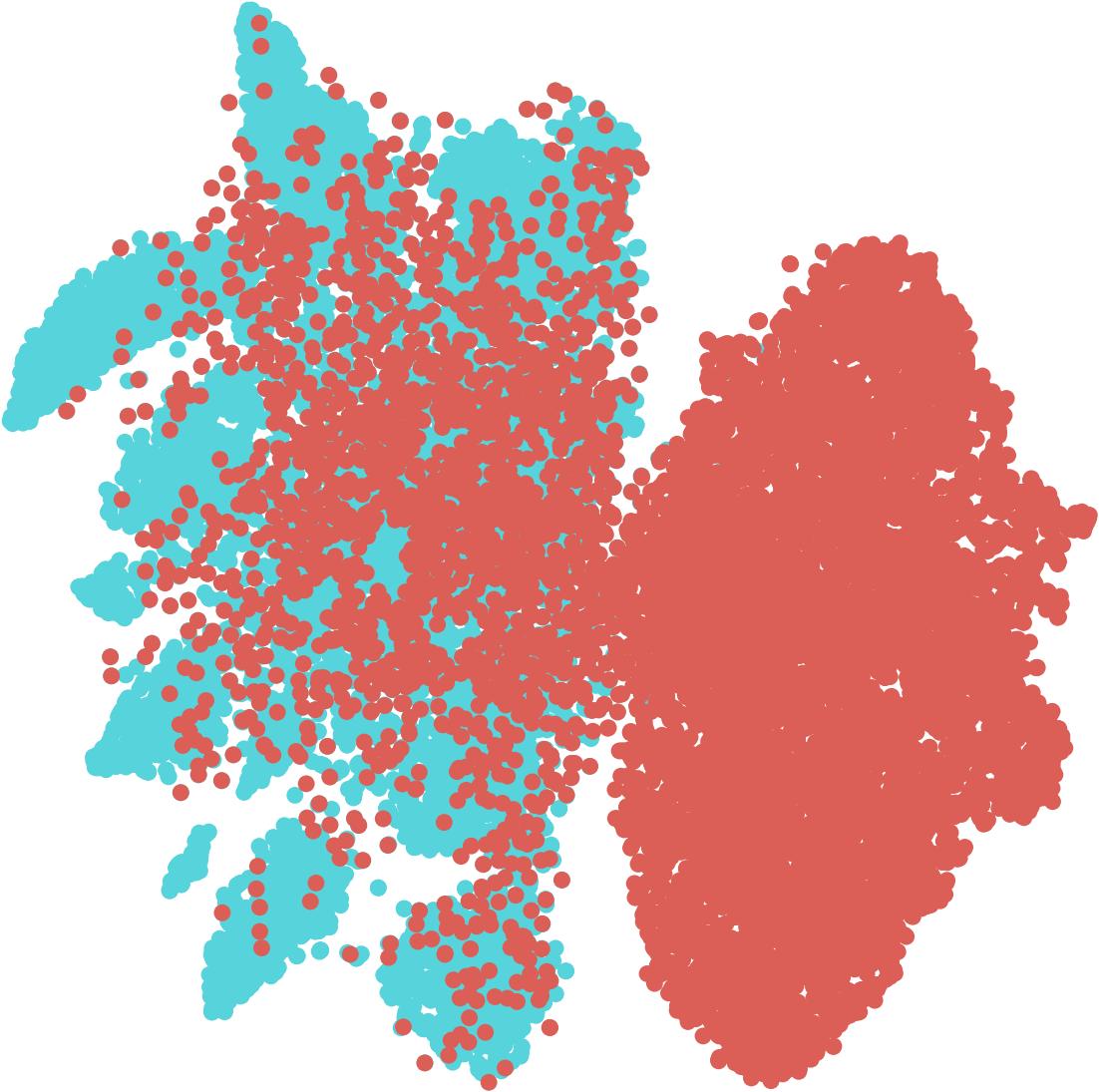} & \includegraphics[height=1in]{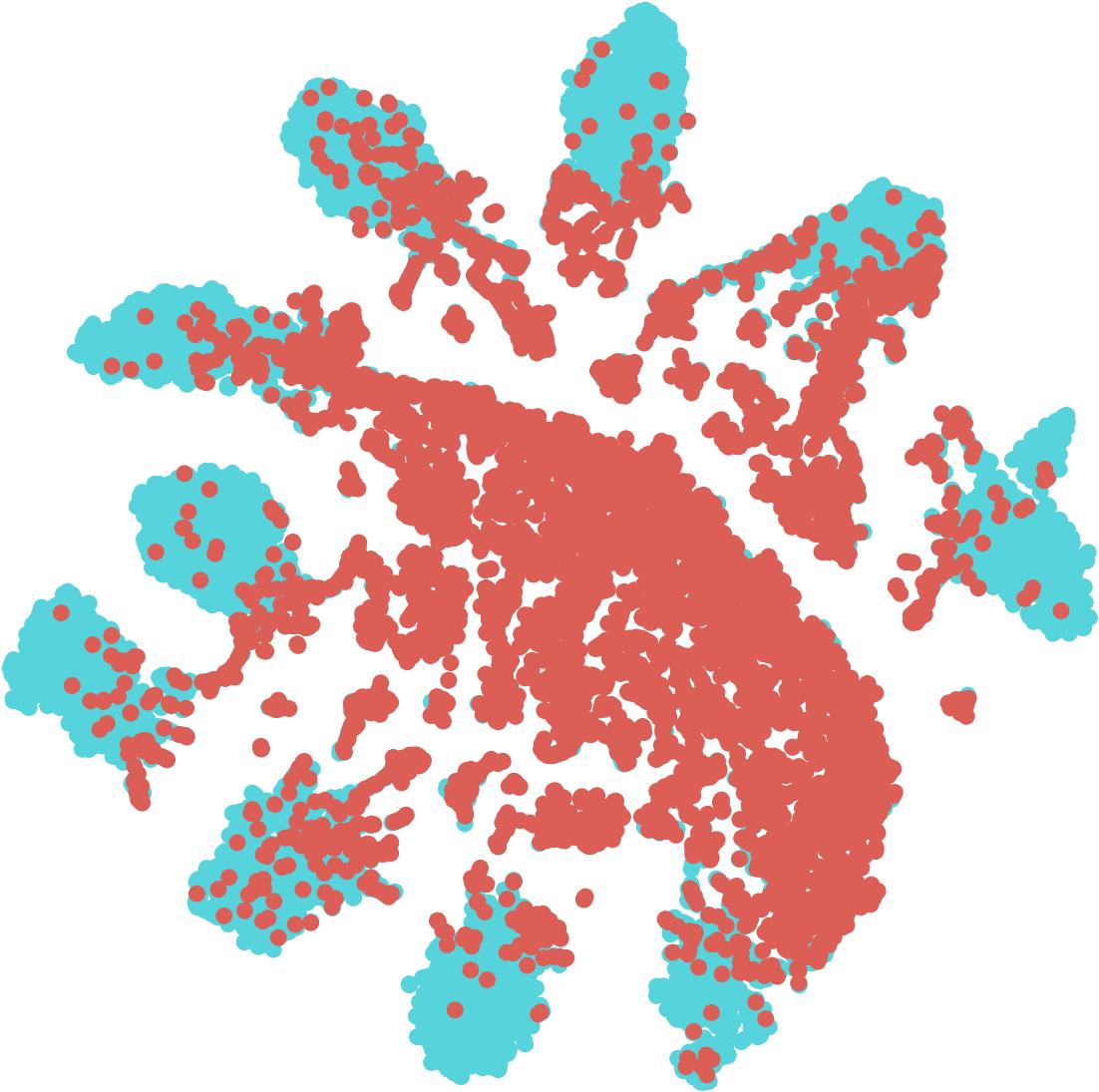}\\
		\\
		SVHN &  
		\includegraphics[height=1in]{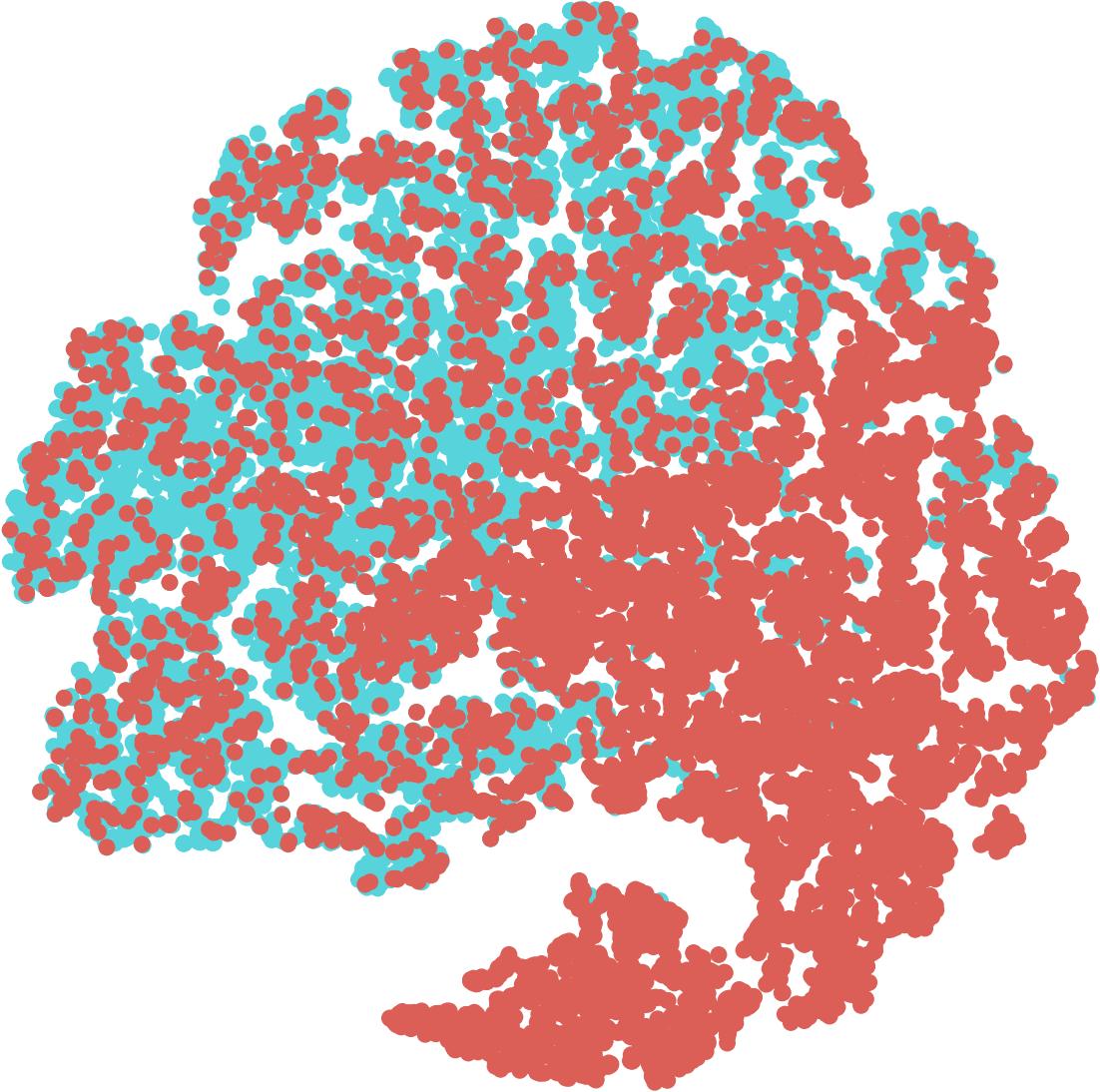} & \includegraphics[height=1in]{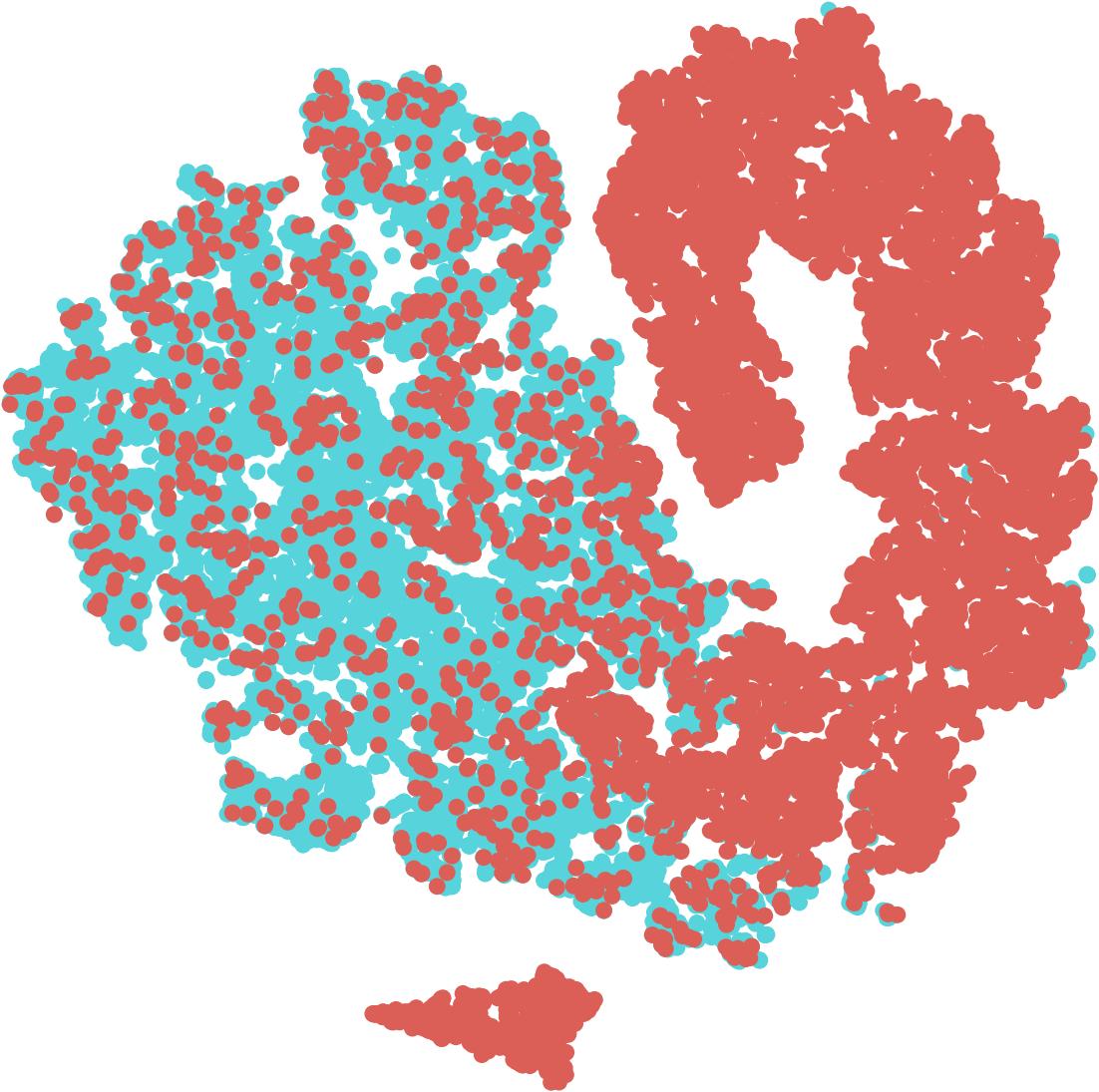} & \includegraphics[height=1in]{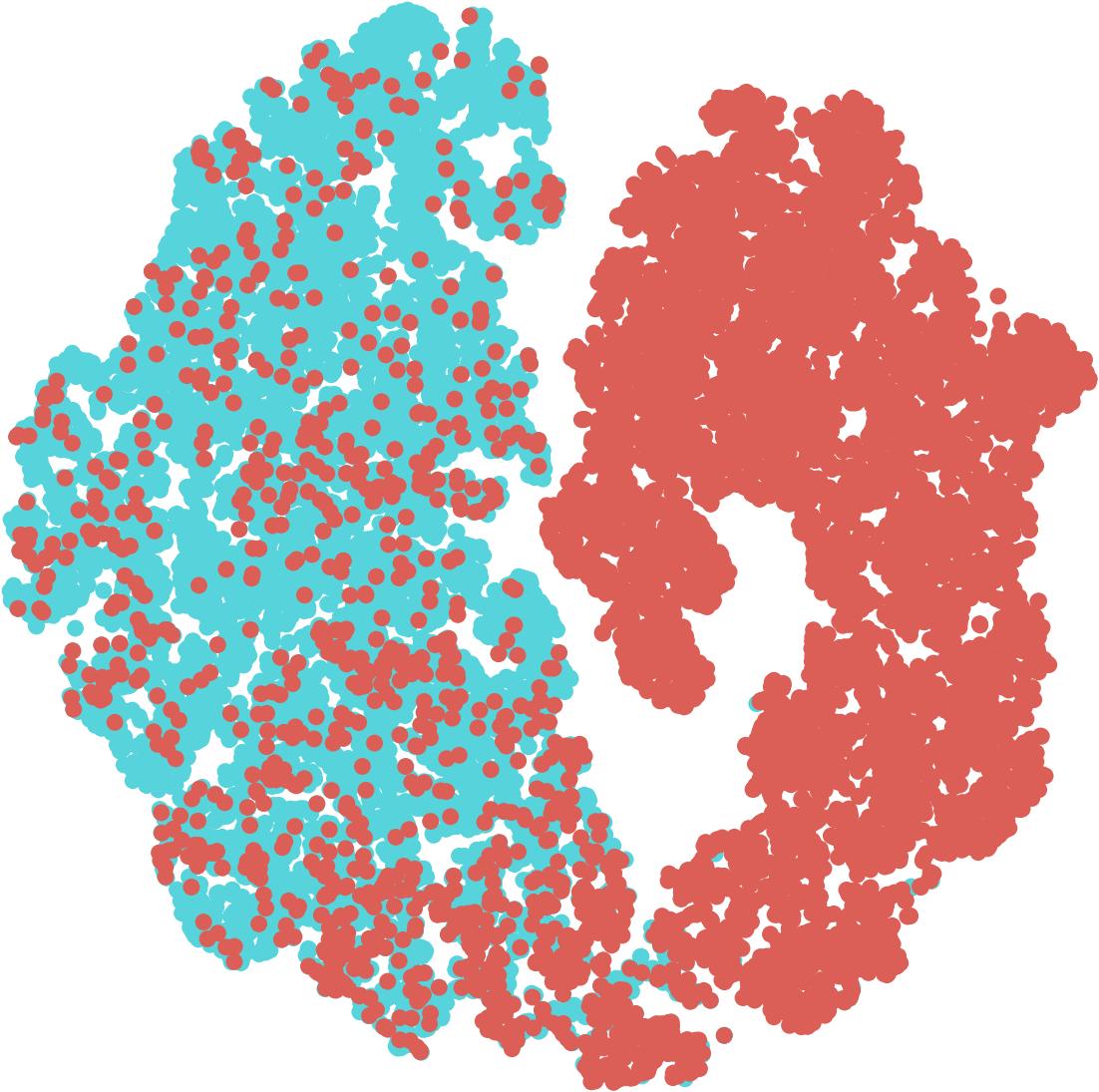} & \includegraphics[height=1in]{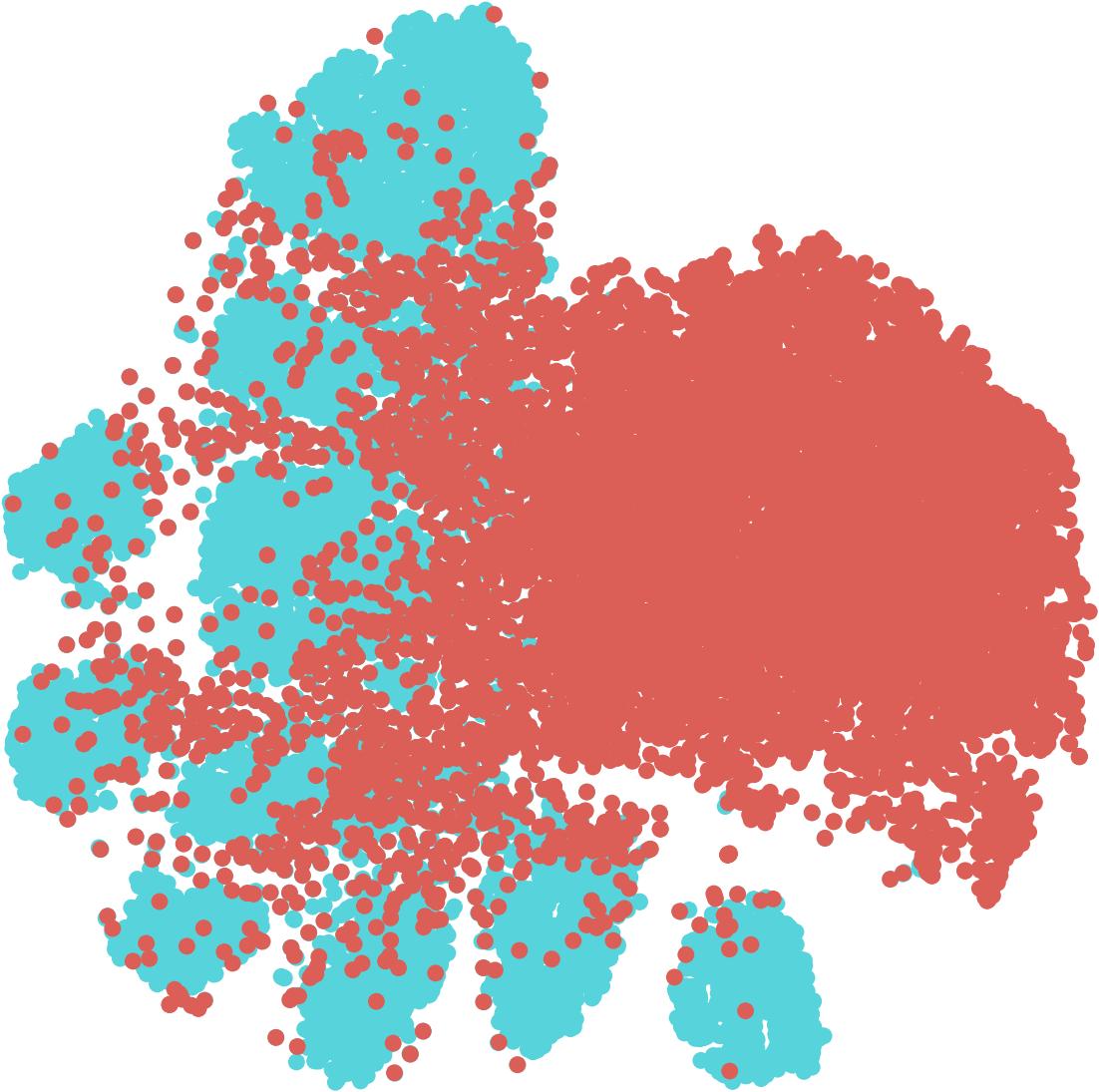} & \includegraphics[height=1in]{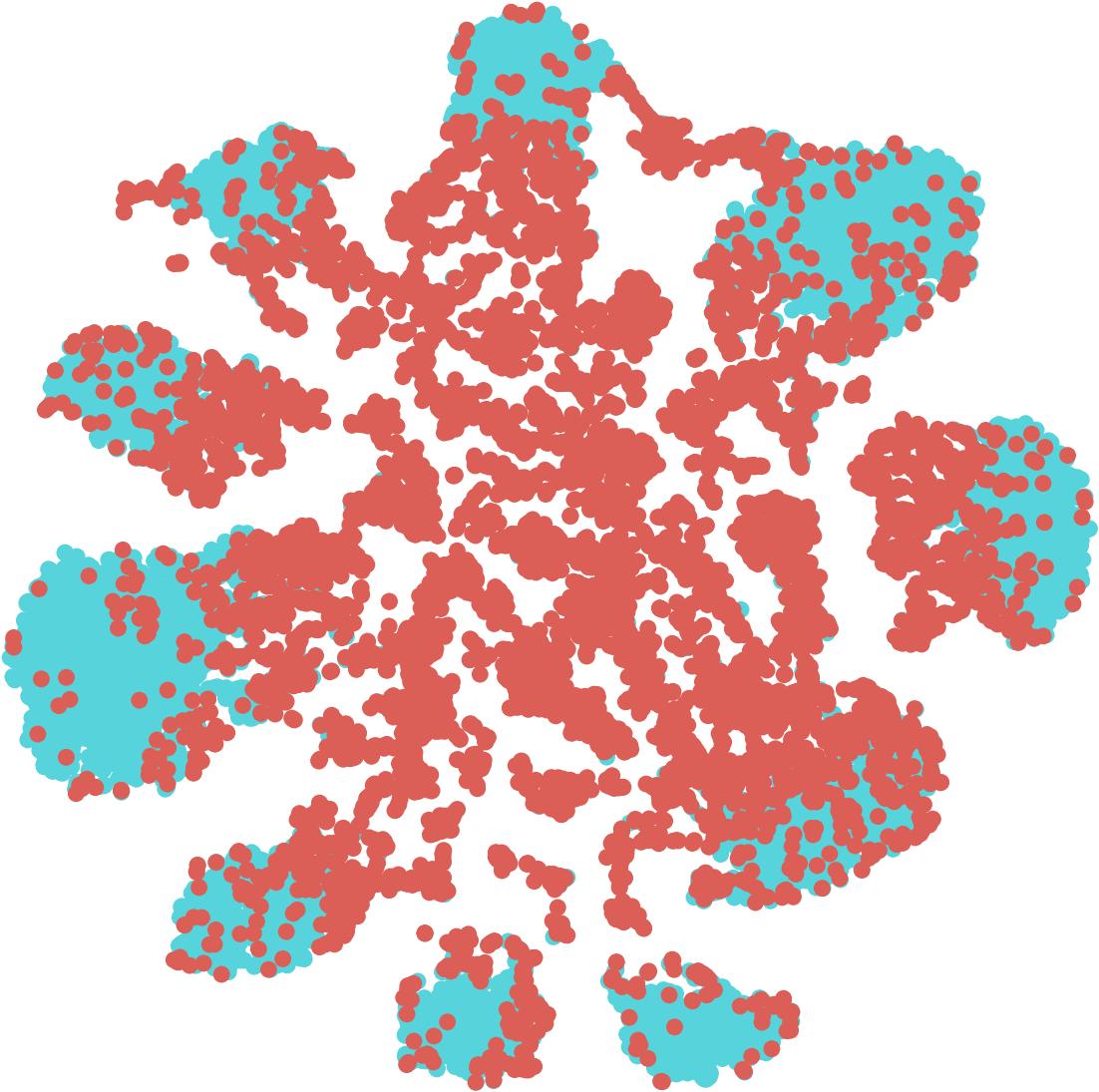}\\
		\\
		CIFAR-100 &  \includegraphics[height=1in]{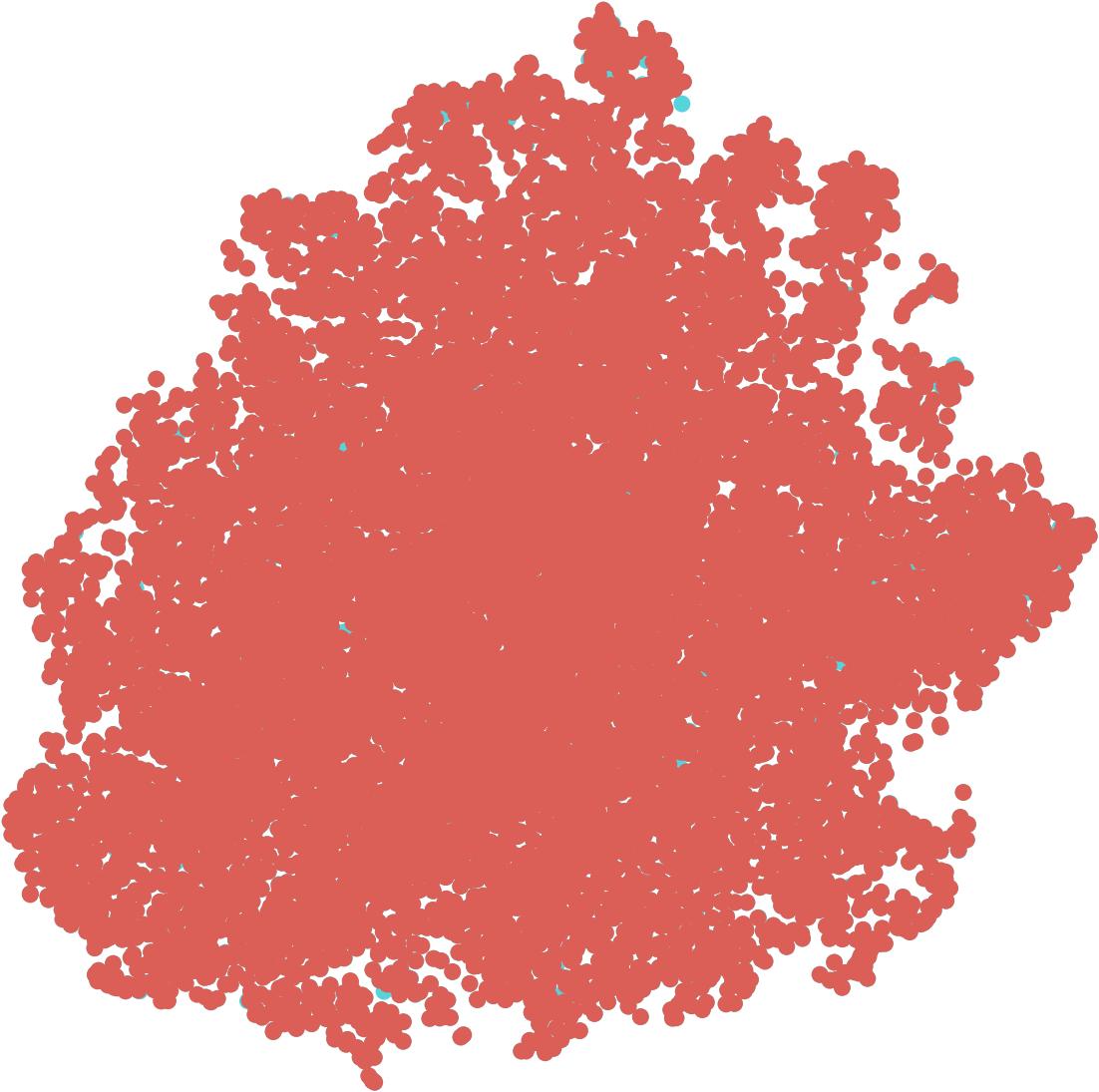} & \includegraphics[height=1in]{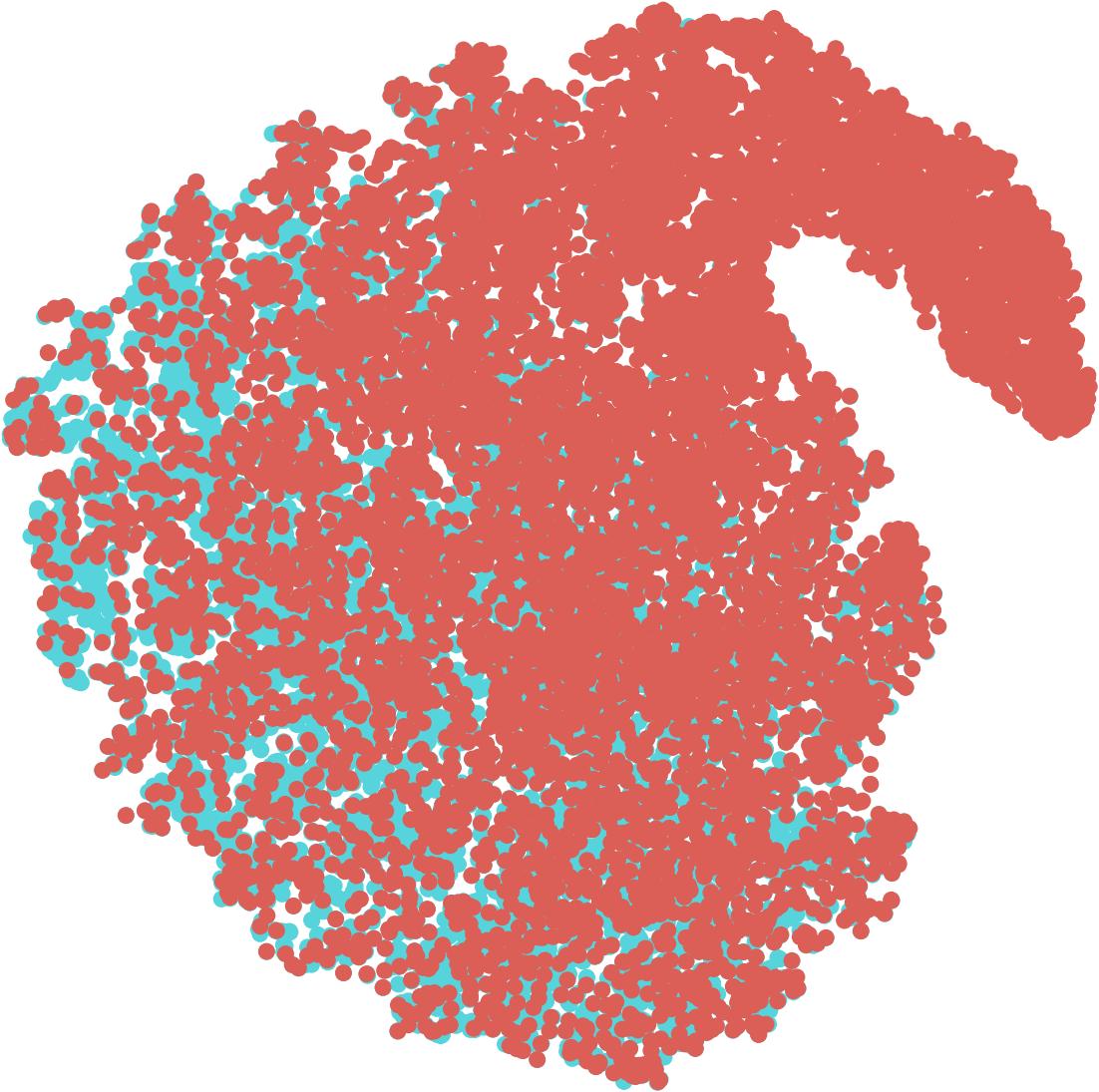} & \includegraphics[height=1in]{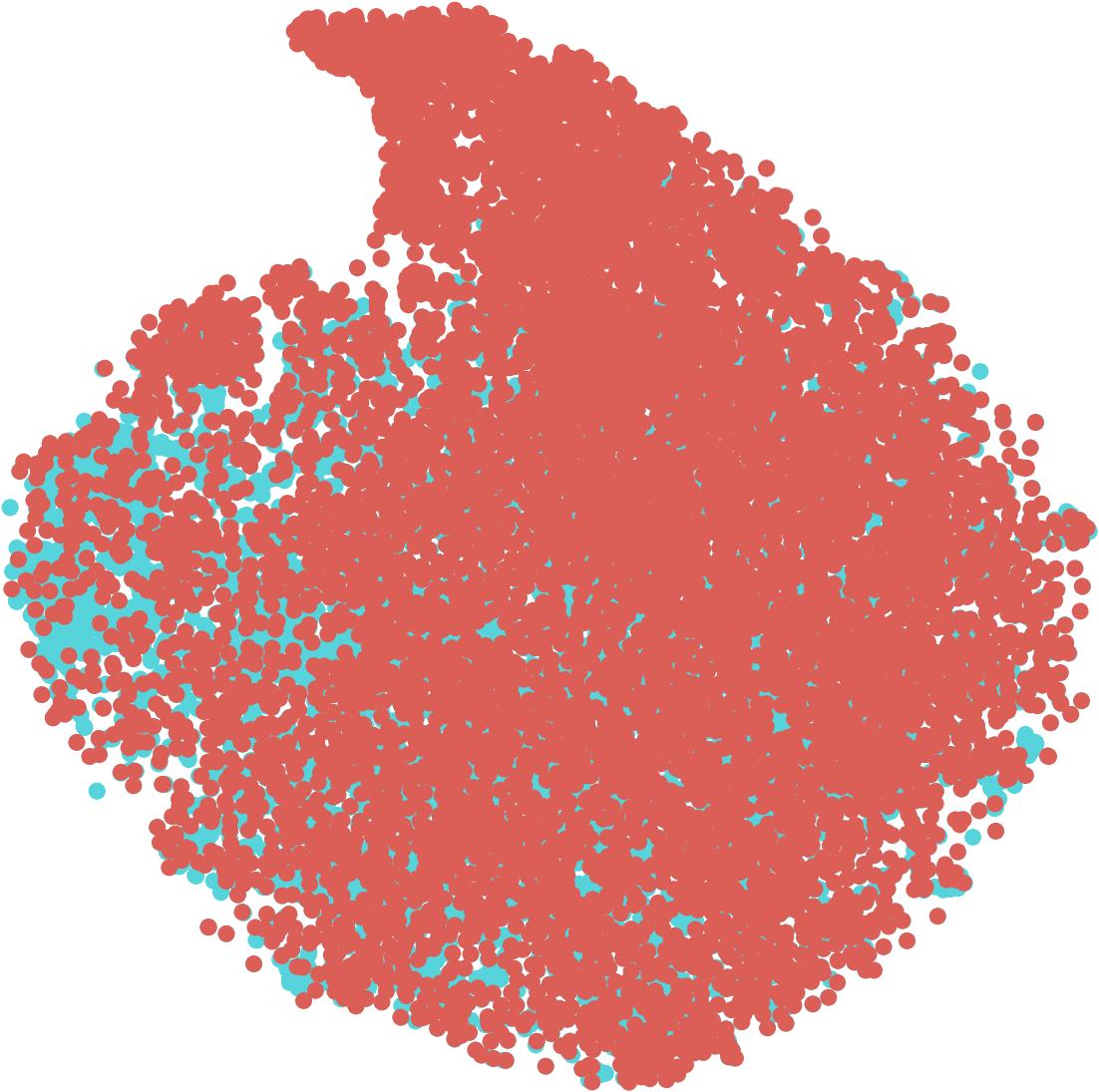} & \includegraphics[height=1in]{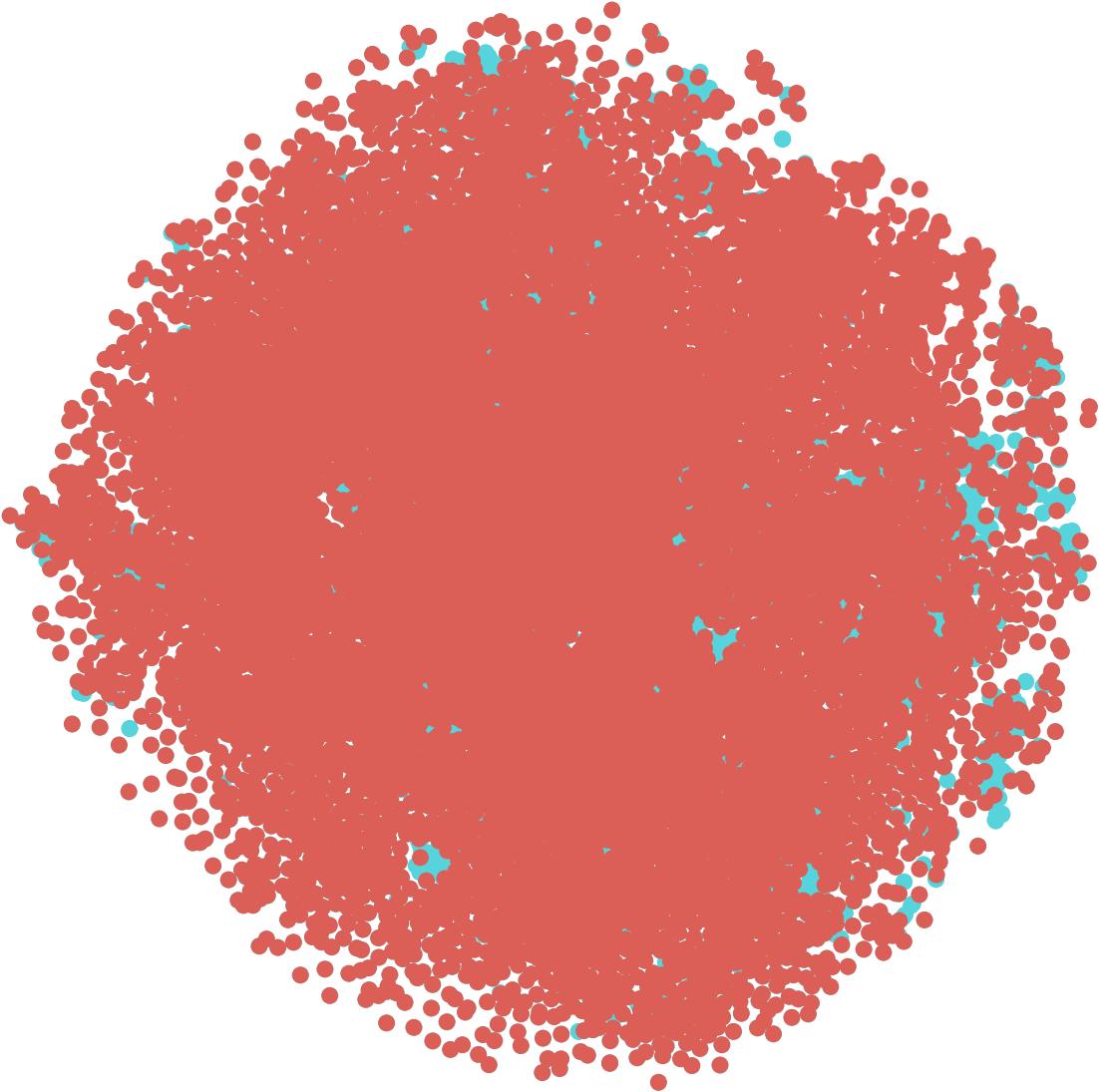} & \includegraphics[height=1in]{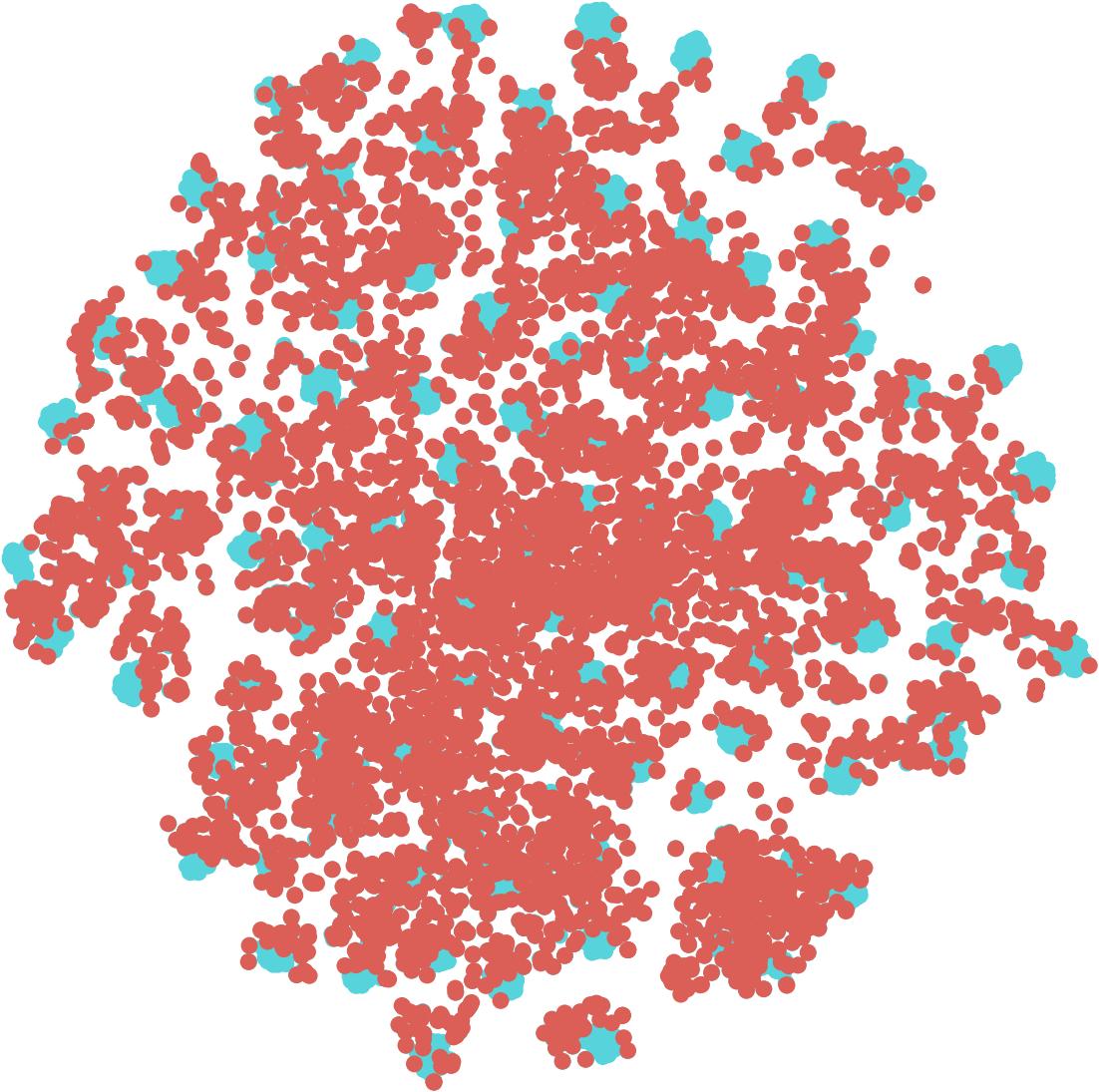}\\
		\bottomrule
	\end{tabular}
\end{table*}

\begin{table*}[t]
	\caption{Comparison of distributions of words generated from each selected hidden block of ResNet-34 under DeepFool attack on the CIFAR-10 dataset and their transformed version on the proposed detector. {\color{black} The first row corresponds to an image classifier based on a ResNet-34 model. 
    The second row corresponds to the proposed adversarial example detector.
    The input of the proposed detector consists of the feature maps produced by $\text{BN}_1$, $\text{Res}_1$, $\text{Res}_2$, $\text{Res}_3$, and $\text{Res}_4$ for detecting adversarial examples fed into the ResNet-34-based Image Classifier.} Each point represents a word corresponding to an adversarial example (red), a benign example of the source class \textit{frog} (blue), and a benign example of the target class \textit{bird} (green).}
	\label{tbl:compare_classifer_detector}
	\centering
        \begin{tabular}{m{2cm}<{\centering}| m{2cm}<{\centering}|m{2cm}<{\centering}|m{2cm}<{\centering}|m{2cm}<{\centering}|m{2cm}<{\centering}|m{2cm}<{\centering}} 
		\toprule
		 & Input & $\text{BN}_1$ & $\text{Res}_1$ & $\text{Res}_2$ & $\text{Res}_3$ & $\text{Res}_4$ \\
		\midrule
		ResNet-34-based Image Classifier & 
		\includegraphics[height=0.8in]{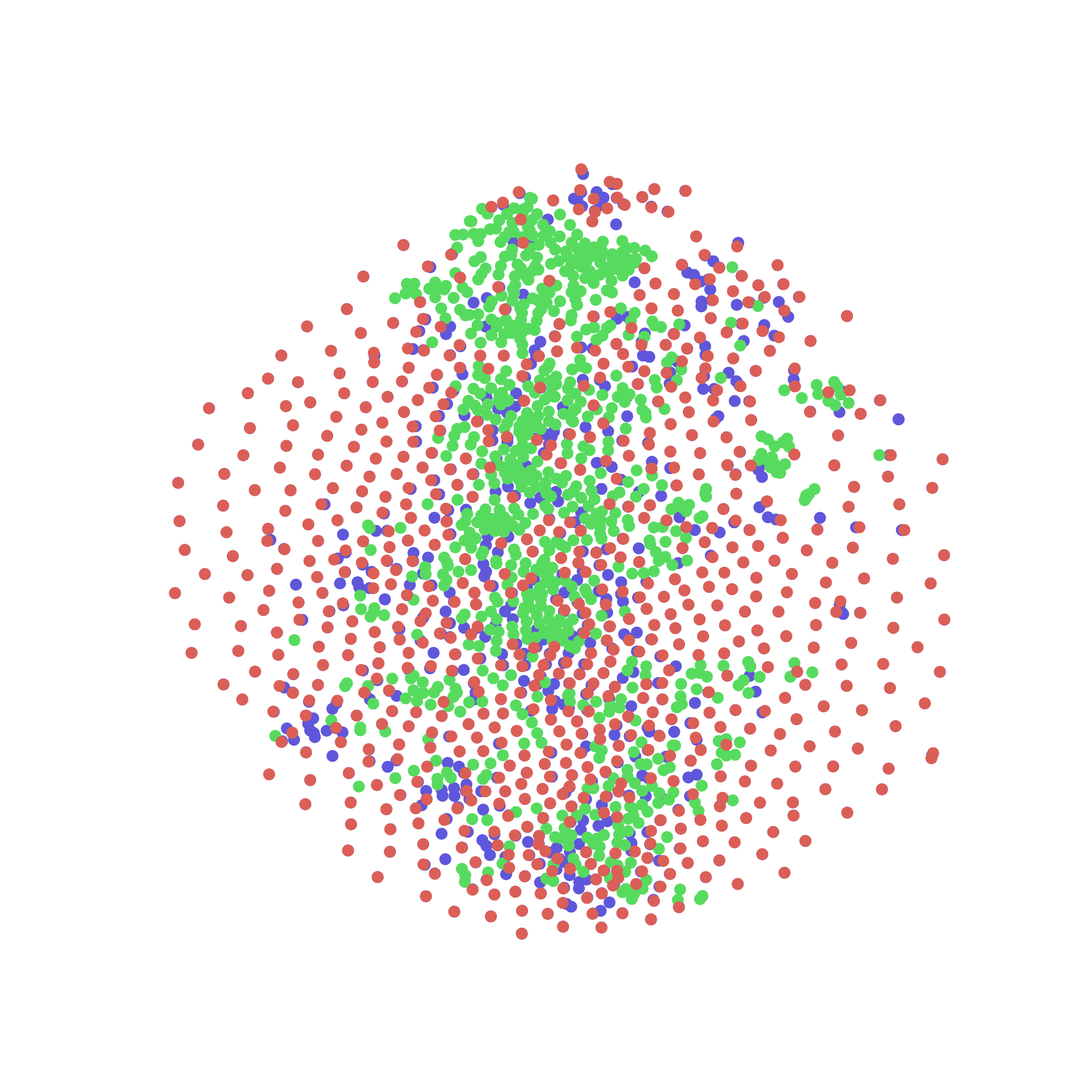} &
		\includegraphics[height=0.8in]{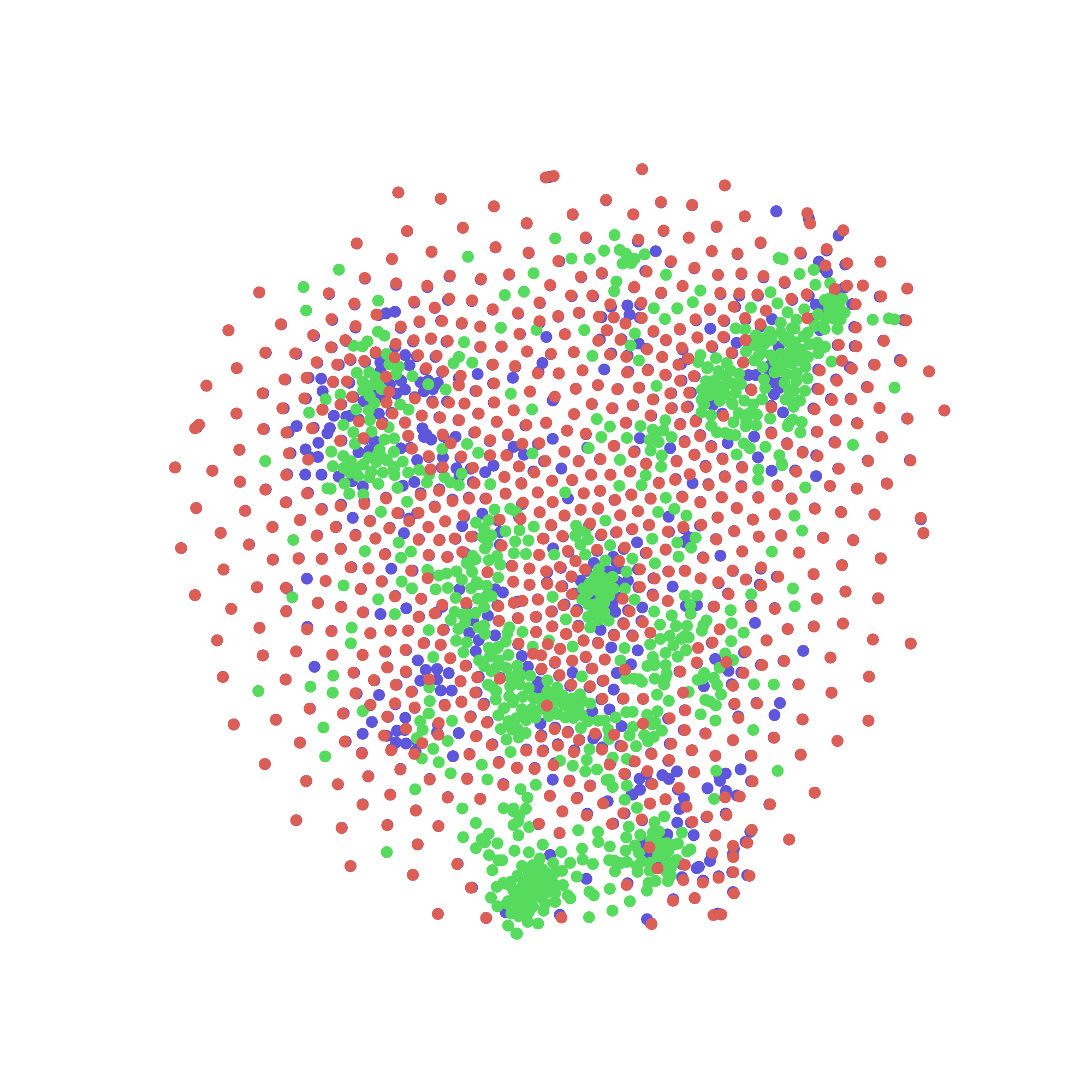} & \includegraphics[height=0.8in]{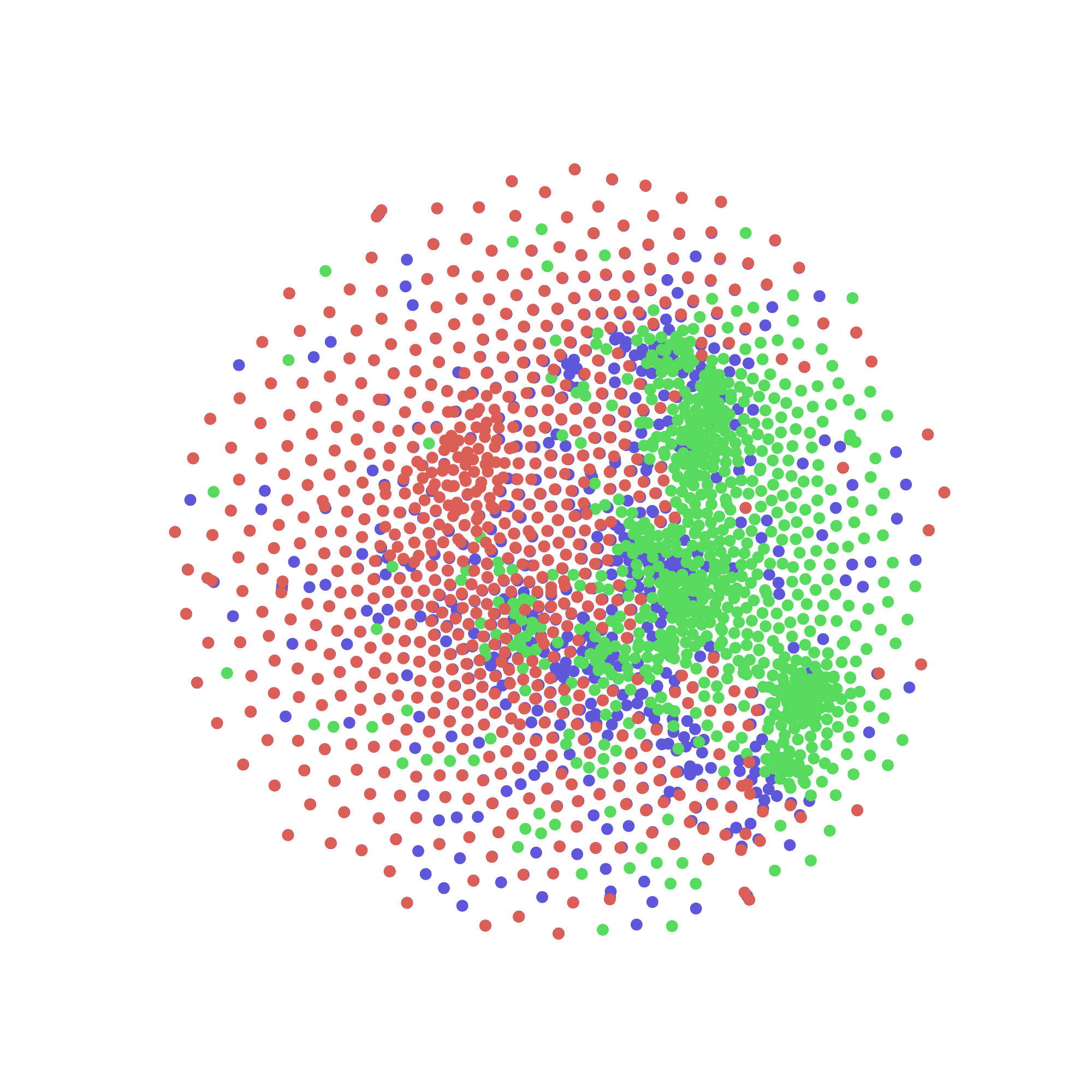} & \includegraphics[height=0.8in]{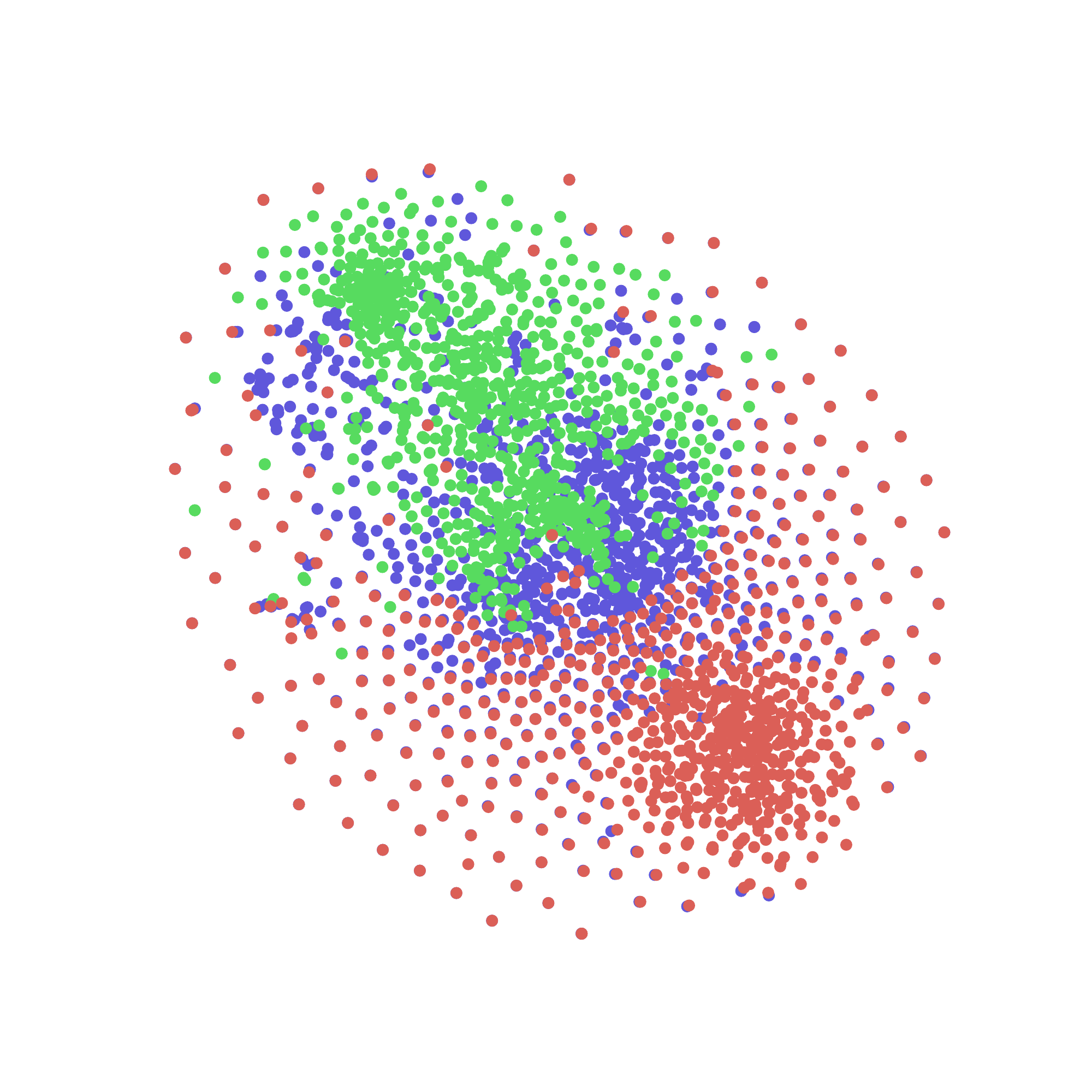} & \includegraphics[height=0.8in]{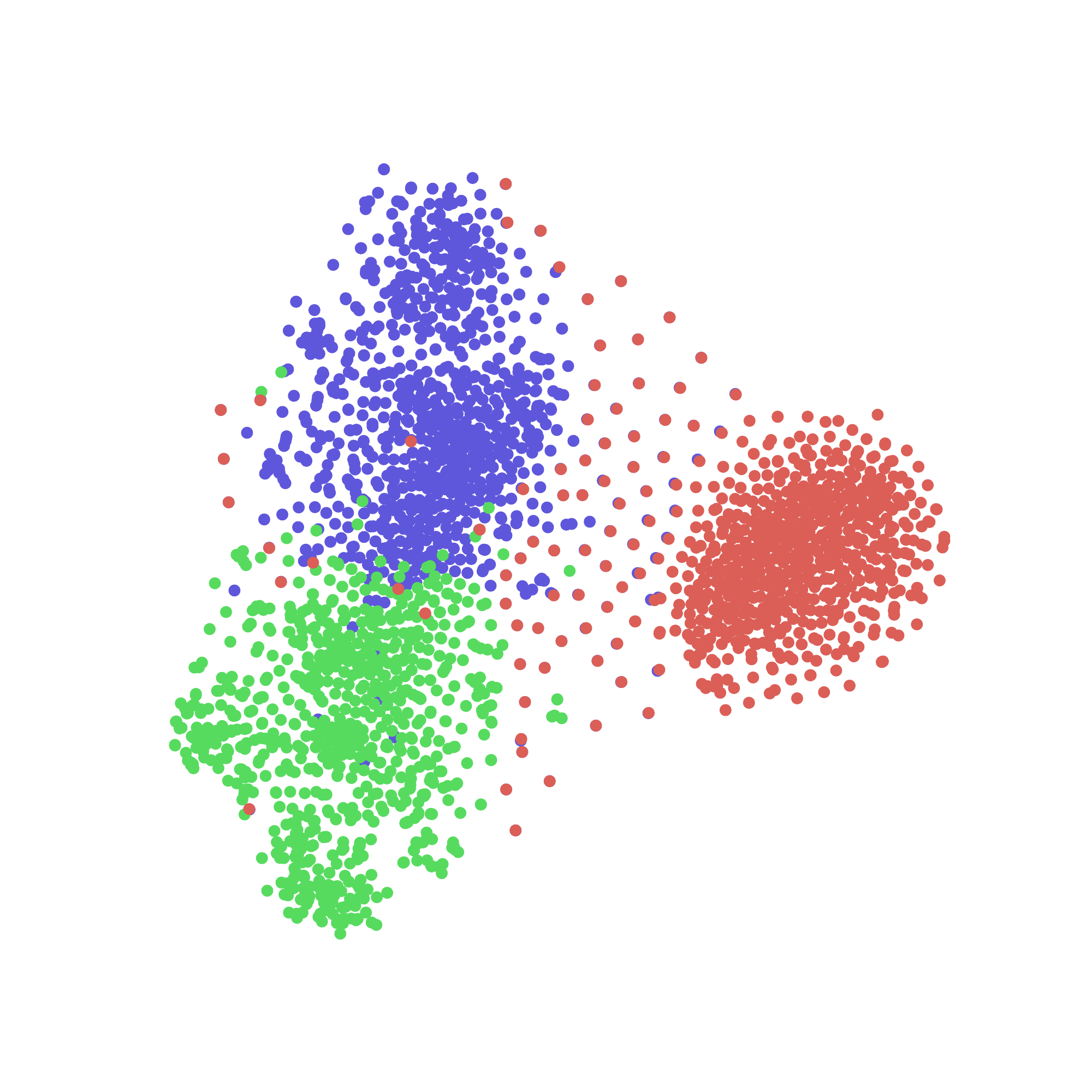} & \includegraphics[height=0.8in]{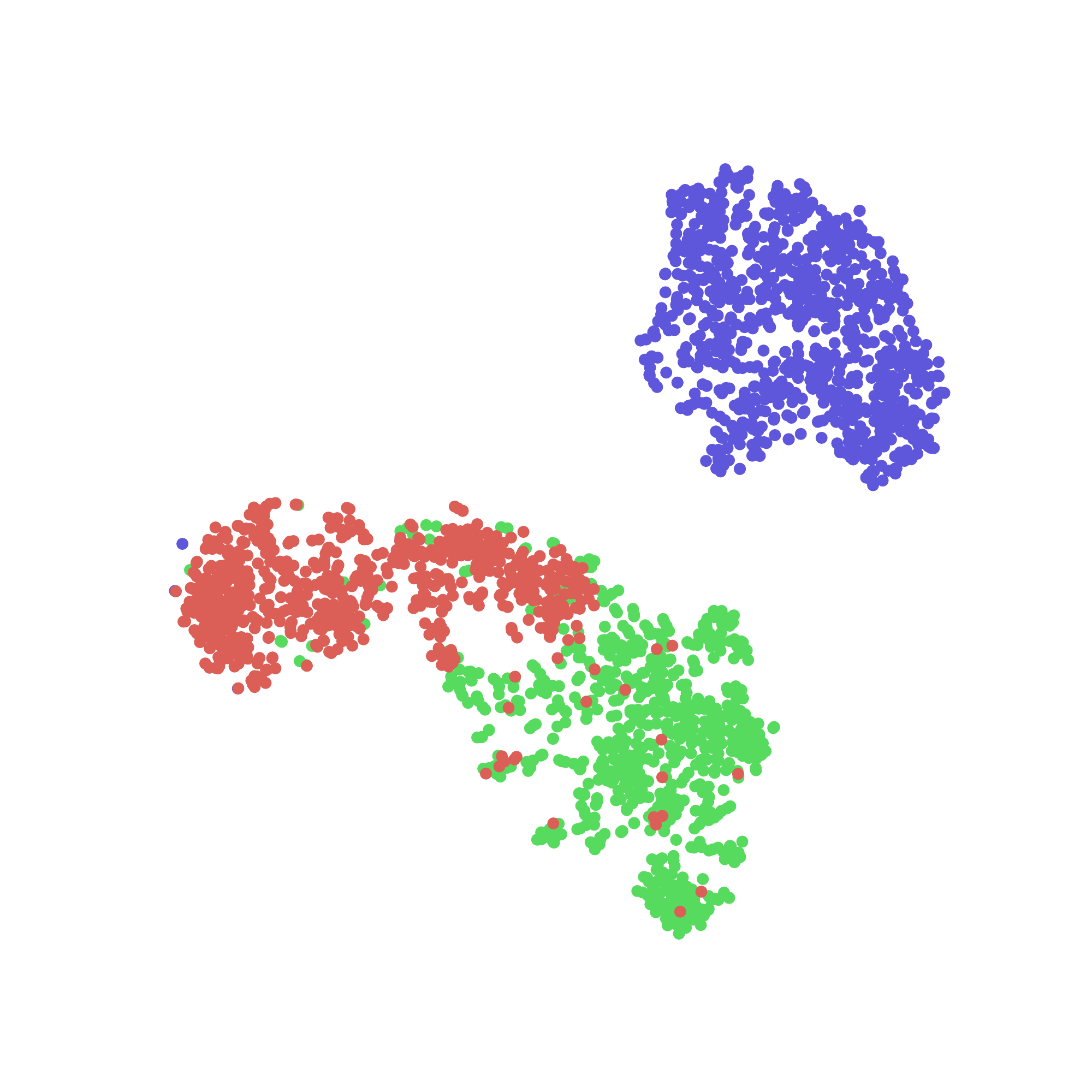}\\
		\hline
		The Proposed Detector &  
		 &
		\includegraphics[height=0.8in]{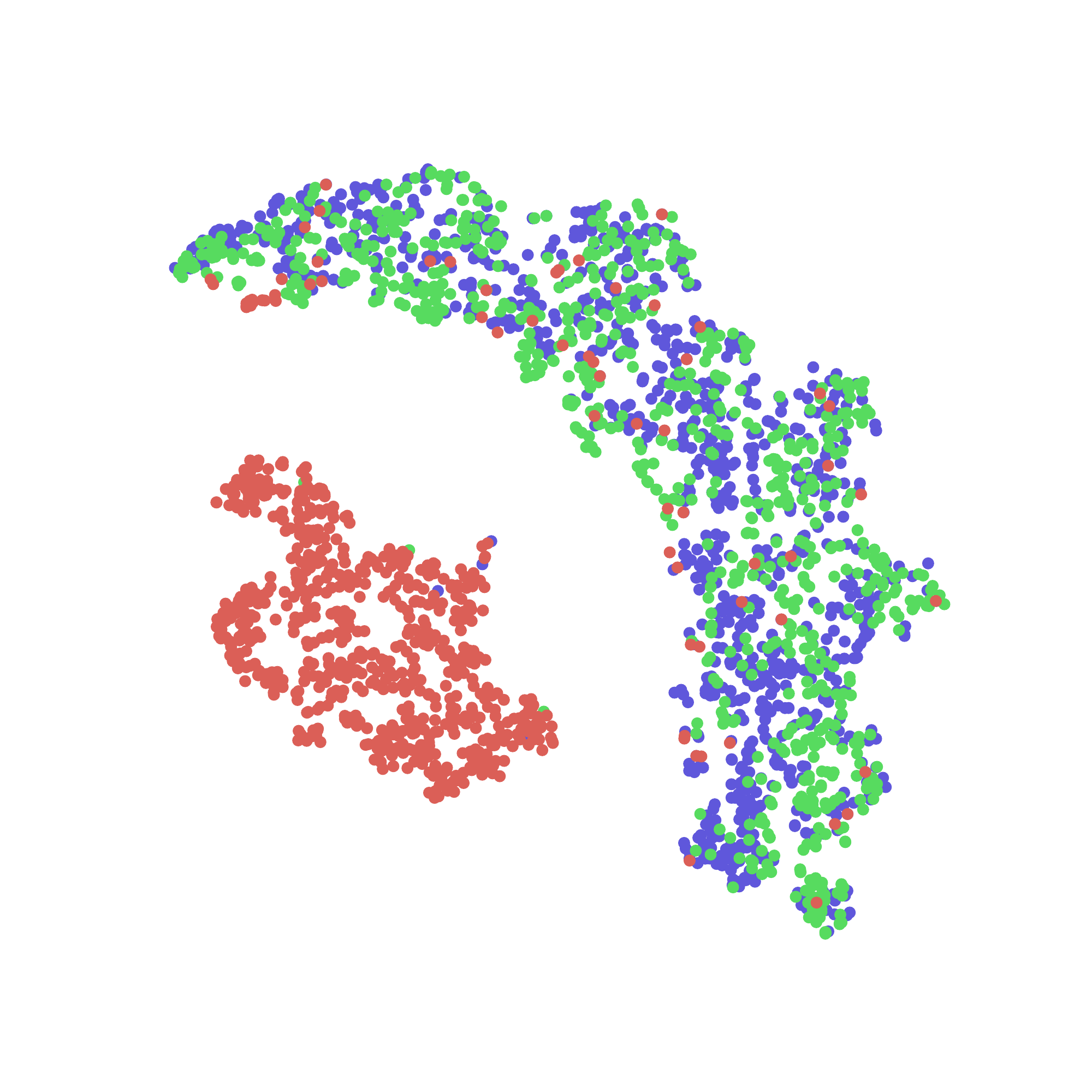} & \includegraphics[height=0.8in]{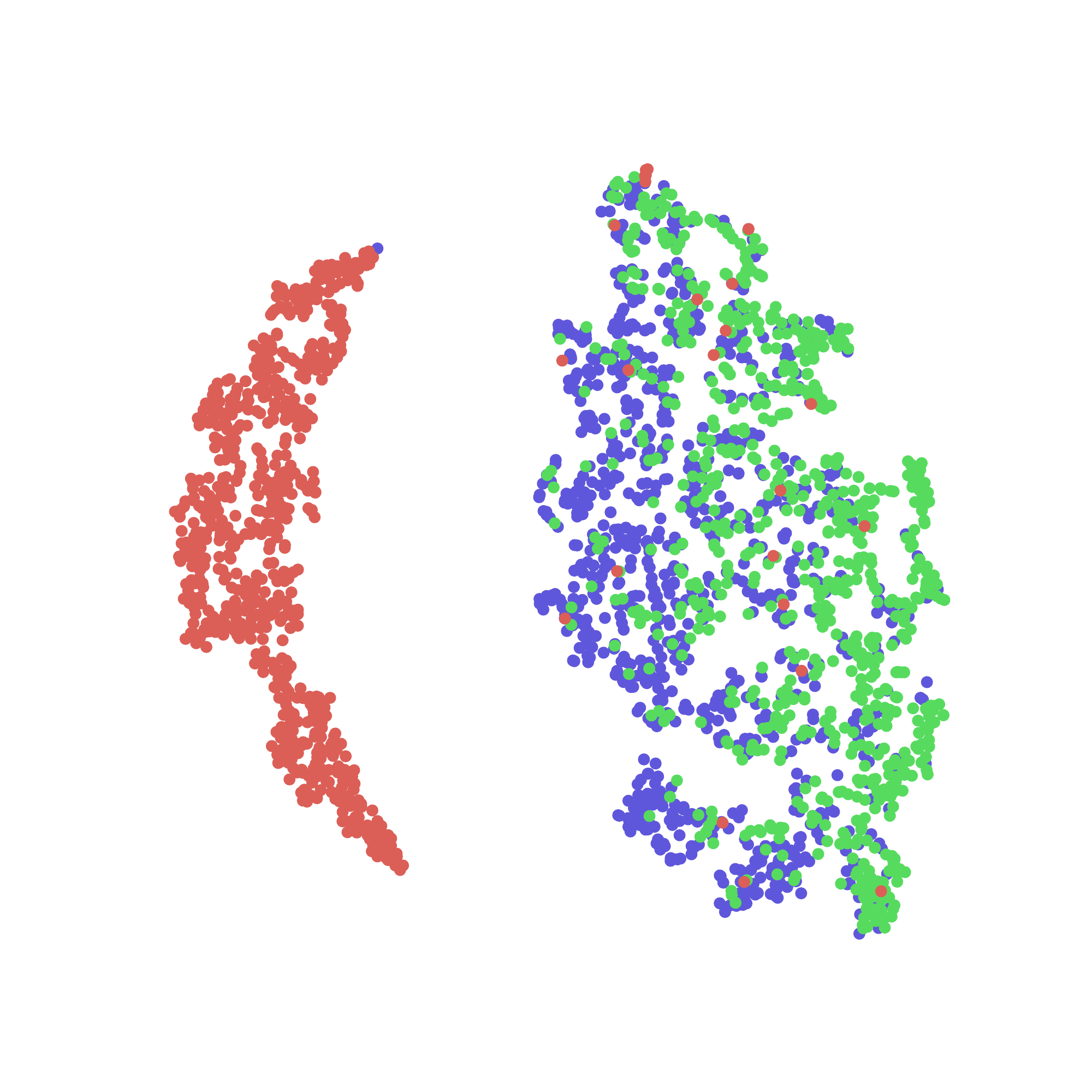} & \includegraphics[height=0.8in]{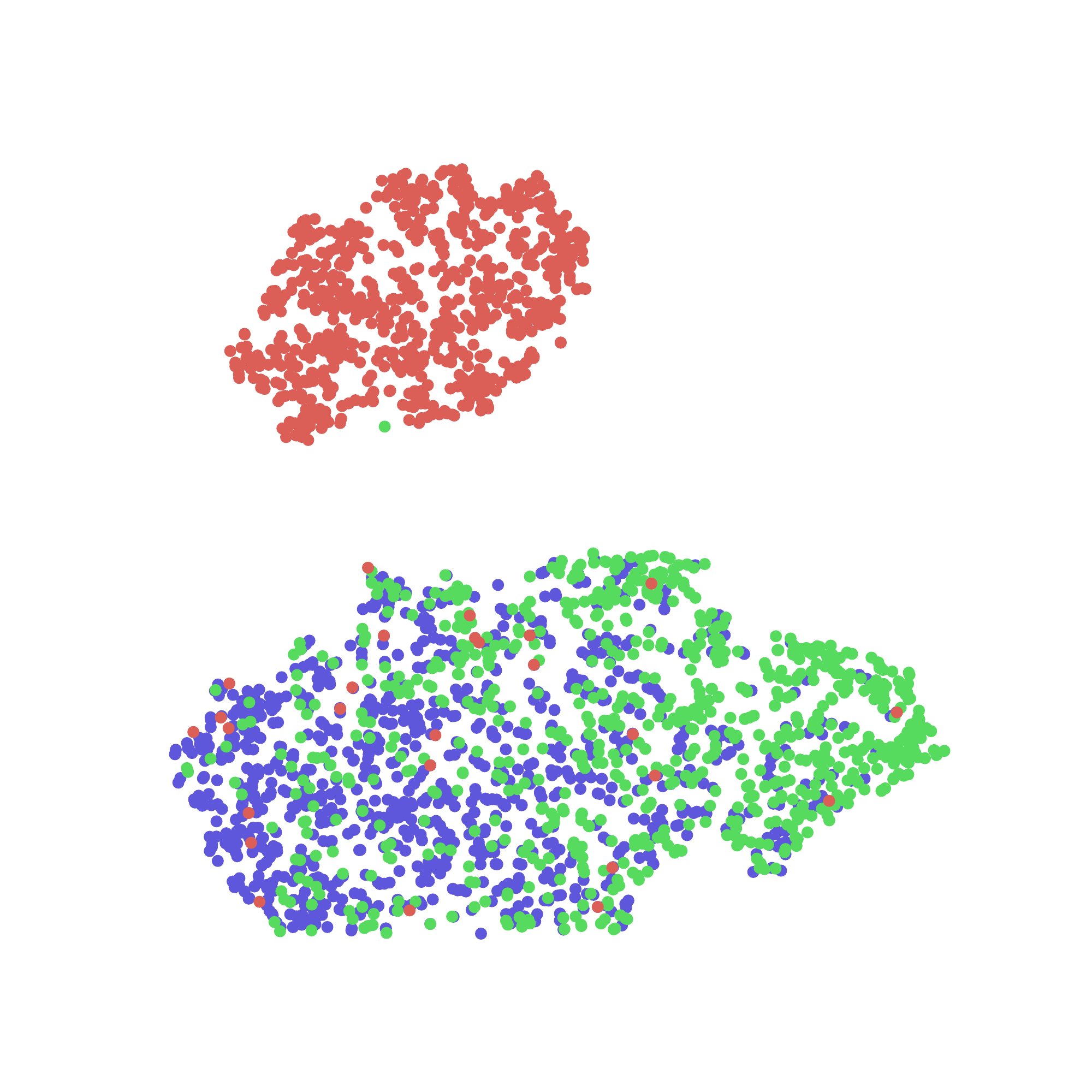} & \includegraphics[height=0.8in]{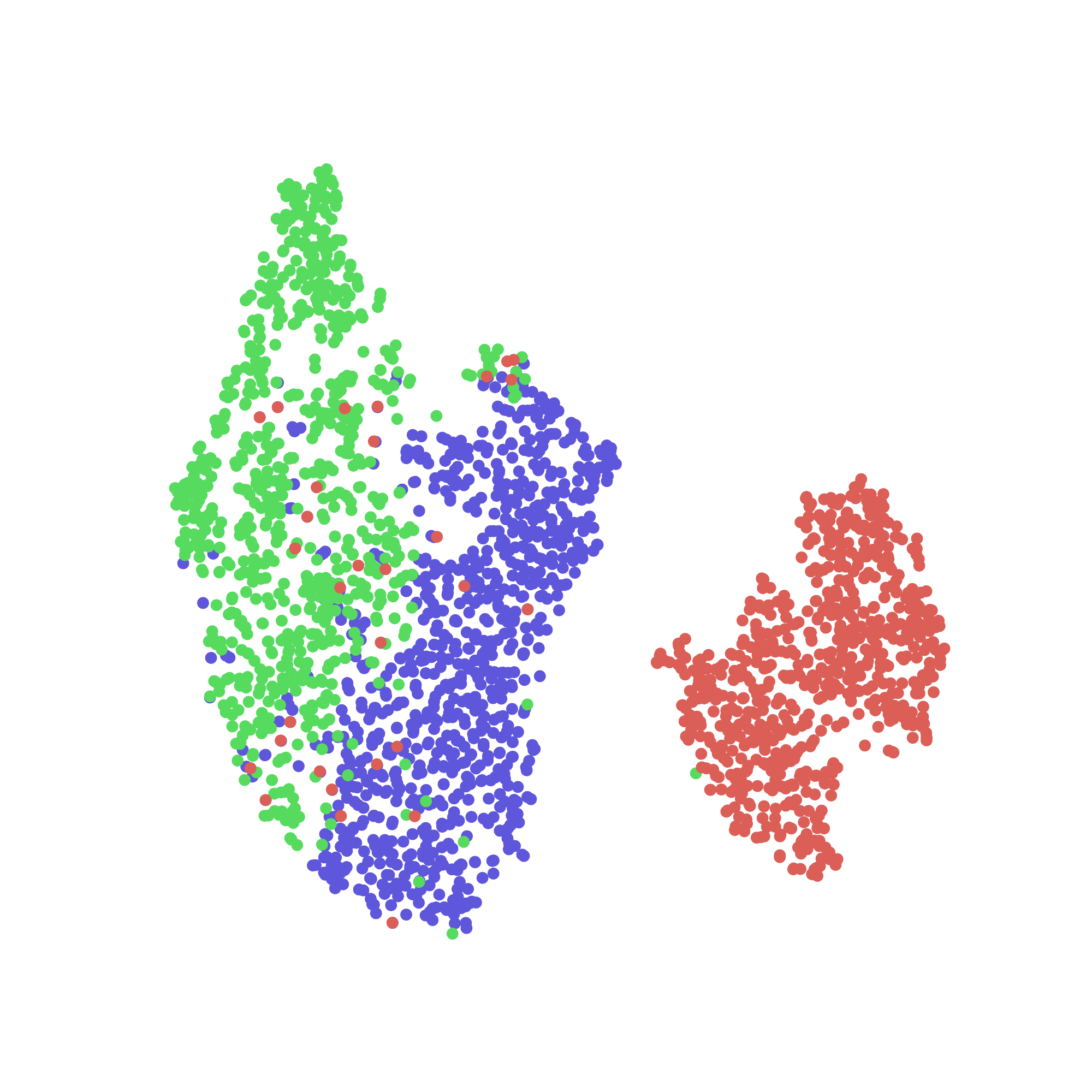} & \includegraphics[height=0.8in]{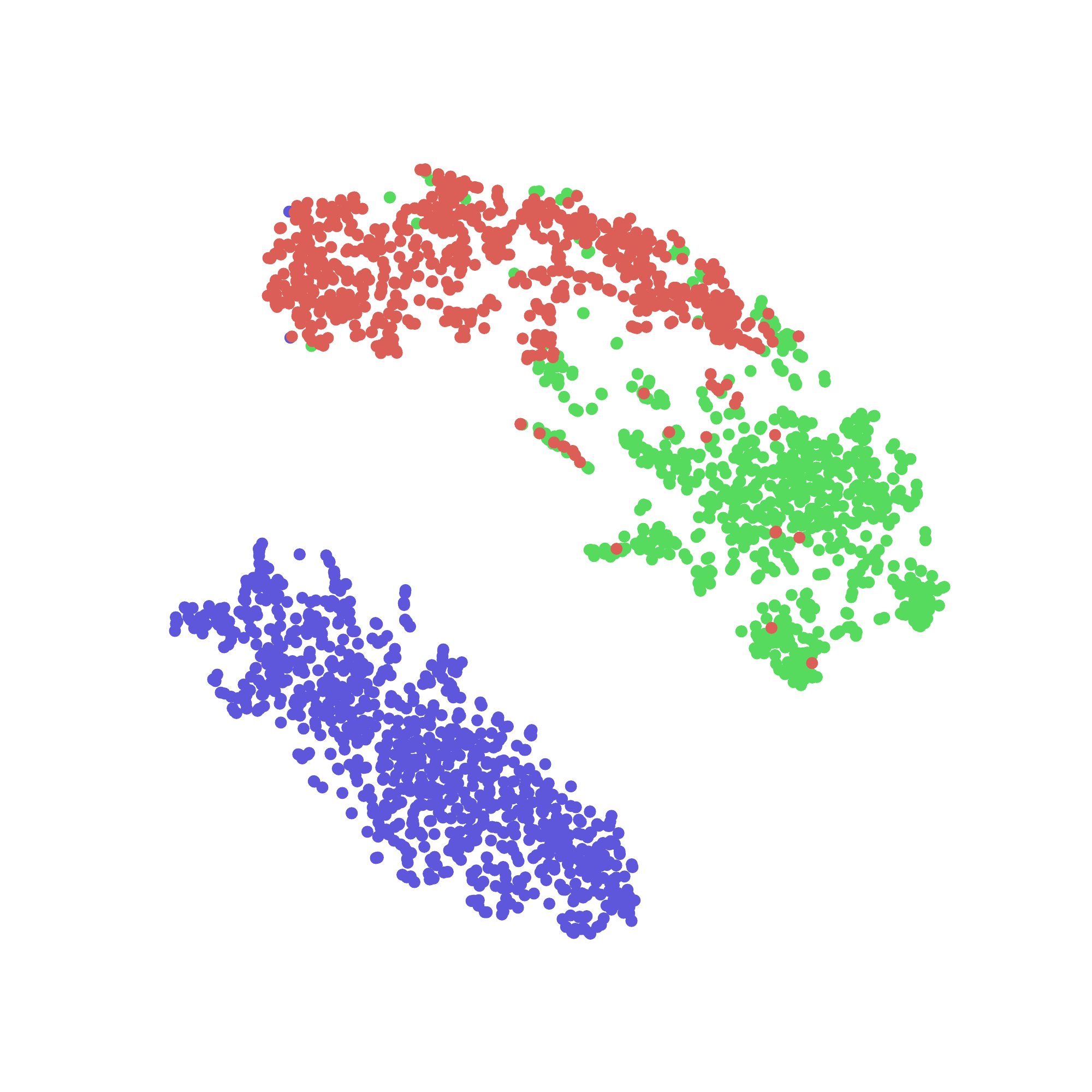}\\
		\bottomrule
	\end{tabular}
\end{table*}

Table~\ref{tbl:table_of_sne_all_adv} shows that, by using the proposed interpretation of feature maps to word vectors, the adversarial and benign examples (i.e., the red and blue points) exhibit different levels of separability at the selected hidden layers of the classifier for adversarial example detection. The CIFAR-100 dataset is less visually separable than the CIFAR-10 and SVHN datasets. For this reason, the attacks on CIFAR-100 are generally harder to detect, as shown in Table~\ref{tab:auc}. Nevertheless, the new detector is able to achieve reasonable detection rates, by exploiting the spatio-temporal characteristics of the feature maps for improved separability, as demonstrated in Table~\ref{tab:auc}.

\begin{figure}[t]
	\centering
    \begin{subfloat}[DeepFool]{\includegraphics[height=1 in]{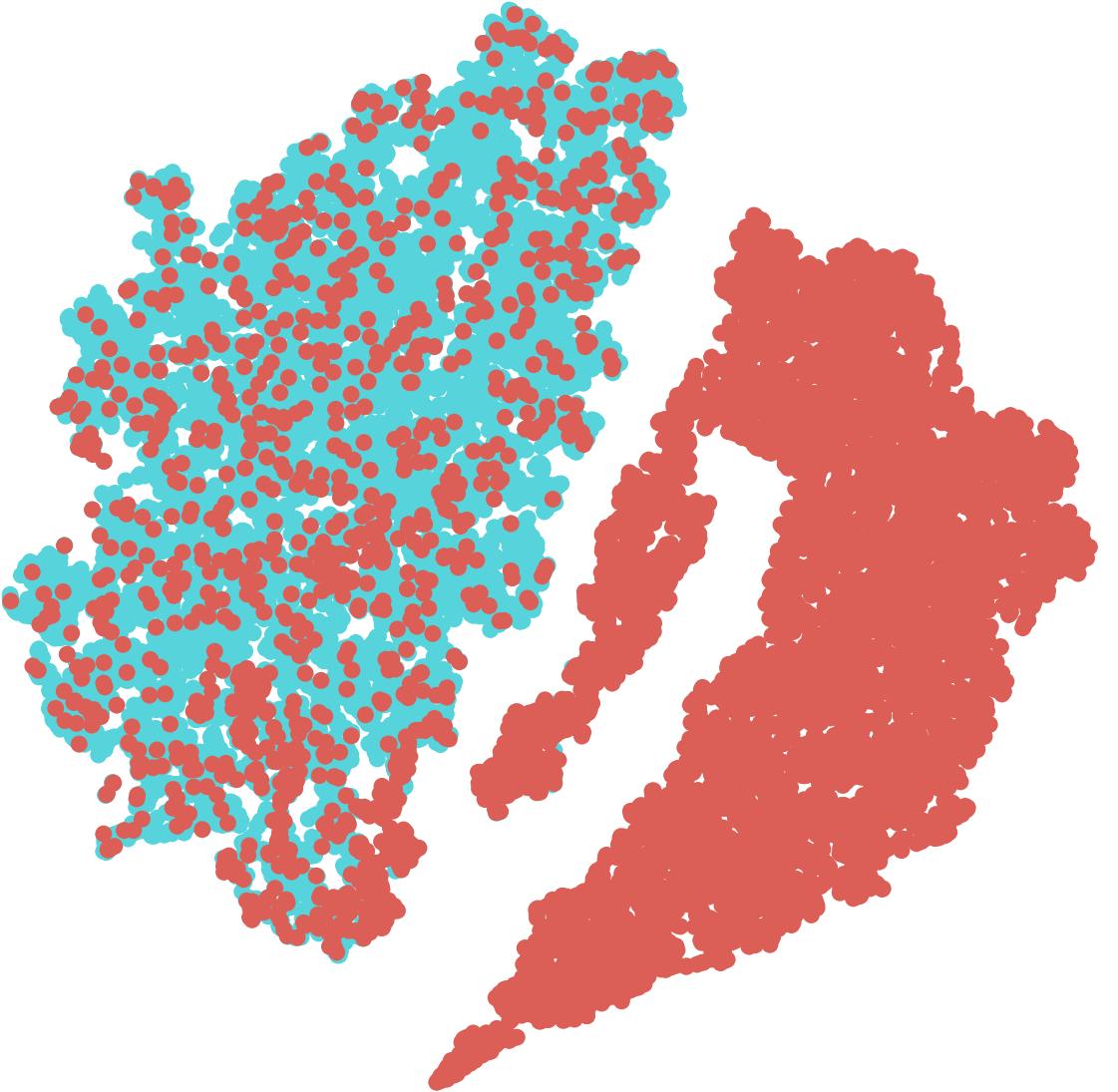}}\end{subfloat} 
    \begin{subfloat}[EAD]{\includegraphics[height=1 in]{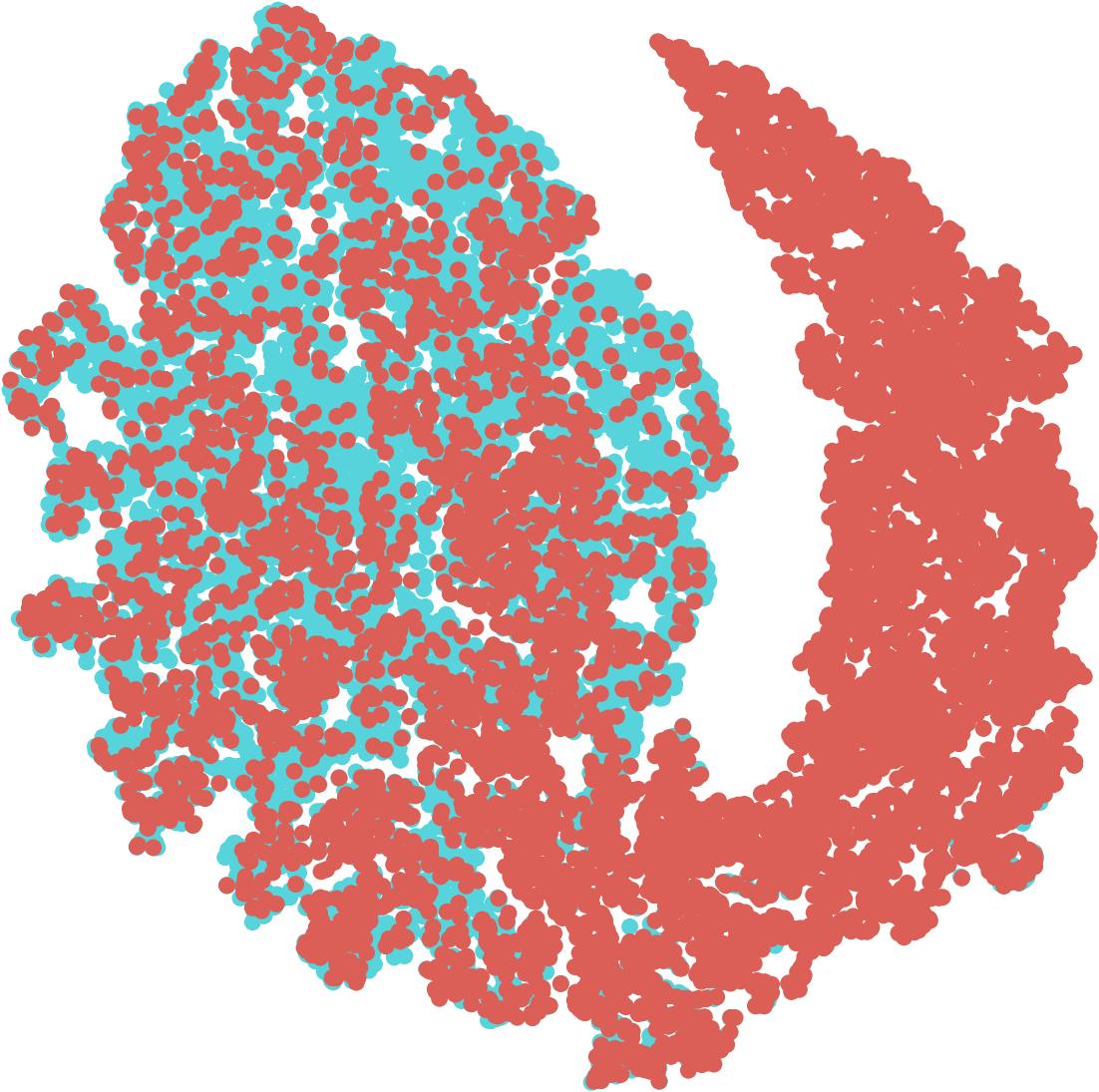}}\end{subfloat}~
	\begin{subfloat}[FGSM (0.1)]{\includegraphics[height=1 in]{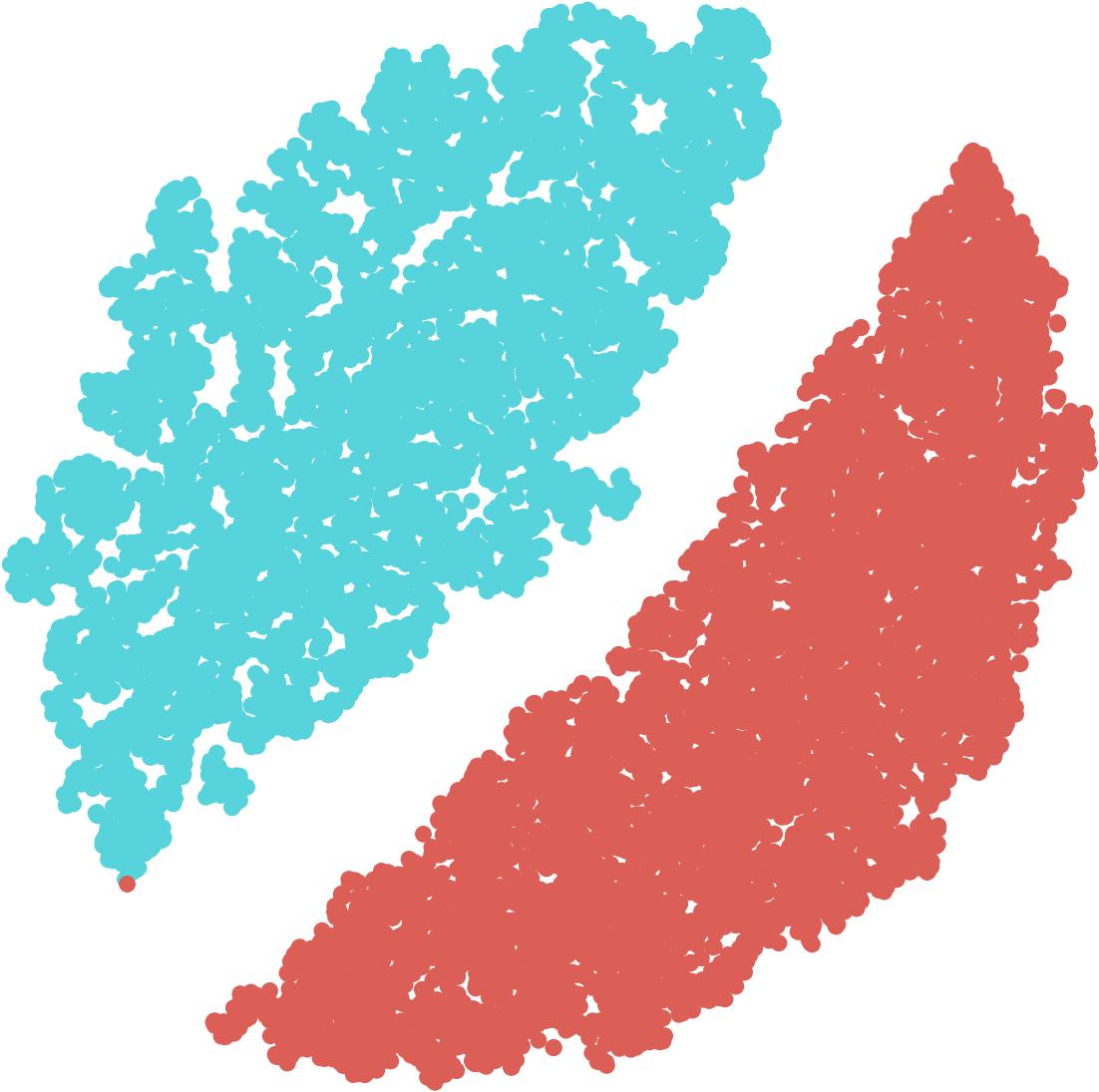}}\end{subfloat} ~
    \begin{subfloat}[Auto (0.02)]{\includegraphics[height=1 in]{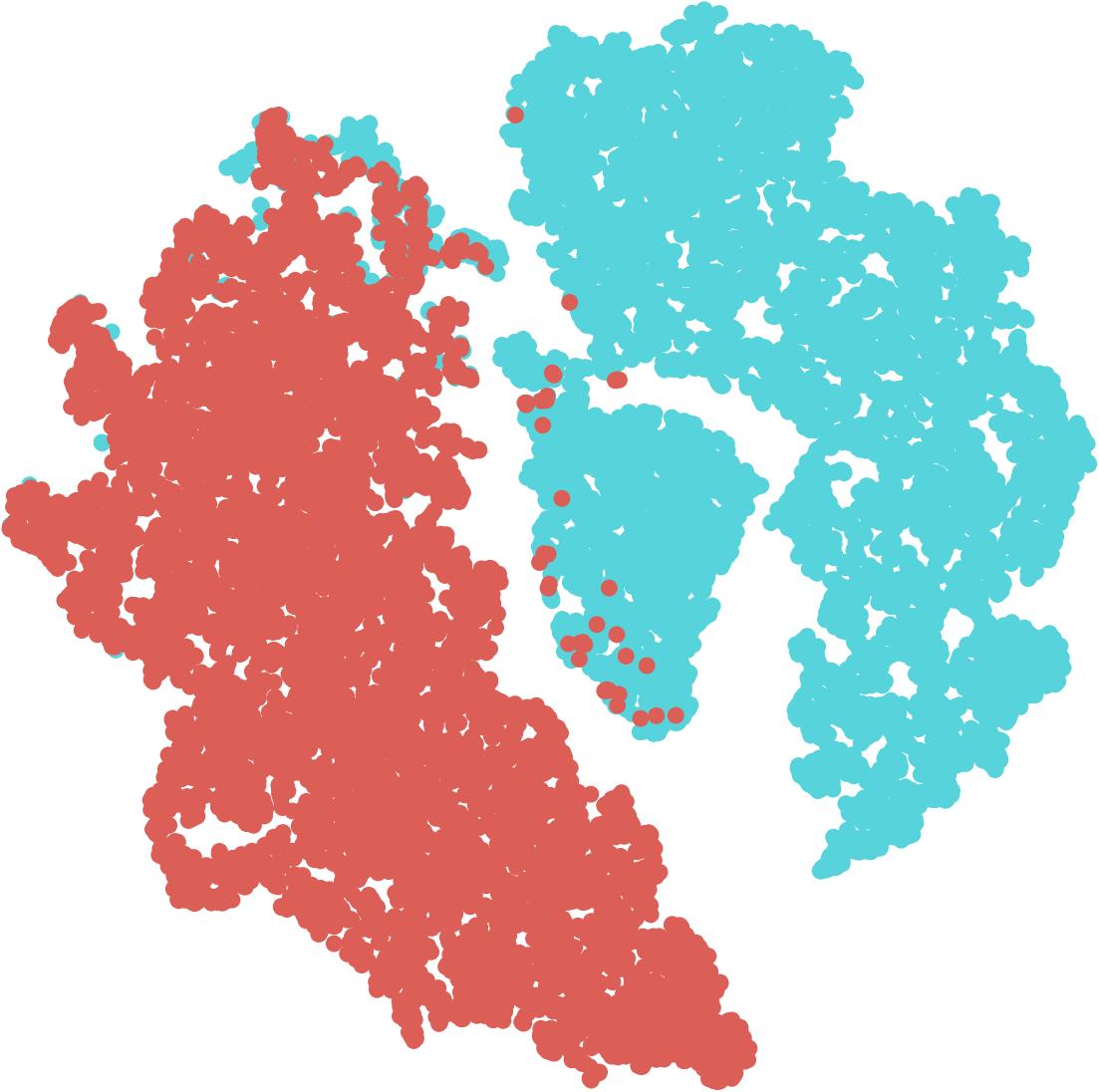}}\end{subfloat} ~
	\begin{subfloat}[Auto (8/255)]{\includegraphics[height=1 in]{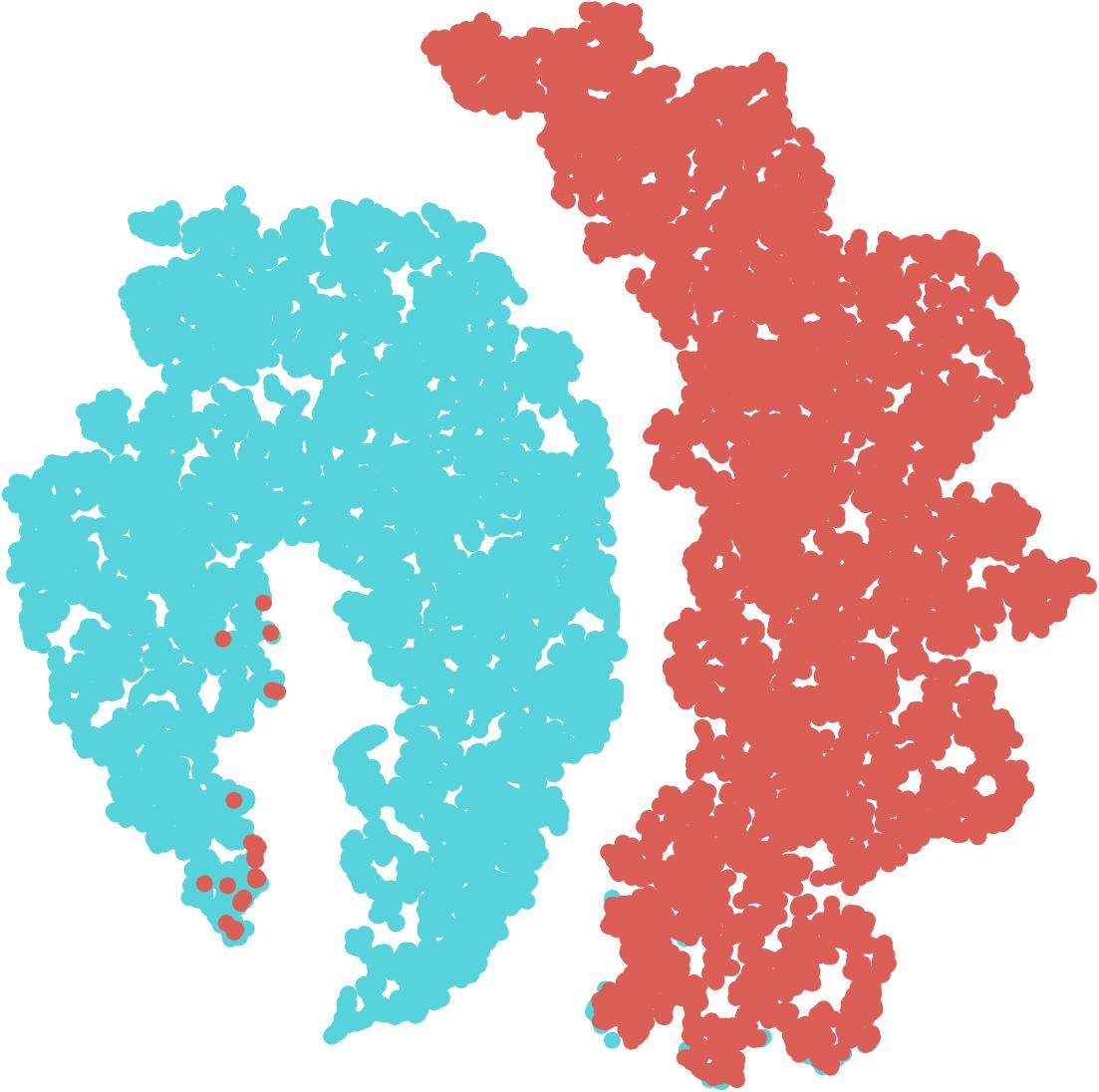}}\end{subfloat} ~
    \begin{subfloat}[JSMA (1,0.1)]{\includegraphics[height=1 in]{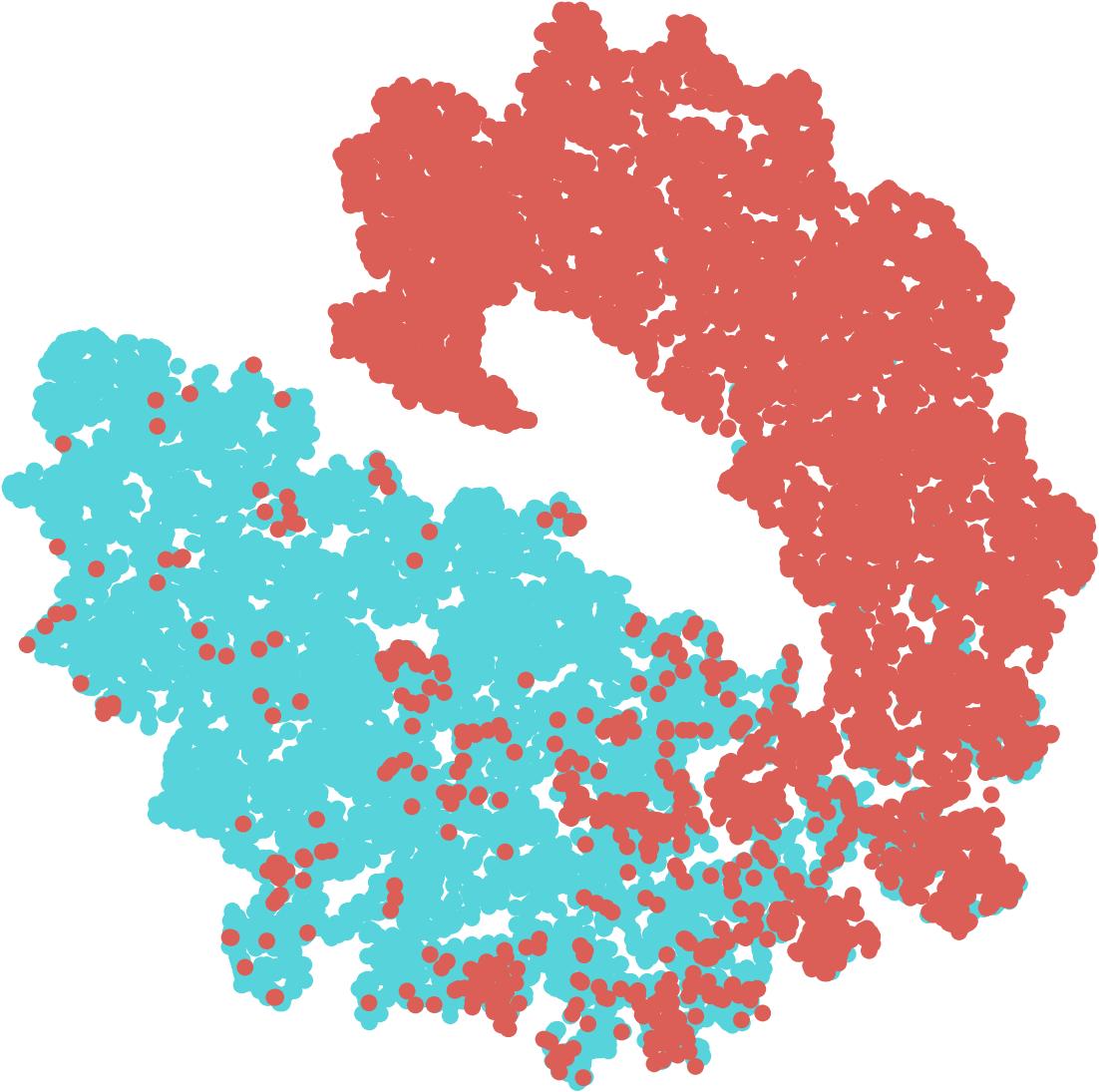}}\end{subfloat} ~
	\begin{subfloat}[PGD (0.02)]{\includegraphics[height=1 in]{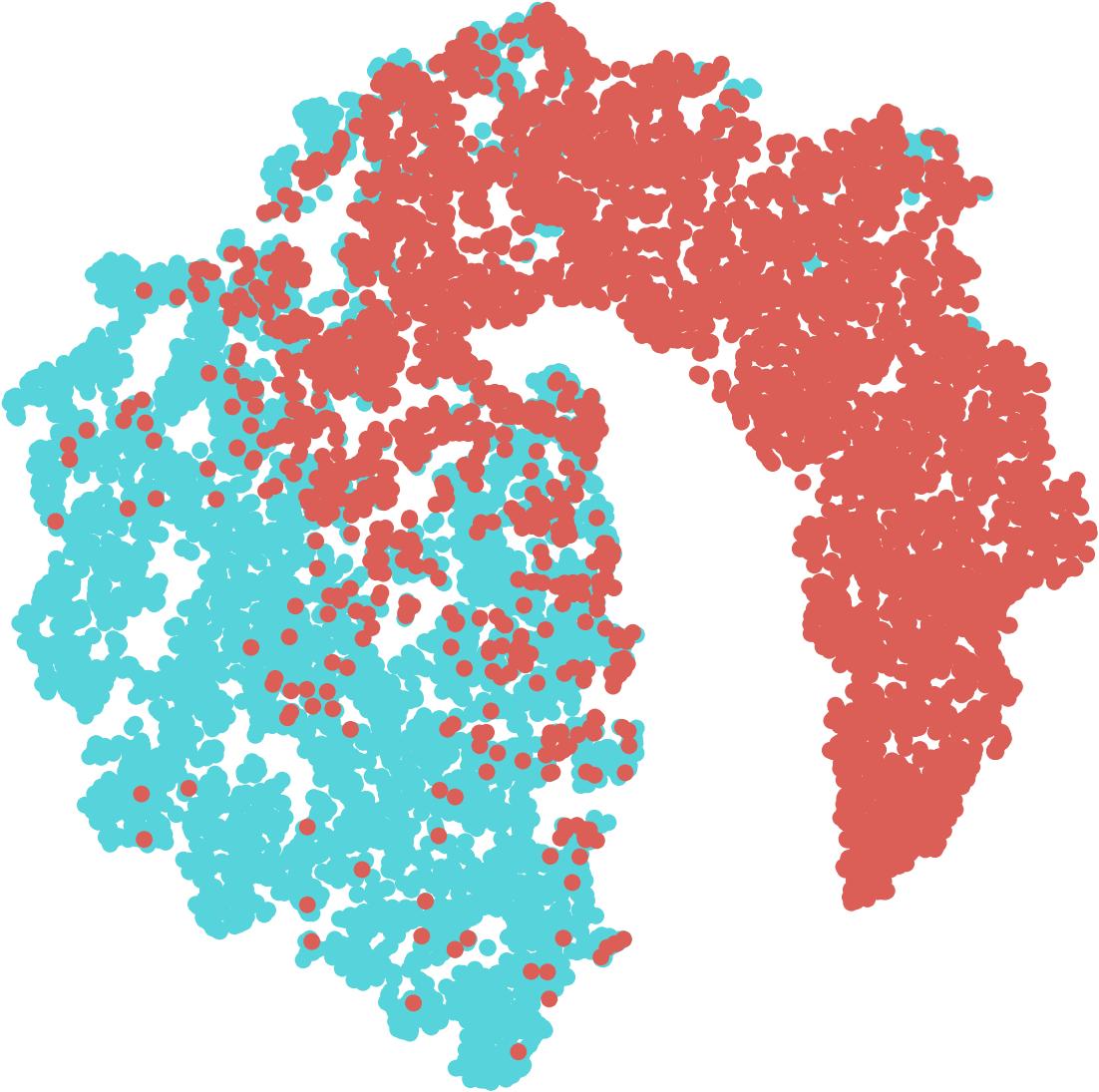}}\end{subfloat}~
	\begin{subfloat}[PGD (8/255)]{\includegraphics[height=1 in]{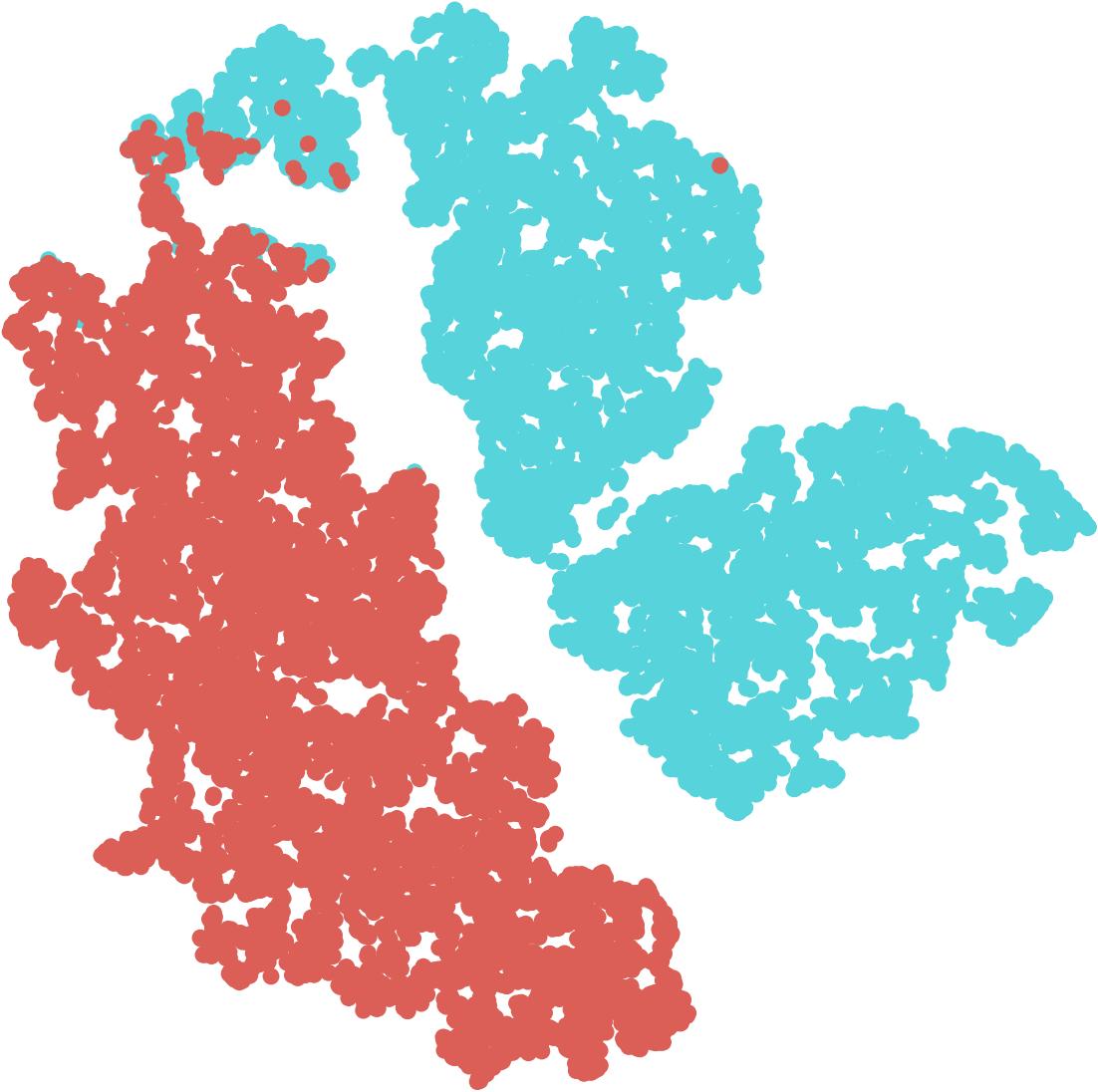}}\end{subfloat}~
    \begin{subfloat}[C\&W]{\includegraphics[height=1 in]{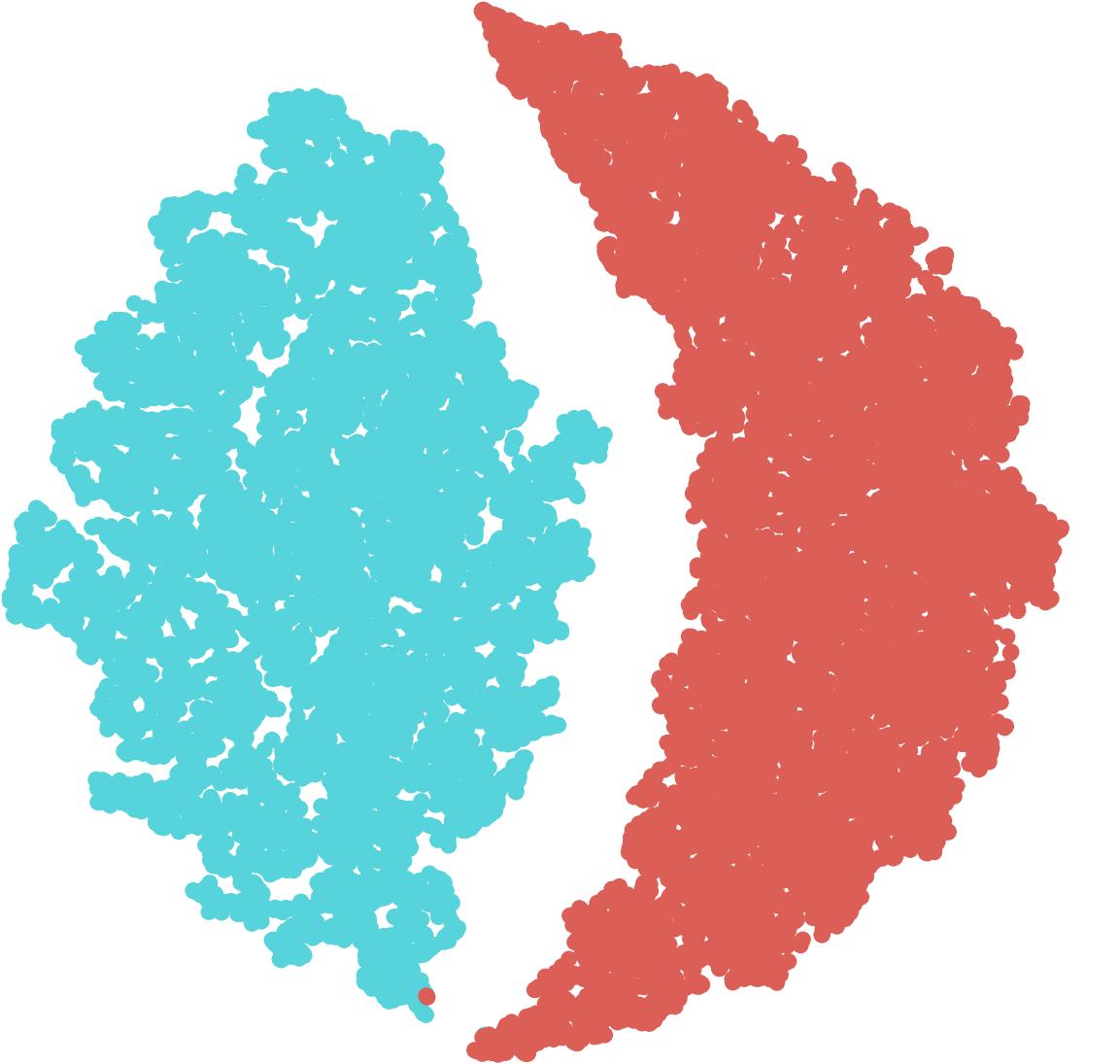}}\end{subfloat}     
	\caption{The t-SNE figures of the words generated at the $\text{Res}_1$ hidden block of the ResNet-34 model on the CIFAR-10 dataset. Red and blue points represent feature maps corresponding to adversarial and benign examples, respectively.}
	\label{fig:CIFAR-10_res1}
\end{figure}

Fig.~\ref{fig:CIFAR-10_res1} shows the separability of adversarial and benign examples of the CIFAR-10 dataset under six considered attacks, by using the word interpretation of feature maps in the proposed detector (more explicitly, the word embedding layer of the detector). 
We see that the adversarial examples (red) generated by the DeepFool and EAD attacks are more difficult to distinguish from the benign examples (blue), as compared to four other attacks. This is because DeepFool and EAD produce less perturbation than the other attacks, making their adversarial examples less distinguishable and reducing the detection capability of the existing detectors.

\begin{figure}[t]
	\centering
	\includegraphics[width=3.0in]{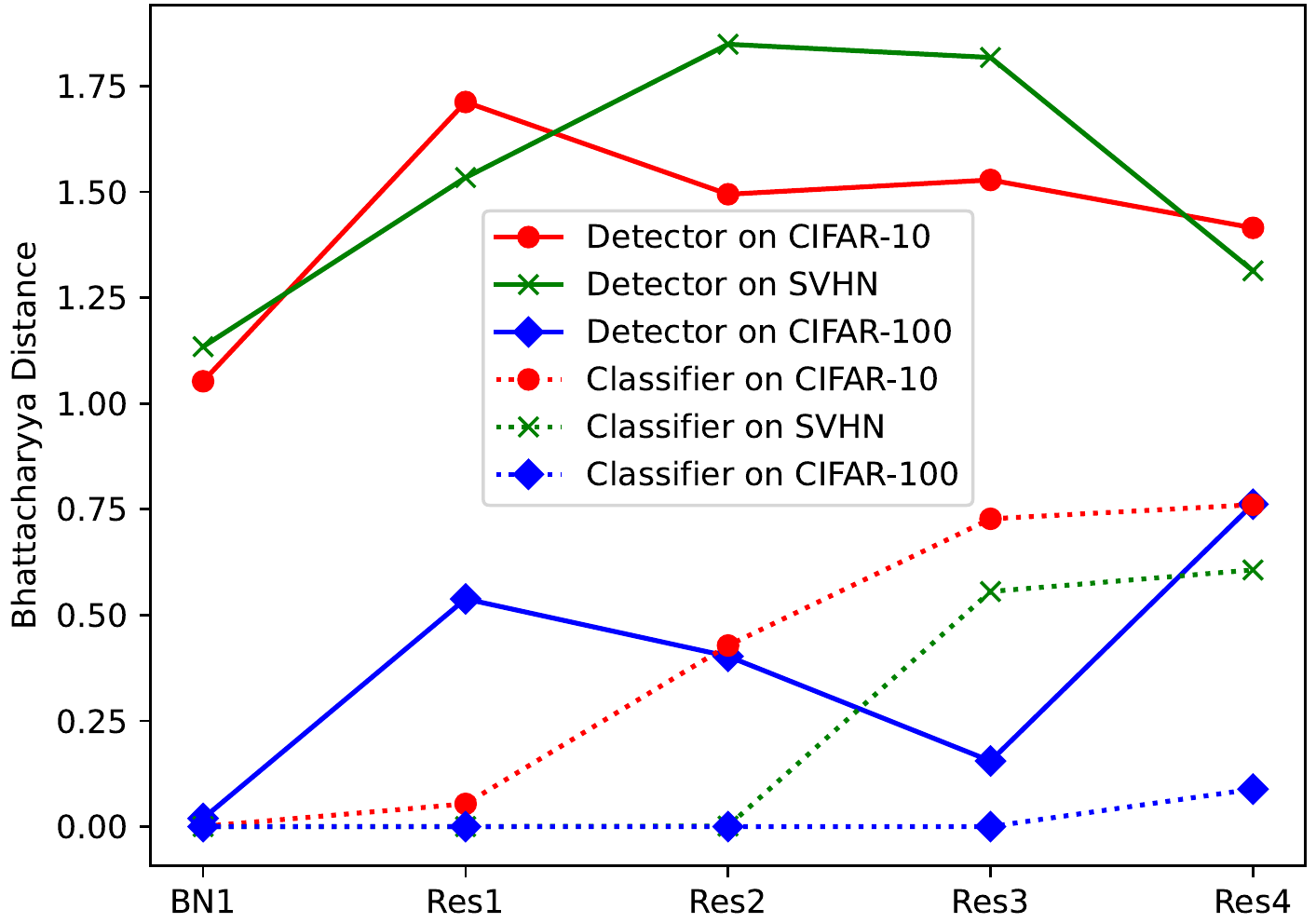}
	\caption{The Bhattacharyya distance between the distribution of adversarial and benign examples across hidden blocks of the ResNet-34 model under DeepFool attacks.}
	\label{fig: Bdist}
\end{figure}

In Fig.~\ref{fig: Bdist}, we use Bhattacharyya distance~\cite{7121017} to quantify the separability of the red and blue clusters in Table~\ref{tbl:table_of_sne_all_adv}. Bhattacharyya distance is a measure of the amount of overlap between two statistical samples or populations. As shown in Fig.~\ref{fig: Bdist}, the Bhattacharyya distances of the proposed detector are much larger than those of the image classifier under attack. The Bhattacharyya distances of the classifier are all close to zero at $\text{BN}_1$ and gradually increase along the remaining hidden blocks until they are big enough to misclassify the perturbed images. The word embedding layer of our detector enlarges the difference between adversarial and benign examples, improving their separability and detection accuracy.

Table~\ref{tbl:compare_classifer_detector} demonstrates the processes of misclassifying \textit{frogs} (the source class) into \textit{birds} (the target class) in the CIFAR-10 dataset, along with the processes of correctly classifying \textit{frogs} and \textit{birds} in the dataset. DeepFool is launched to perturb frog images to be misclassified into birds. The t-SNE figures are plotted for the feature maps produced by the hidden blocks of the ResNet-34 image classifier under the DeepFool attack; see the second row of Table~\ref{tbl:compare_classifer_detector}. 
The t-SNE figures are also plotted for the word vectors generated by the Word Embedding Layer of the new detector based on the feature maps; see the third row of the table. 

We see in Table~\ref{tbl:compare_classifer_detector} that DeepFool is effective in misleading the classifier to misclassify frog images to bird images since adversarial examples (i.e., perturbed frog images) are not separable from unperturbed bird images and are distantly separate from unperturbed frog images at the last hidden block of the ResNet-34 model, i.e., Res$_4$. 
The adversarial examples are not separable from the benign (unperturbed) examples at the earlier hidden blocks of the ResNet-34 model. 

We also see in Table~\ref{tbl:compare_classifer_detector} that the adversarial examples can be effectively separated from both of the (unperturbed) frog and bird classes, after the feature maps from the hidden blocks of the ResNet-34 model are interpreted as word vectors in the proposed detector. 
The only exception is the word vector corresponding to the last hidden block of the ResNet-34, i.e., Res$_4$.
This is due to the fact that the feature maps of Res$_4$ are directly fed to the global average pooling layer of the Word Embedding Layer with no feature extracted; see Fig.~\ref{fig: arch}. Nevertheless, the word vectors based on the feature maps do contribute to the detection of adversarial examples after becoming part of a sentence, as shown in Section~\ref{sec: ablation}.

\subsection{Ablation Study}
\label{sec: ablation}
To help understand the proposed detector, we conduct two ablation studies. The first is to modify the word embedding layer by masking out some of the generated word vectors. The second ablation study is to replace the sentiment analyzer with different neural network architectures.

\subsubsection{Ablation of the Word Embedding Layer}
{\color{black}
Table~\ref{tab: ablation_words} evaluates the impact of the number of selected hidden layers of the ResNet-34 model on the detection performance of the proposed detector in regards to the adversarial examples generated by the latest attacks (i.e., DeepFool and AutoAttack). The second column indicates the number of selected hidden layers. The third column lists the selected hidden layers.}

As shown in Table~\ref{tab: ablation_words}, the detection of adversarial examples improves steadily with the increasing number of selected hidden layers under the proposed adversarial example detector, across all three considered datasets. This is due to the fact that more observations of the transformations of the image representation in the classifier (or, in other words, a longer sentence) provide richer information to distinguish the adversarial and benign examples. This validates our design of interpreting a series of hidden-layer feature maps as a sentence for effective adversarial example detection.  

As also shown in Table~\ref{tab: ablation_words}, the inclusion (or direct use) of the input image does not always contribute to the improvement of adversarial example detection. As a matter of fact, the table consistently shows that  the adversarial example detection depends primarily on the feature maps produced by the later hidden layers of the DNN under attack, where a perturbed image is increasingly transformed to a different class of natural images in the image classifier. 

\begin{table}[t]
	\caption{\color{black}The AUC score (\%) under different numbers of  hidden layers selected in the ResNet-34-based image classifier for adversarial example detection by the proposed detector. DeepFool and AutoAttack (0.02) are the attack models.}
	\label{tab: ablation_words}
    \renewcommand{\arraystretch}{1}
    \renewcommand\tabcolsep{4pt}
	\centering
	\begin{tabular}{l|c|c|c c c c}
		\hline
		&No.&Word(s)&CIFAR-10&SVHN&CIFAR-100&Avg.\\
		\hline   
		&~&Img& 94.99 & 97.72 &79.71 & 90.81 \\
	    &~&$\text{BN}_1$& 95.37 & 97.34 &76.58 & 89.76 \\
	    &~&$\text{Res}_1$& 96.27 &	97.63 &85.88 & 93.26\\
	    &~1&$\text{Res}_2$& 96.07 &97.70&	83.97 & 92.58\\
	   &~ &$\text{Res}_3$& 96.93 &	\textbf{98.35}& 79.89 & 91.72\\
	    &~&$\text{Res}_4$& \textbf{97.66} &97.62&\textbf{88.90} & 94.73\\
		\cline{2-7}
		\multirow{5}{*}{\rotatebox[origin=c]{90}{DeepFool}}&~&Img, $\text{BN}_1$& 95.43 &96.89 &77.38 & 89.90\\
	    &~&$\text{BN}_1$, $\text{Res}_1$& 96.31 &97.45 &85.49  &93.08\\
	    &~2&$\text{Res}_1$, $\text{Res}_2$& 96.24 &	97.78 &85.48  &93.17\\
	    &~&$\text{Res}_2$, $\text{Res}_3$& 96.88 &98.46&84.91  &93.42\\
	    &~&$\text{Res}_3$, $\text{Res}_4$& \textbf{98.98} &	\textbf{99.34} & \textbf{91.93} &96.75\\
		\cline{2-7}
       &~ &Img, $\cdots$, $\text{Res}_1$& 96.18 &97.43 &84.71  &92.77\\
	    &~3&$\text{BN}_1$, $\cdots$, $\text{Res}_2$& 96.29 &97.69 &85.92  &93.30\\
	    &~&$\text{Res}_1$, $\cdots$, $\text{Res}_3$& 97.47 &98.39 &86.74  &94.20\\
	   &~ &$\text{Res}_2$, $\cdots$, $\text{Res}_4$& \textbf{99.31} &\textbf{99.45} &\textbf{93.33}  &97.36\\
		\cline{2-7}
		&~&Img, $\cdots$, $\text{Res}_2$& 96.07 &97.71 &85.38 &93.05\\
	    &~4&$\text{BN}_1$, $\cdots$, $\text{Res}_3$& 97.28 &98.30 &86.21  &93.93\\
	   &~ &$\text{Res}_1$, $\cdots$, $\text{Res}_4$& \textbf{99.40} &\textbf{99.44} &\textbf{94.12} &97.65\\
	    \cline{2-7}
		&~5&Img, $\cdots$, $\text{Res}_3$& 97.36 &98.39 &86.31  &94.02\\
	    &~&$\text{BN}_1$, $\cdots$, $\text{Res}_4$& \textbf{99.37} &\textbf{99.53} &\textbf{94.38} &\textbf{97.83}\\
	    \cline{2-7}
		&~6&Img, $\cdots$, $\text{Res}_4$& \textbf{99.39} &	99.50 &93.59  &97.49\\
		\hline
        &~ & Img & 99.96 & 99.92 & 99.85 & 99.91  \\
       & ~ & $	\text{BN}_1$ & 99.88 & 99.95 & \textbf{99.95} & 99.93  \\
        &~ & $\text{Res}_1$ & \textbf{99.98} & \textbf{99.96} & 99.94 & 99.96  \\
        &1 & $\text{Res}_2$ & 99.97 & 99.92 & \textbf{99.95} & 99.95  \\
        &~ & $\text{Res}_3$ & 99.87 & 99.70 & 99.81 & 99.79  \\
        &~ & $\text{Res}_4$ & 94.74 & 94.33 & 99.12 & 96.06  \\
                \cline{2-7}
        &~ & Img, $\text{BN}_1$ & 99.86 & 99.95 & 99.62 & 99.81  \\
        &~ & $\text{BN}_1$, $\text{Res}_1$ & \textbf{99.99} & \textbf{99.96} & 99.96 & 99.97  \\
        \multirow{5}{*}{\rotatebox[origin=c]{90}{AutoAttack}}&2 & $\text{Res}_1$, $\text{Res}_2$ & \textbf{99.99} & 99.94 & 99.96 & 99.96  \\
        &~ & $\text{Res}_2$, $\text{Res}_3$ & \textbf{99.99} & \textbf{99.96} & 99.95 & 99.97  \\
        &~ & $\text{Res}_3$, $\text{Res}_4$ & 99.88 & 99.48 & \textbf{99.97} & 99.78  \\
                \cline{2-7}
        &~ & Img, $\cdots$, $\text{Res}_1$ & \textbf{99.99} & \textbf{99.97} & 99.96 & 99.97  \\
        &3 & $\text{BN}_1$, $\cdots$, $\text{Res}_2$ & \textbf{99.99} & 99.95 & 99.98 & 99.97  \\
        &~ & $\text{Res}_1$, $\cdots$, $\text{Res}_3$ & \textbf{99.99} & \textbf{99.97} & 99.98 & 99.98  \\
        &~ & $\text{Res}_2$, $\cdots$, $\text{Res}_4$ & 99.98 & 99.93 & \textbf{99.99} & 99.97  \\
                \cline{2-7}
        &~ & Img, $\cdots$, $\text{Res}_2$ & \textbf{99.99} & 99.96 & 99.98 & 99.98  \\
        &4 & $\text{BN}_1$, $\cdots$, $\text{Res}_3$ & \textbf{99.99} & \textbf{99.97} & 99.98 & 99.98  \\
        &~ & $\text{Res}_1$, $\cdots$, $\text{Res}_4$ & \textbf{99.99} & \textbf{99.97} & 100 & \textbf{99.99}  \\
                \cline{2-7}
        &5 & Img, $\cdots$, $\text{Res}_3$ & \textbf{99.99} & \textbf{99.97} & 99.98 & 99.98  \\
        &~ & $\text{BN}_1$, $\cdots$, $\text{Res}_4$ & \textbf{99.99} & \textbf{99.97} & \textbf{100} & \textbf{99.99}  \\
                \cline{2-7}
        &6 & Img, $\cdots$, $\text{Res}_4$ & \textbf{99.99} & \textbf{99.97} & \textbf{100} & \textbf{99.99}  \\
                \hline
	\end{tabular}
\end{table}

\subsubsection{Ablation of Sentiment Analyzer}The sentiment analyzer employs $n$-gram convolutional kernels to extract features. An $n$-gram convolutional kernel convolute $n$ words each time.
{\color{black}
We first investigate the impact of the number of different $n$-gram convolutional kernels on the performance of adversarial example detection. The latest typical attacks, AutoAttack and DeepFool, are used to produce adversarial examples. 
Table~\ref{tab: n_gram} shows that the best detection accuracy is achieved when all four types of $n$-gram convolutional kernels are implemented in the convolutional layer of the sentiment analyzer, except for AutoAttack on the CIFAR-10 dataset. The reason is that an $n$-gram convolutional kernel attempts to learn local features when $n$ is small, or learn global features when $n$ is large. With all types of $n$-gram convolutional kernels used, features are captured in all scales.}

\begin{table}[t]
	\caption{\color{black}The impact of $n$-gram convolutional kernels in the sentiment analyzer on the AUC score (\%), where DeepFool and AutoAttack (0.02) are the attack models. ResNet-34 is used to build the image classifier.}
	\label{tab: n_gram}
	\centering
	\begin{tabular}{c|c|c|c|c|c c c }
		\hline
		Attack&\multicolumn{4}{c}{$n$-gram}\vline&CIFAR-10&SVHN&CIFAR-100\\
		&1&2&3&4&&&\\
		\hline   
		&\checkmark&\checkmark&&&99.40 &99.44 &94.31  \\
		&\checkmark& &\checkmark&& 99.42 &99.47 &94.19  \\
		&\checkmark& & &\checkmark & 99.39 &99.46 &94.41\\
		DeepFool&&\checkmark&\checkmark&& 99.42 &99.48 &94.37  \\
		&&\checkmark&&\checkmark& 99.37&99.40 &94.38  \\
		&&&\checkmark&\checkmark& 99.42 &99.39 &94.28  \\
		&\checkmark&\checkmark&\checkmark&& 99.40 &99.49 &93.95 \\
		&&\checkmark&\checkmark&\checkmark& 99.39 &	99.45 &94.03 \\
		&\checkmark&\checkmark&\checkmark&\checkmark& \textbf{99.43}&\textbf{99.53}&\textbf{94.52} \\
		\hline 
		&\checkmark&\checkmark&&& 99.99& 99.97& 99.99 \\
		&\checkmark& &\checkmark& & 99.99& 99.97& 99.99 \\
		&\checkmark& & &\checkmark & 99.99& 99.97& 99.99 \\
		AutoAttack&&\checkmark&\checkmark&& 99.99& 99.97& 99.99 \\
		&&\checkmark&&\checkmark& 100& 99.96& 100 \\
		&&&\checkmark&\checkmark& 100& 99.97& 99.99 \\
		&\checkmark&\checkmark&\checkmark& & 100& 99.97& 100 \\
		&&\checkmark&\checkmark&\checkmark& 100& 99.97& 100 \\
		&\checkmark&\checkmark&\checkmark&\checkmark& 99.99& 99.97& 100 \\
		\hline 
	\end{tabular}
\end{table}

\begin{figure}[t]
	\centering
	\begin{subfloat}[CNN-based Detector]{\includegraphics[height=1 in]{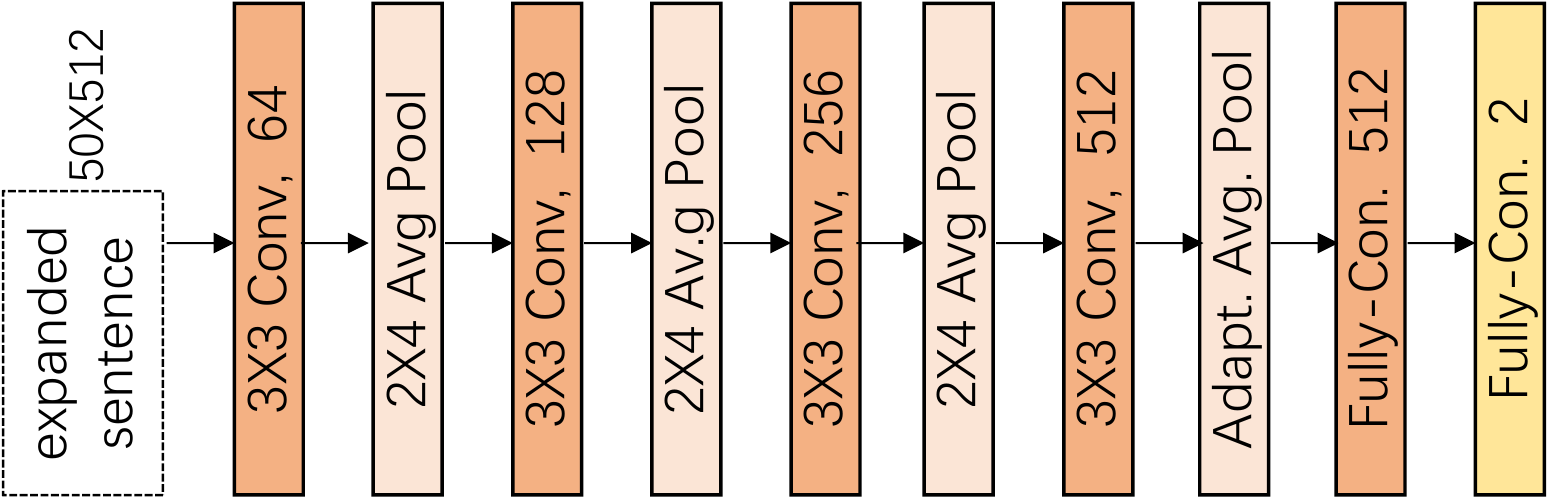}}\end{subfloat}
	\begin{subfloat}[BiLSTM-based Detector]{\includegraphics[height=1 in]{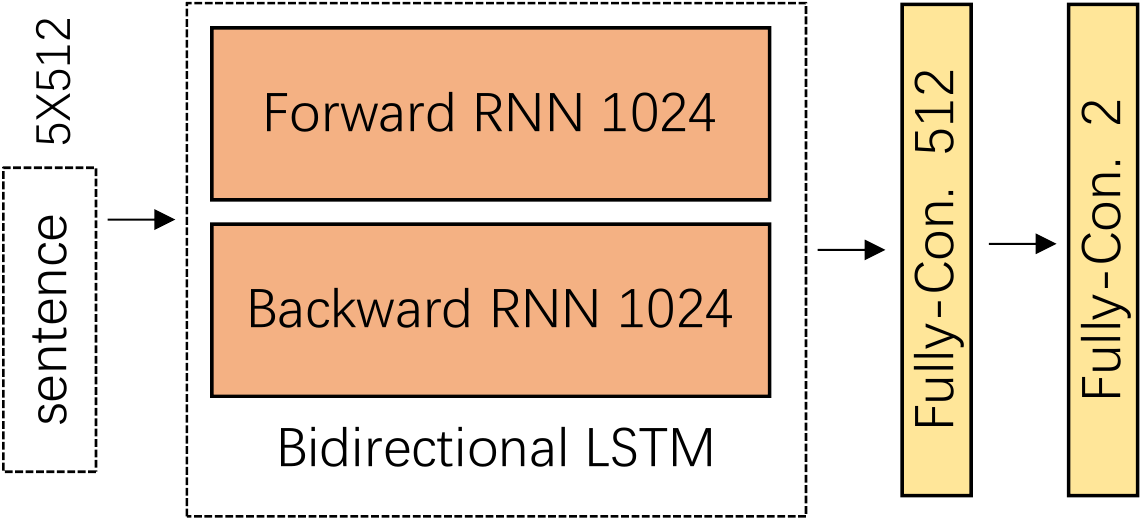}}\end{subfloat}
	\begin{subfloat}[TextRCNN-based Detector]{\includegraphics[height=1 in]{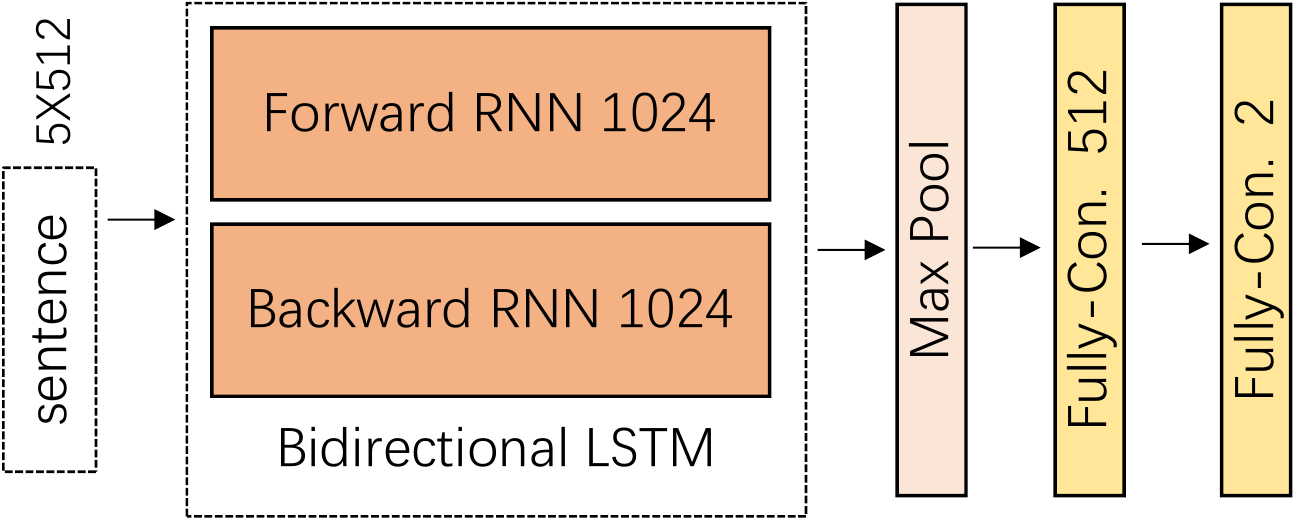}}\end{subfloat}
	\caption{The architectures of neural networks used as the alternative sentiment analyzers.}
	\label{fig: alterNN}
\end{figure}

We proceed to validate our selection of TextCNN, compared to other commonly used sentiment analyzers: CNN~\cite{8998186}, bidirectional long short-term memory network (BiLSTM)~\cite{9107432}, and recurrent convolutional neural networks for text classification (TextRCNN)~\cite{DBLP:conf/aaai/LaiXLZ15}. 
The structures of the alternative sentiment analyzers are illustrated in Fig.~\ref{fig: alterNN}. 
\begin{itemize}
    \item \textbf{CNN-based sentiment analyzer:} The CNN-based sentiment analyzer comprises four convolutional layers, three max-pooling layers, an adapted average-pooling layer, and two fully-connected layers. The hyper-parameters of the convolutional layer are: The kernel size is $3\times3$, the padding size is 1, and the stride size is 1. The hyper-parameters of the max-pooling layer are: The kernel size is $2\times4$, and the stride size is $2\times 4$. The output size of the adapted average-pooling layer is $2\times2$. We replicate the word vectors from the word embedding layer ten times to construct an expanded sentence input to the CNN. 
    
    \item \textbf{BiLSTM-based sentiment analyzer:} BiLSTM consists of a bidirectional LSTM, followed by two fully-connected layers. The BiLSTM outputs two vectors with a length of 1024, which are concatenated and fed into the first fully-connected layer. 
    
    \item \textbf{TextRCNN-based sentiment analyzer:} TextRCNN shares the same structure as BiLSTM, except for an additional 1D max-pooling layer between the BiLSTM and the first fully-connected layer. The 1D max-pooling layer is used to choose the maximum value of each position among all vectors produced by the BiLSTM. 
    
    \item \textbf{TextCNN-based sentiment analyzer:} TextCNN consists of an $n$-gram convolutional layer, a global max-pooling layer, and a fully-connected layer. The $n$-gram convolutional layer contains 1-, 2-, 3-, and 4-gram convolutional kernels, each with 100 instances. The global max-pooling layer outputs four vectors with a length of 100 per vector (the output  of an instance of an $n$-gram convolutional kernel is reduced to a scalar by the global max-pooling layer, and the outputs of the 100 instances are concatenated into a $1\times100$ vector), which are then concatenated into a $1\times 400$ vector and fed to the fully-connected layer. 
\end{itemize}
DeepFool is used to launch the attack, since it is one of the currently most hard-to-detect attacks~\cite{DBLP:conf/cvpr/CohenSG20}. 

{\color{black}
Table~\ref{tab: ablation_nn} compares the proposed TextCNN-based detector and the above alternative Sentiment analyzers. It is observed that the proposed TextCNN-based detector outperforms its CNN-based, BLSTM-based, and TextRCNN-based counterparts in all considered datasets (i.e., CIFAR-10, SVHN, and CIFAR-100), despite the fact that the number of its learnable parameters is more than halved compared to the CNN-based detector (i.e., \textbf{$2.06\times 10^6$} in the TextCNN-based detector vs. $4.15\times 10^6$ in the CNN-based detector). Moreover, TextCNN requires a substantially lower number of model parameters than the other considered sentiment analyzer structures, e.g., by an order of magnitude, as compared to BiLSTM and TextRCNN. The TextCNN is not only superior in adversarial example detection, but also much more computationally efficient. Our sentence interpretation of feature maps and subsequent adoption of TextCNN-based sentiment analysis is effective in detecting adversarial perturbations on images.}

\begin{table}[t]
	\caption{AUC scores (\%) when detecting DeepFool adversarial examples by using different sentiment analyzers.}
	\label{tab: ablation_nn}
	\centering
	\begin{tabular}{m{8em}|m{5em} m{4.5em} m{2.5em} m{5em}} 
		\hline
		Detector&No. Param.&CIFAR-10&SVHN&CIFAR-100\\
		\hline
		CNN-based&$4.15\times 10^6$&99.32&99.51&93.63\\
		BiLSTM-based &$4.04\times 10^7$&99.37&99.51&93.99\\
		TextRCNN-based &$4.04\times 10^7$&99.38&\textbf{99.53}&94.34\\
		TextCNN (ours)&$2.06\times 10^6$&\textbf{99.43}&\textbf{99.53}&\textbf{94.52}\\
		\hline   
	\end{tabular}
\end{table}

\subsection{Generalization of the New Detector}
We further assess the generalization capability of the proposed TextCNN-based adversarial example detector, where the detector is trained on perturbed examples generated by one attack model and tested on examples perturbed by another attack model. The generalization ability is important in situations where the detector has no knowledge of attacks in prior.  

 \begin{table*}[t]
	\caption{\color{black}Evaluation of the generalization of the new detector when  detecting adversarial examples generated by other attack algorithms. The metric is the AUC score (\%). The image classifier is ResNet-34, which achieves the classification accuracy of 93.72\%, 95.97\%, and 75.29\% on benign images in the CIFAR-10, SVHN, and CIFAR-100 datasets, respectively.}
	\label{tab:generalization resnet}
    \renewcommand\tabcolsep{4pt}
	\footnotesize
	\centering
	\begin{tabular}{l | l|c c c c c c c c c c c c}
		\hline
		&Train Against&\multicolumn{12}{c}{Attacking Algorithms}\\ 
		&&\rotatebox[origin=c]{-90}{FGSM (0.1)} &\rotatebox[origin=c]{-90}{FGSM (8/255)} &\rotatebox[origin=c]{-90}{JSMA (1,0.1)} &\rotatebox[origin=c]{-90}{JSMA (0.8,0.3)} &\rotatebox[origin=c]{-90}{DeepFool}&\rotatebox[origin=c]{-90}{C\&W}&\rotatebox[origin=c]{-90}{PGD (0.02)} &\rotatebox[origin=c]{-90}{PGD (8/255)} &\rotatebox[origin=c]{-90}{EAD}& \rotatebox[origin=c]{-90}{Auto (8/255)} & \rotatebox[origin=c]{-90}{Auto (0.02)}&\rotatebox[origin=c]{-90}{Avg.}\\
		\hline
		\multirow{11}{*}{\rotatebox[origin=c]{90}{CIFAR-10}}
        &FGSM (0.1)&100& 100& 94.44 & 95.56& 93.70 & 100 & \textbf{79.27} & 98.36 &80.80 & 99.38 & 96.41&94.36\\
        \cline{2 - 14}
        &FGSM (8/255)& 100& 100& 94.90 & 96.04& 94.03 & 100& 
        84.18& 99.30 & \textbf{81.99} & 99.79& 98.27&95.32\\
        \cline{2 - 14}
        &JSMA (1,0.1)&100 & 96.18&100 & 99.99&96.98 & 99.96 & \textbf{69.80}& 89.50& 95.96&95.19&88.85&93.86\\
        \cline{2 - 14}
        &JSMA (0.8,0.3)& 100& 96.92& 100& 99.99& 97.39& 99.99& \textbf{78.27}& 92.60& 96.46& 96.59&93.39&95.60\\
        \cline{2 - 14}
	    &DeepFool&100 & 99.93&99.40 & 99.61&99.43 & 100& \textbf{96.06}& 99.45&99.06&99.88&99.67&\textbf{99.32}\\
        \cline{2 - 14}
        &C\&W&100 & 100&95.31 & 96.37&93.89 & 100& 81.73& 98.93&\textbf{81.71}& 99.64& 97.29& 94.99\\
        \cline{2 - 14}
		&PGD (0.02)&97.69 & 100&93.44 & 95.28& \textbf{93.20}& 99.78& 99.75& 100 &93.50& 100& 99.99&97.51\\
        \cline{2 - 14}
        &PGD (8/255)& 100& 100& 93.17& 94.82& 94.79& 100& 95.15& 100& \textbf{84.07}& 100& 99.98&96.54\\
        \cline{2 - 14}
		&EAD& 99.89& 98.69& 99.96& 99.95&99.11 &99.61 & \textbf{88.93}& 97.58 &99.47 & 99.41 & 98.90&98.32\\
        \cline{2 - 14}
        &Auto (8/255)& 100& 100& 93.42& 94.75& 94.86& 100& 91.90& 99.99& \textbf{84.42}& 100& 99.96&96.30\\
        \cline{2 - 14}
        &Auto (0.02)& 100& 100& 94.32& 95.96& 96.35& 100& 98.85& 99.99 & \textbf{90.32} & 100& 99.99& 97.80\\
		\hline
		\multirow{11}{*}{\rotatebox[origin=c]{90}{SVHN}}
        &FGSM (0.1)& 100& 99.09& 93.76& 93.53 &94.59 &99.96 & \textbf{77.94} & 96.86&79.71&97.52&91.00&93.09\\
        \cline{2 - 14}
        &FGSM (8/255)& 100& 99.90& 94.86& 94.78& 95.62& 100&84.79&99.17&\textbf{82.18}& 99.53 &96.62& 95.22\\
        \cline{2 - 14}
        &JSMA (1,0.1)& 99.48& 87.74& 100& 100& 97.24& 95.97& \textbf{77.30}& 87.06&98.69&92.34&92.48& 93.48\\
        \cline{2 - 14}
        &JSMA (0.8,0.3)& 99.75& 92.58& 100& 100& 98.26& 98.39& \textbf{82.87}& 93.37&98.92&96.32&95.60& 96.01\\
        \cline{2 - 14}
	    &DeepFool& 100& 98.20& 99.78& 99.69& 99.53& 99.97& \textbf{94.07}& 98.24&98.68&99.03&98.63& \textbf{98.71}\\
        \cline{2 - 14}
        &C\&W&100 & 99.78&94.74 & 94.67& 95.45& 100& 83.90& 98.87&\textbf{81.86}&99.23&95.77& 94.93\\
        \cline{2 - 14}
		&PGD (0.02)&98.05 & 100& 99.45& 97.60& 97.49& 99.05& 99.56& 99.99&\textbf{90.82}&99.99&99.88& 98.35\\
        \cline{2 - 14}
        &PGD (8/255)& 99.98& 100& 94.59& 94.45& 95.88& 99.99& 94.84& 100& \textbf{82.53}&99.98&99.94& 96.56\\
        \cline{2 - 14}
		&EAD& 99.76& 93.93& 99.99& 99.91&99.02 & 99.05& \textbf{88.90}& 96.65&99.73&98.72&98.10& 97.61\\
        \cline{2 - 14}
        &Auto (8/255)& 99.97& 100& 93.06& 92.84& 96.17& 99.98& 92.98& 99.98& \textbf{84.84}& 100& 99.87&96.34\\
        \cline{2 - 14}
        &Auto (0.02)& 98.94& 99.77& 98.68& 98.73& 96.42& 99.43& 97.26& 99.94& \textbf{91.64}& 99.90& 99.97&98.24\\
		\hline
		\multirow{11}{*}{\rotatebox[origin=c]{90}{CIFAR-100}}
        &FGSM (0.1)& 100& 99.94& 98.07& 98.61 &64.88 & 100&76.66&98.04&\textbf{59.15}& 98.86&95.60&89.98\\
        \cline{2 - 14}
        &FGSM (8/255)& 100& 100& 98.08& 98.50 & 66.86& 100& 81.06& 99.28& \textbf{60.17}& 99.76& 98.09& 91.07\\
        \cline{2 - 14}
        &JSMA (1,0.1)& 100& 99.50& 99.99& 100& \textbf{66.17}&99.98 & 76.90& 96.36&71.36&96.95&89.55& 90.61\\
        \cline{2 - 14}
        &JSMA (0.8,0.3)& 100& 98.79& 100& 100& \textbf{68.53}& 99.97& 78.16& 95.79& 73.91& 94.24& 86.78& 90.56\\
        \cline{2 - 14}
	    &DeepFool& 99.80& 99.13& 96.11& 97.16& 94.52& 98.96& 97.74& 91.65&91.72&\textbf{34.77}&39.70& 85.57\\
        \cline{2 - 14}
        &C\&W& 100& 99.98& 98.21& 98.57& 65.88& 100& 79.66& 98.72&\textbf{59.95}&99.37&96.73& 90.64\\
        \cline{2 - 14}
		&PGD (0.02)&99.82 & 100&97.79 & 97.80&88.39 & 99.88&99.50 & 99.94&\textbf{84.11}&100&99.98&\textbf{97.02}\\
        \cline{2 - 14}
        &PGD (8/255)& 100& 100& 97.34& 97.80& 71.65& 100& 94.47& 100& \textbf{63.79}&99.96&99.76&93.16\\
        \cline{2 - 14}
		&EAD& 99.98& 99.43&98.69 & 99.79& 92.46&97.72 & 94.18& 94.80 &94.39&48.03&\textbf{45.04}&87.68\\
        \cline{2 - 14}
        &Auto (8/255)& 100& 100& 96.61& 97.12& 70.15& 100& 90.99& 99.98& \textbf{62.52}&100&99.97&92.48\\
        \cline{2 - 14}
        &Auto (0.02)& 100& 100& 96.35& 97.36& 72.43& 100& 95.85& 99.98& \textbf{64.20}&100&100&93.29\\
		\hline
	\end{tabular}
\end{table*}

\begin{table*}[!h]
	\caption{\color{black}Evaluation of the generalization of the new detector when detecting adversarial examples generated by other attack algorithms. The used metric is the AUC score (\%). The image classifier is Inception-V3, which achieves the classification accuracy of 93.83\%, 95.79\%, and 73.01\% on benign images in the CIFAR-10, SVHN, and CIFAR-100 datasets, respectively.}
	\label{tab:generalization inception}
    \renewcommand\tabcolsep{4pt}
	\footnotesize
	\centering
	\begin{tabular}{l | l|c c c c c c c c c c c c}
		\hline
		&Train Against&\multicolumn{8}{c}{Attacking Algorithms}\\
		&&\rotatebox[origin=c]{-90}{FGSM (0.1)} &\rotatebox[origin=c]{-90}{FGSM (8/255)} &\rotatebox[origin=c]{-90}{JSMA (1,0.1)} &\rotatebox[origin=c]{-90}{JSMA (0.8,0.3)} &\rotatebox[origin=c]{-90}{DeepFool}&\rotatebox[origin=c]{-90}{C\&W}&\rotatebox[origin=c]{-90}{PGD (0.02)} &\rotatebox[origin=c]{-90}{PGD (8/255)} &\rotatebox[origin=c]{-90}{EAD}& \rotatebox[origin=c]{-90}{Auto (8/255)} & \rotatebox[origin=c]{-90}{Auto (0.02)}&\rotatebox[origin=c]{-90}{Avg.}\\
		\hline
		\multirow{11}{*}{\rotatebox[origin=c]{90}{CIFAR-10}}
        &FGSM (0.1)&100 &99.46 &97.47 &97.63 &93.99 &84.27&\textbf{67.29}&81.20&90.22&93.45&87.94&90.27 \\
        \cline{2 - 14}
        &FGSM (8/255)&100 &100 &97.94 &98.11 &94.48 &\textbf{87.51} &88.65&99.39&89.14&99.92&99.36&95.86 \\
        \cline{2 - 14}
        &JSMA (1,0.1)&100 &98.66 &99.98 &99.98 &97.48 &97.19 &\textbf{73.72} &85.05 &98.27&93.21&88.55&93.83\\
        \cline{2 - 14}
        &JSMA (0.8,0.3)&100 &99.39 &99.98 &99.98 &97.82 &97.24 &\textbf{75.65} &88.45 &98.25&95.57&91.61&94.90\\
        \cline{2 - 14}
	    &DeepFool&100 &99.96 &99.69 &99.70 &99.22 &98.67 &\textbf{93.95} &98.77 &99.05&99.87&99.56&\textbf{98.95}\\
        \cline{2 - 14}
        &C\&W&99.64 &98.98 &99.59 &99.63 &98.63 &99.36 &\textbf{89.60} &95.07 &98.79&97.75&97.11&97.65\\
        \cline{2 - 14}
		&PGD (0.02)&99.94 &99.94 &98.97 &99.00 &97.41 &97.41 &98.63 &99.78 &\textbf{95.47}&99.82&99.67&98.73\\
        \cline{2 - 14}
        &PGD (8/255)&100 &100 &97.53 &97.74 &94.38 &89.30 &95.65 &99.95 &\textbf{88.46}&99.98&99.84& 96.62\\
        \cline{2 - 14}
		&EAD&100 &99.80 &99.90 &99.91 &99.06 &99.05 &\textbf{88.54} &95.97 &99.29&99.16&98.23&98.08\\
        \cline{2 - 14}
        &Auto (8/255)&100 &100 &97.78 &97.95 &94.75 &90.60 &94.58 &99.91 &\textbf{89.28}&100&99.93&96.80\\
        \cline{2 - 14}
        &Auto (0.02)&100 &100 &98.46 &98.61 &96.04 &94.57 &96.54 &99.90 &\textbf{92.23}&100&99.96&97.85\\
		\hline
		\multirow{11}{*}{\rotatebox[origin=c]{90}{SVHN}}
        &FGSM (0.1)&100 &96.53 &96.75 &96.86 &95.91 &87.38 &\textbf{72.08}&83.83&90.26&91.81&85.93&90.67 \\
        \cline{2 - 14}
        &FGSM (8/255)&100 &99.79 &95.45 &95.45 &95.11&\textbf{70.94} &71.61&96.03&82.44&98.34&90.77&90.54 \\
        \cline{2 - 14}
        &JSMA (1,0.1)&99.87 &95.81 &99.95 &99.95 &98.27 &97.88 &\textbf{78.65} &89.69 &98.85&96.39&93.88&95.38\\
        \cline{2 - 14}
        &JSMA (0.8,0.3)&99.81 &94.88 &99.96 &99.95 &98.10 &97.89 &\textbf{78.55} &88.48 &98.80&95.80&93.40&95.06\\
        \cline{2 - 14}
	    &DeepFool&99.98 &99.17 &99.71 &99.73 &99.32 &98.81 &\textbf{89.96} &97.90 &98.78&99.10&97.99&\textbf{98.22}\\
        \cline{2 - 14}
        &C\&W&97.32 &88.43 &98.93 &99.05 &95.71 &99.26 &\textbf{80.34} &84.43 &97.87&93.22&93.02&93.42\\
        \cline{2 - 14}
		&PGD (0.02)&99.16 &96.97 &99.04 &99.12 &97.96 &98.16 &\textbf{95.89} &98.48 &96.86&98.58&98.57&98.07\\
        \cline{2 - 14}
        &PGD (8/255)&99.92 &99.89 &97.67 &97.83 &97.02 &85.47 &\textbf{85.09} &99.81 &88.48&99.89&99.14&95.47\\
        \cline{2 - 14}
		&EAD &99.07 &93.68 &99.66 &99.66 &98.28 &98.94 &\textbf{83.75} &92.24&99.12&96.57&95.59&96.05\\
        \cline{2 - 14}
        &Auto (8/255)&99.89 &99.83 &97.64 &97.82 &96.86 &85.20 &\textbf{83.21} &99.64 &88.56&99.93&99.20&95.25\\
        \cline{2 - 14}
        &Auto (0.02)&99.61 &99.35 &98.70 &98.90 &97.69 &91.97 &\textbf{90.76} &99.58 &92.87&99.88&99.74&97.19\\
		\hline
		\multirow{11}{*}{\rotatebox[origin=c]{90}{CIFAR-100}}
        &FGSM (0.1)&100 &96.57 &95.72 &95.39 &59.48 &59.68&63.40 &75.97&56.65&\textbf{32.54}&26.71&69.28 \\
        \cline{2 - 14}
        &FGSM (8/255)&100 &100 &99.14 &99.14 &64.96 &75.66 &82.96&99.02&\textbf{62.98}&84.48&69.22&85.23 \\
        \cline{2 - 14}
        &JSMA (1,0.1)&100 &99.33 &99.97 &99.98 &68.67 &84.88 &82.15 &94.74 &70.38&60.83&\textbf{52.82}&83.07\\
        \cline{2 - 14}
        &JSMA (0.8,0.3)&100 &99.13 &99.97 &99.99 &70.47 &86.76 &82.66 &94.43 &71.84&55.80&\textbf{48.86}&82.72\\
        \cline{2 - 14}
	    &DeepFool&93.72 &95.44 &91.66 &92.06 &89.52 &96.32 &82.00 &76.57 &86.18&\textbf{73.50}&92.17&88.10\\
        \cline{2 - 14}
        &C\&W&96.81 &96.13 &91.86 &92.32 &83.37 &96.60 & 85.01&83.46 &80.45&\textbf{29.75}&31.22&78.82\\
        \cline{2 - 14}
		&PGD (0.02)&99.90 &99.70 &98.68 &98.63 &80.06 &97.25 &96.77 &99.06 &77.08&57.45&\textbf{55.38}&87.27\\
        \cline{2 - 14}
        &PGD (8/255)&100 &99.99 &99.42 &99.41 &67.49 &85.26 &92.82 &99.89 &\textbf{65.72}&92.41&85.78&\textbf{89.84}\\
        \cline{2 - 14}
		&EAD&98.01 &96.71 &95.43 &95.79 &90.56 &96.03 &82.29 &78.96 &89.73&\textbf{29.43}&31.19&80.38\\
        \cline{2 - 14}
        &Auto (8/255)&100 &99.98 &99.32 &99.29 &\textbf{57.64} &72.57 &85.08 &99.58 &58.24&100&99.94&88.33\\
        \cline{2 - 14}
        &Auto (0.02)&99.98 &99.96 &99.06 &99.01 &58.90 &73.71 &87.89 &99.59 &\textbf{57.91}&100&99.98&88.73\\
		\hline
	\end{tabular}
\end{table*}

{\color{black}
Tables~\ref{tab:generalization resnet} and~\ref{tab:generalization inception} show the proposed detector has good generalization ability, with an average detection rate of more than 90\% in most cases (see the last columns of the tables). The detector demonstrates effective generalization in detecting attacks launched by various recent attack models (such as AutoAttack, FGSM, PGD, EAD, C\&W, and JSMA) when trained against DeepFool on the CIFAR-10 and SVHN datasets. Consistent results are observed when the image classifier is ResNet-34 or Inception-V3. 

In the presence of unseen attacks, the detector achieves an average AUC score of more than 98\% on the CIFAR-10 and SVHN datasets. When trained against PGD on the CIFAR-100 dataset, the proposed detector generalizes well to detect unseen attacks, with an average AUC score of over 93\% for the ResNet-34 image classifier and over 87\% for the Inception-V3 image classifier. However, when the unseen attack is AutoAttack, the detection performance of the detector trained against EAD falls below 50\%. To this end, an ensemble solution with detectors trained against DeepFool and PGD is recommended to ensure adequate generalization in detecting unseen attacks, especially for classification tasks with a large number of classes, e.g., CIFAR-100.

It is also noticed that the proposed detector performs better on the CIFAR-10 dataset than it does on the CIFAR-100 dataset, because its detection performance relies on the feature maps produced by the image classifier. The classification accuracy of the ResNet-34 (and Inception-V3) image classifier drops about 20\% when switching from CIFAR-10 to CIFAR-100; in other words, the feature maps of the image classifiers capture more comprehensive features from the CIFAR-10 dataset than they do from the CIFAR-100 dataset. 
}

\begin{figure*}[t]
	\centering
	\begin{subfloat}[CIFAR-10]{\includegraphics[width=2.3in]{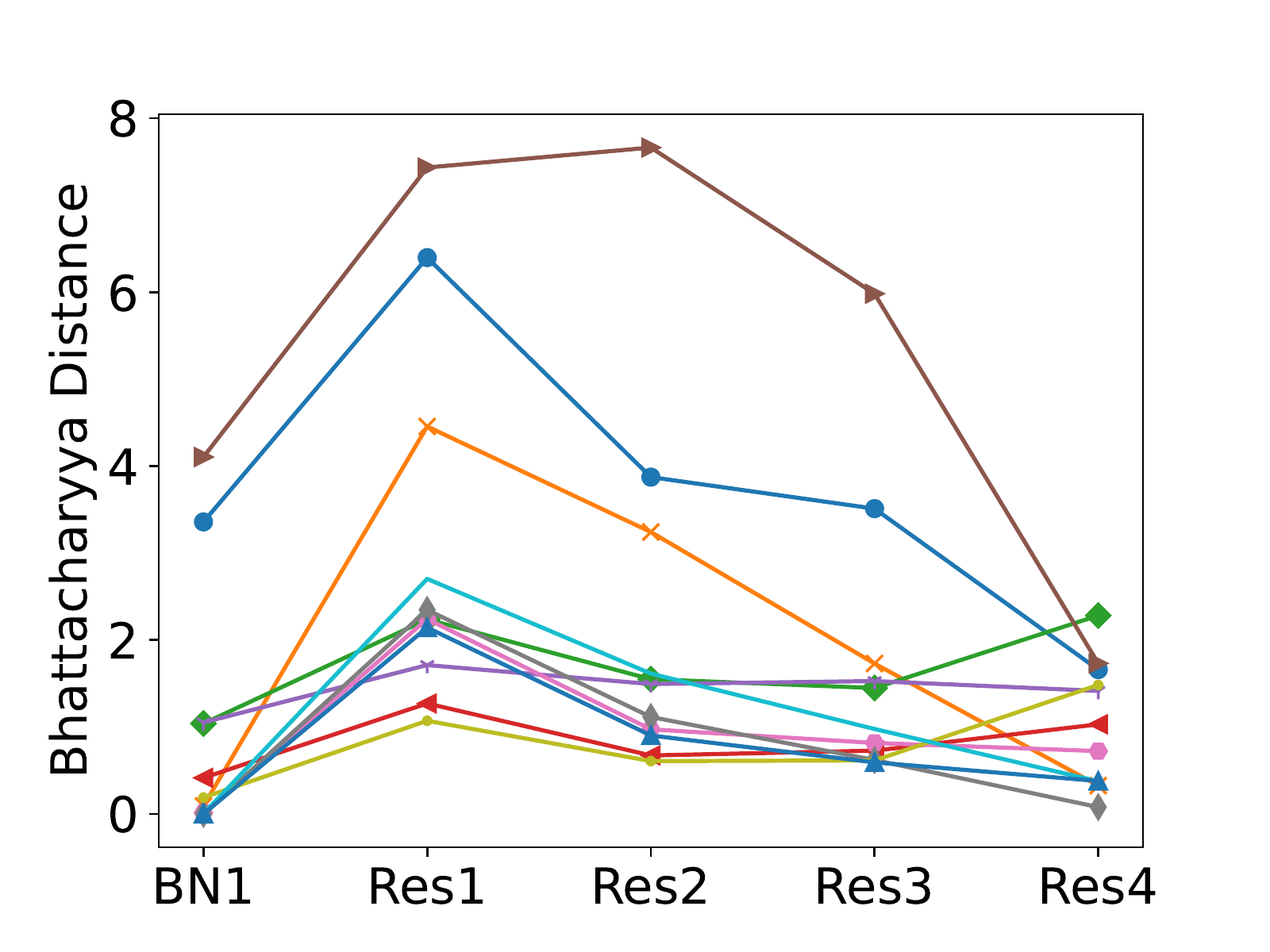}}\end{subfloat}
	\begin{subfloat}[SVHN]{\includegraphics[width=2.3in]{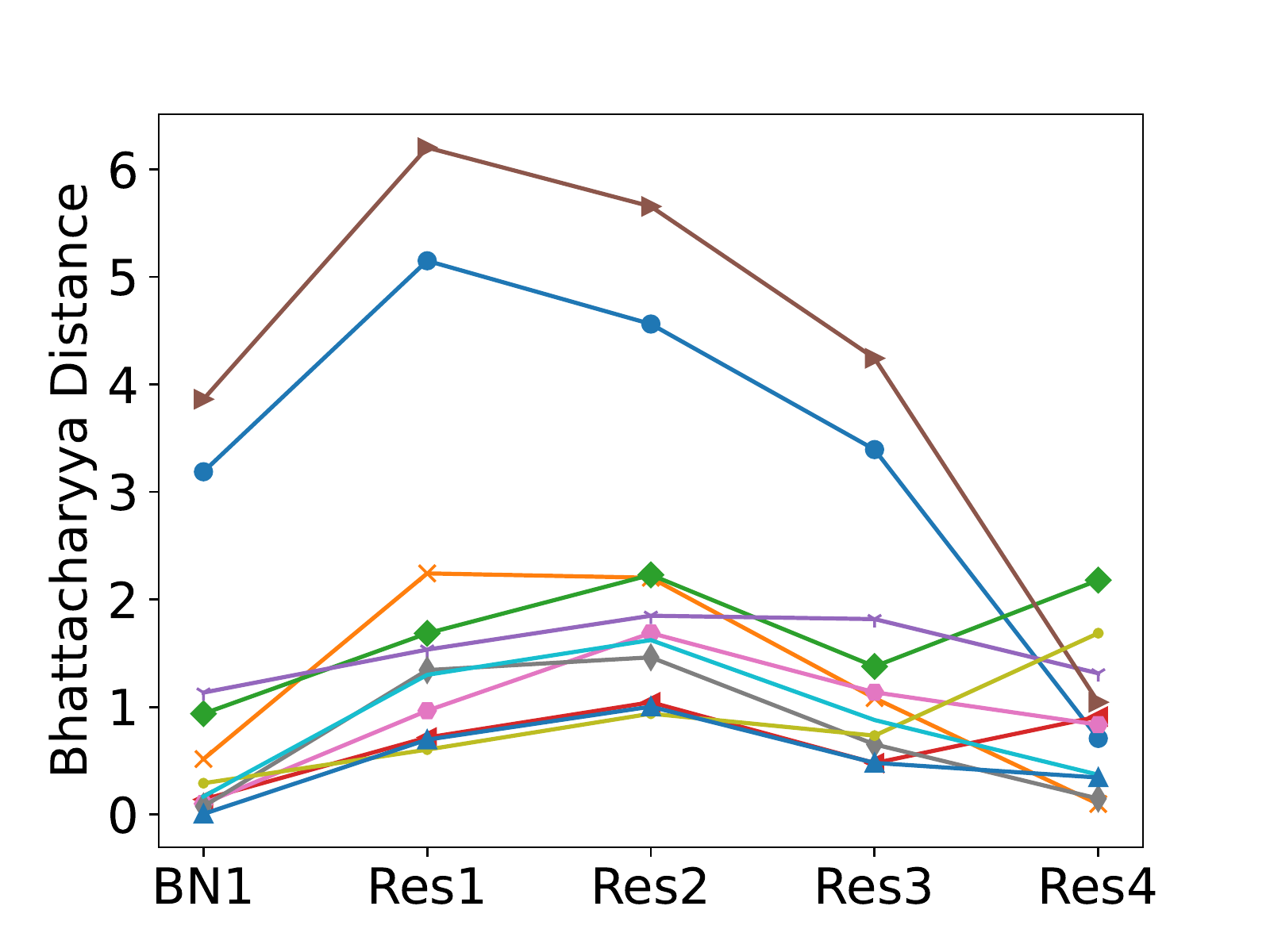}}	\end{subfloat}
	\begin{subfloat}[CIFAR-100]{\includegraphics[width=2.3in]{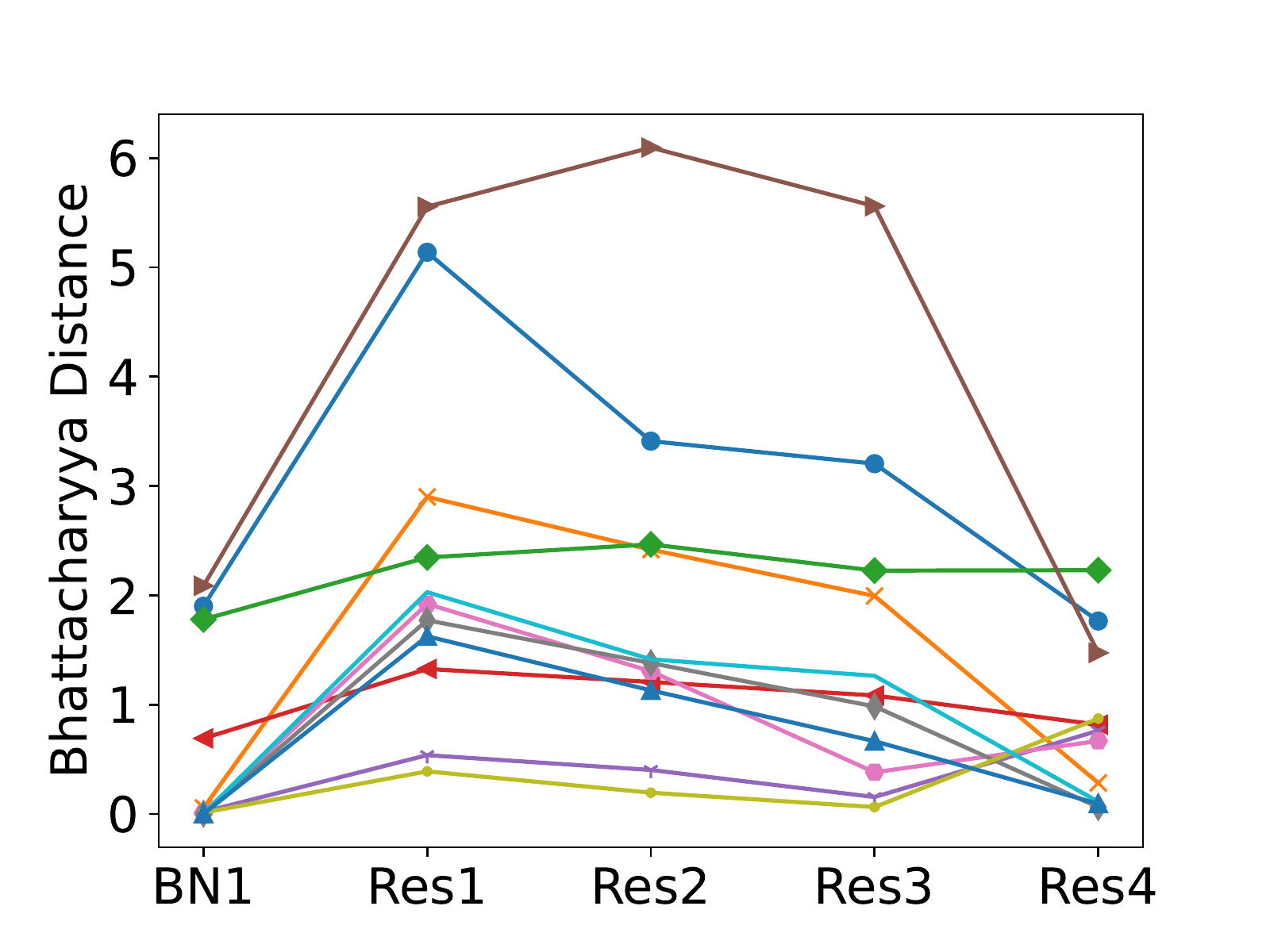}}\end{subfloat}
     \begin{subfloat}{\includegraphics[width=7in]{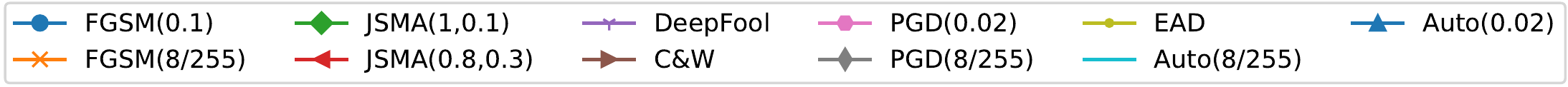}}\end{subfloat}
	\caption{\color{black}The Bhattacharyya distances between the distributions of adversarial and benign examples. The proposed detector trained on adversarial examples generated by DeepFool is used to produce word vectors. The image classifier is ResNet-34. }
	\label{fig: BdistOnAllSets}
\end{figure*}

Fig.~\ref{fig: BdistOnAllSets} reveals that all attacks can cause the perturbed images to deviate from their unperturbed version to different degrees (as can be measured by the Bhattacharyya distance). This leads to the perturbed images being misclassified. Amongst all the considered attack models, DeepFool and PDG (8/255) are the most representative attacks on the 10-class dataset (i.e., CIFAR-10 and SVHN) and 100-class dataset (i.e., CIFAR-100), respectively. As a result, the proposed adversarial example detector trained against DeepFool or PDG (8/255) can be generalized effectively to detect attacks launched by the other attack models, as shown in Table~\ref{tab:generalization resnet}.

\subsection{Defense against White-box Attacks}
\label{sec: adapted}
Last but not least, we consider a relatively rare yet more threatening situation where an attacker can access the gradient of the detector's loss function and adapts the adversarial examples to fool both the image classifier and the proposed detector. 
We use PDG as the benchmark attack (with the perturbation budget $\epsilon = 8/255$, the iteration step size $\alpha = 2/255$, and the iteration number 20 \cite{DBLP:conf/iclr/MadryMSTV18}), since PGD can iteratively refine its perturbation to an image based on the gradients of the loss functions of both the image classifier and the detector with respect to the image.

When performing the adapted PGD attacks, the attacker can take two strategies to combine the image classifier's loss function with the proposed detector's loss function for the generation of new adversarial examples. 
\begin{itemize}
    \item The first strategy is to alternate between minimizing the loss functions of the classifier and the detector~\cite{DBLP:conf/nips/TramerCBM20}. That is, the attacker updates the adversarial examples based on the gradient of the classifier's loss function in odd-numbered steps, and  based on the gradient of the detector's loss function in even-numbered steps.
    \item The second strategy is to linearly combine the two loss functions into one by replacing \eqref{eq: pgd} with~\cite{DBLP:conf/iclr/MetzenGFB17}
\begin{align}
\mathbf{x}^{i+1} =& \textstyle \prod_{\mathbf{x} + \mathcal{S}_{\epsilon}}\bigg(\mathbf{x}^{i} + \alpha\Big((1-\sigma)\,\text{sign}\big(\nabla_{\mathbf{x}}\mathcal{L}_c(\mathbf{x},\mathbf{y})\big) \nonumber \\&+\sigma\, \text{sign}\big(\nabla_\mathbf{x} \mathcal{L}_d(\mathbf{x},\mathbf{y}_d)\big)\Big)\bigg),
\label{eqn: weighted_loss}
\end{align}
where $\mathbf{y}_d$ is the ground-truth class of the input image~$\mathbf{x}$, i.e., adversarial or benign; and $\sigma\in [0,1]$ is a weighting coefficient to balance between the image classifier's loss function $\mathcal{L}_c(\cdot,\cdot)$ and the detector's loss function $\mathcal{L}_d(\cdot,\cdot)$.
    
\end{itemize}
To detect the adapted attacks, the proposed detector is trained on the original adversarial examples produced by PGD (0.02) and then improves its detection by training again on newly generated PGD adversarial examples.

Table~\ref{tab: adpt_alternative_pgd} shows that the proposed detector remains highly effective under the adapted PGD attack based on the first strategy of alternating minimization of the classifier's and detector's loss functions. Specifically, when the attacker refines an adversarial example against the pre-trained detector for 20 iterations (i.e., one epoch), the detection accuracy (i.e., the ratio of detected adversarial examples)
drops to 50.67\%. Nonetheless, our detector improves its accuracy to 95.74\%, after being trained against adversarial examples for five epochs.

Table~\ref{tab: adpt_comb_updated_pgd} shows that the proposed detector is effective under the adapted PGD attack using the second strategy of combined classifier's and detector's loss function. The adversarial example detector is so robust after being trained five epochs in the first strategy that it can achieve the detection accuracy of over 98\% for $\sigma<1$. When the attacker optimizes the adversarial examples solely on the loss function of the detector, i.e., $\sigma=1$, the detection accuracy drops from 98.15\% to 93.04\%. Nevertheless, the  attack success rate of the newly generated adversarial examples drops dramatically to only 7.79\%. The conclusion drawn is that the proposed detector can effectively withstand white-box attacks.

\begin{table}[t]
	\caption{The adapted PDG attack on the CIFAR-10 dataset using the classifier's and the new detector's loss functions in an alternating manner, i.e., $\sigma=0$ in odd-numbered steps and $\sigma=1$ in even-numbered steps. }
	\label{tab: adpt_alternative_pgd}
	\centering
	\begin{tabular}{c|c | c c}
		\hline
		{No. epochs}&Attack success rate (\%) &\multicolumn{2}{c}{Detection}\\
		&&Accuracy (\%) &AUC\\
		\hline   
		0&67.73&50.67&0.9007\\
		1&58.67&93.57&0.9801\\
		2&60.20&91.45&0.9826\\
		3&62.10&93.52&0.9912\\
		4&61.38&94.79&0.9923\\
		5&60.77&95.74&0.9906\\
	    \hline
	\end{tabular}
\end{table}

\begin{table}[t]
	\caption{The adapted PDG attack on the CIFAR-10 using the combined classifier's and detector's loss function.} 
	\label{tab: adpt_comb_updated_pgd}
	\centering
	\begin{tabular}{c|c | c c}
		\hline
		$\sigma$&Attack success rate (\%) &\multicolumn{2}{c}{Detection}\\
		&&Accuracy (\%) &AUC\\
		\hline   
		0&96.29&	98.12&	0.9999\\
		0.2&68.25&	98.94&	0.9998\\
		0.4&61.19&	98.74&	0.9997\\
		0.6&55.55&	98.45&	0.9995\\
		0.8&48.01&	98.15&	0.9991\\
		1&7.69&	93.04&	0.9810\\
	    \hline
	\end{tabular}
\end{table}

\section{Conclusion}
\label{sec: conclusion}
In this paper, we proposed a new adversarial example detector by recasting the adversarial image detection as a text sentiment analysis problem and performing a binary classification using a TextCNN model. Extensive tests demonstrated the superiority of the detector in detecting various latest attacks on three popular datasets. The new detector also demonstrated a strong generalization ability by accurately detecting adversarial examples generated by unknown attacks, and is resistant to white-box attacks in situations where the gradients of the detector are exposed. The new detector only has about 2 million parameters, and takes shorter than 4.6 milliseconds to detect an adversarial example generated by the latest attack models using a Tesla K80 GPU card.

%
%

\ifCLASSOPTIONcaptionsoff
  \newpage
\fi

\bibliographystyle{IEEEtran}
\bibliography{dnndetector}

\begin{thebibliography}{10}
\providecommand{\url}[1]{#1}
\csname url@samestyle\endcsname
\providecommand{\newblock}{\relax}
\providecommand{\bibinfo}[2]{#2}
\providecommand{\BIBentrySTDinterwordspacing}{\spaceskip=0pt\relax}
\providecommand{\BIBentryALTinterwordstretchfactor}{4}
\providecommand{\BIBentryALTinterwordspacing}{\spaceskip=\fontdimen2\font plus
\BIBentryALTinterwordstretchfactor\fontdimen3\font minus
  \fontdimen4\font\relax}
\providecommand{\BIBforeignlanguage}[2]{{%
\expandafter\ifx\csname l@#1\endcsname\relax
\typeout{** WARNING: IEEEtran.bst: No hyphenation pattern has been}%
\typeout{** loaded for the language `#1'. Using the pattern for}%
\typeout{** the default language instead.}%
\else
\language=\csname l@#1\endcsname
\fi
#2}}
\providecommand{\BIBdecl}{\relax}
\BIBdecl

\bibitem{9013065}
D.~J.Miller \emph{et~al.}, ``Adversarial learning targeting deep neural network
  classification: A comprehensive review of defenses against attacks,''
  \emph{Proc. IEEE}, vol. 108, no.~3, pp. 402--433, 2020.

\bibitem{DBLP:conf/iclr/MadryMSTV18}
A.~Madry \emph{et~al.}, ``Towards deep learning models resistant to adversarial
  attacks,'' in \emph{Proc. {ICLR}, Vancouver, BC, Canada, April 30 - May 3,
  2018}.

\bibitem{8844593}
A.~Chernikova \emph{et~al.}, ``Are self-driving cars secure? {Evasion} attacks
  against deep neural networks for steering angle prediction,'' in \emph{Proc.
  {SPW} 2019}, 2019, pp. 132--137.

\bibitem{9252132}
Y.~Zhong \emph{et~al.}, ``Towards transferable adversarial attack against deep
  face recognition,'' \emph{IEEE Trans. Info. Forensics Security}, vol.~16, pp.
  1452--1466, 2021.

\bibitem{DBLP:conf/iclr/Ma0WEWSSHB18}
X.~Ma \emph{et~al.}, ``Characterizing adversarial subspaces using local
  intrinsic dimensionality,'' in \emph{Proc. {ICLR}, Vancouver, BC, Canada,
  April 30 - May 3, 2018}.

\bibitem{DBLP:journals/corr/abs-1803-04765}
N.~Papernot \emph{et~al.}, ``Deep k-nearest neighbors: Towards confident,
  interpretable and robust deep learning,'' \emph{CoRR}, vol. abs/1803.04765,
  2018.

\bibitem{DBLP:conf/cvpr/CohenSG20}
G.~Cohen \emph{et~al.}, ``Detecting adversarial samples using influence
  functions and nearest neighbors,'' in \emph{Proc. {CVPR}, Seattle, WA, USA,
  June 13-19, 2020}.

\bibitem{DBLP:conf/nips/LeeLLS18}
K.~Lee \emph{et~al.}, ``A simple unified framework for detecting
  out-of-distribution samples and adversarial attacks,'' in \emph{Proc.
  {NeurIPS}, Montr{\'{e}}al, Canada, December 3-8, 2018}, pp. 7167--7177.

\bibitem{DBLP:journals/corr/abs-2209-00005}
Z.~He \emph{et~al.}, ``Be your own neighborhood: Detecting adversarial example
  by the neighborhood relations built on self-supervised learning,''
  \emph{CoRR}, vol. abs/2209.00005, 2022.

\bibitem{LUO2022108383}
W.~Luo, C.~Wu, L.~Ni, N.~Zhou, and Z.~Zhang, ``Detecting adversarial examples
  by positive and negative representations,'' \emph{Applied Soft Computing},
  vol. 117, p. 108383, 2022.

\bibitem{DBLP:journals/corr/GoodfellowSS14}
I.~J. Goodfellow \emph{et~al.}, ``Explaining and harnessing adversarial
  examples,'' in \emph{Proc. {ICLR}, San Diego, CA, USA, May 7-9, 2015}.

\bibitem{DBLP:conf/eurosp/PapernotMJFCS16}
N.~Papernot \emph{et~al.}, ``The limitations of deep learning in adversarial
  settings,'' in \emph{Proc. {Euro S{\&}P}, Saarbr{\"{u}}cken, Germany, March
  21-24, 2016}.

\bibitem{DBLP:conf/cvpr/Moosavi-Dezfooli16}
S.~Moosavi{-}Dezfooli,  \emph{et~al.}, ``Deepfool: {A} simple and accurate
  method to fool deep neural networks,'' in \emph{Proc. {CVPR}, Las Vegas, NV,
  USA, June 27-30, 2016}.

\bibitem{DBLP:conf/aaai/ChenSZYH18}
P.~Chen \emph{et~al.}, ``{EAD:} elastic-net attacks to deep neural networks via
  adversarial examples,'' in \emph{Proc. {AAAI}, New Orleans, Louisiana, USA,
  February 2-7, 2018}.

\bibitem{DBLP:conf/icml/Croce020a}
F.~Croce \emph{et~al.}, ``Reliable evaluation of adversarial robustness with an
  ensemble of diverse parameter-free attacks,'' in \emph{Proc. {ICML}, 13-18
  July 2020, Virtual Event}, vol. 119, pp. 2206--2216.

\bibitem{7121017}
A.~Mohammadi \emph{et~al.}, ``Improper complex-valued bhattacharyya distance,''
  \emph{IEEE Trans. Neural Netw. Learning Sys.}, vol.~27, no.~5, pp.
  1049--1064, 2016.

\bibitem{DBLP:conf/sp/Carlini017}
N.~Carlini \emph{et~al.}, ``Towards evaluating the robustness of neural
  networks,'' in \emph{Proc. {SP}, San Jose, CA, USA, May 22-26, 2017}.

\bibitem{8297086}
G.~Jin \emph{et~al.}, ``Deep saliency map estimation of hand-crafted
  features,'' in \emph{Proc. {ICIP}}, 2017, pp. 4262--4266.

\bibitem{DBLP:conf/icml/Croce020}
F.~Croce \emph{et~al.}, ``Minimally distorted adversarial examples with a fast
  adaptive boundary attack,'' in \emph{Proc. {ICML}, {\normalfont July 13-18,
  2020}}, vol. 119, pp. 2196--2205.

\bibitem{DBLP:conf/eccv/AndriushchenkoC20}
M.~Andriushchenko \emph{et~al.}, ``Square attack: {A} query-efficient black-box
  adversarial attack via random search,'' in \emph{Proc. {ECCV}, {\normalfont
  August 23-28, 2020}}, vol. 12368, pp. 484--501.

\bibitem{8936536}
J.~Y. Choi \emph{et~al.}, ``Ensemble of deep convolutional neural networks with
  gabor face representations for face recognition,'' \emph{IEEE Trans. Image
  Process.}, vol.~29, pp. 3270--3281, 2020.

\bibitem{8804390}
V.~Santhanam \emph{et~al.}, ``A generic improvement to deep residual networks
  based on gradient flow,'' \emph{IEEE Trans. Neural Netw. Learning Sys.},
  vol.~31, no.~7, pp. 2490--2499, 2020.

\bibitem{9462227}
K.~Alrawashdeh \emph{et~al.}, ``Defending deep learning based anomaly detection
  systems against white-box adversarial examples and backdoor attacks,'' in
  \emph{Proc. {ISTAS}}, 2020, pp. 294--301.

\bibitem{DBLP:conf/emnlp/Kim14}
Y.~Kim, ``Convolutional neural networks for sentence classification,'' in
  \emph{Proc. {EMNLP}, Doha, Qatar, October 25-29, 2014}, pp. 1746--1751.

\bibitem{8788581}
D.~Deng \emph{et~al.}, ``Sparse self-attention {LSTM} for sentiment lexicon
  construction,'' \emph{IEEE/ACM Trans. Audio, Speech, Language Process.},
  vol.~27, no.~11, pp. 1777--1790, 2019.

\bibitem{8241844}
L.~Yu \emph{et~al.}, ``Refining word embeddings using intensity scores for
  sentiment analysis,'' \emph{IEEE/ACM Trans. Audio, Speech, Language
  Process.}, vol.~26, no.~3, pp. 671--681, 2018.

\bibitem{DBLP:conf/nips/MikolovSCCD13}
T.~Mikolov \emph{et~al.}, ``Distributed representations of words and phrases
  and their compositionality,'' in \emph{Proc. Advances in Neural Information
  Processing Systems 26, Lake Tahoe, Nevada, United States, December 5-8,
  2013}, pp. 3111--3119.

\bibitem{DBLP:journals/corr/AbadiABBCCCDDDG16}
M.~Abadi \emph{et~al.}, ``Tensorflow: Large-scale machine learning on
  heterogeneous distributed systems,'' \emph{CoRR}, vol. abs/1603.04467, 2016.

\bibitem{Papernot2016CleverHans}
N.~Papernot \emph{et~al.}, ``Cleverhans v2.1.0: An adversarial machine learning
  library,'' in \emph{Proc. {USENIX})}, August 2018.

\bibitem{https://doi.org/10.48550/arxiv.1807.01069}
M.~I. Nicolae \emph{et~al.}, ``Adversarial robustness toolbox v1.0.0,'' 2018.

\bibitem{rauber2017foolboxnative}
J.~Rauber \emph{et~al.}, ``Foolbox native: Fast adversarial attacks to
  benchmark the robustness of machine learning models in pytorch, tensorflow,
  and jax,'' \emph{J. Open Source Softw.}, vol.~5, no.~53, p. 2607, 2020.

\bibitem{DBLP:journals/corr/abs-2010-01950}
H.~Kim, ``Torchattacks : {A} pytorch repository for adversarial attacks,''
  \emph{CoRR}, vol. abs/2010.01950, 2020.

\bibitem{DBLP:journals/tip/JingSYYN17}
L.~Jing \emph{et~al.}, ``Multi-label classification by semi-supervised singular
  value decomposition,'' \emph{{IEEE} Trans. Image Process.}, vol.~26, no.~10,
  pp. 4612--4625, 2017.

\bibitem{9064929}
A.~Chatzimparmpas \emph{et~al.}, ``{t-viSNE}: Interactive assessment and
  interpretation of {t-SNE} projections,'' \emph{IEEE Trans. Visualization
  Computer Graphics}, vol.~26, no.~8, pp. 2696--2714, 2020.

\bibitem{8998186}
S.~Yu \emph{et~al.}, ``Understanding convolutional neural networks with
  information theory: An initial exploration,'' \emph{IEEE Trans. Neural Netw.
  Learning Sys.}, vol.~32, no.~1, pp. 435--442, 2021.

\bibitem{9107432}
L.~Huang \emph{et~al.}, ``Audio replay spoof attack detection by joint
  segment-based linear filter bank feature extraction and attention-enhanced
  densenet-bilstm network,'' \emph{IEEE/ACM Trans. Audio, Speech, Language
  Process.}, vol.~28, pp. 1813--1825, 2020.

\bibitem{DBLP:conf/aaai/LaiXLZ15}
S.~Lai \emph{et~al.}, ``Recurrent convolutional neural networks for text
  classification,'' in \emph{Proc. {AAAI}, January 25-30, 2015, Austin, Texas,
  {USA}}, 2015, pp. 2267--2273.

\bibitem{DBLP:conf/nips/TramerCBM20}
F.~Tram{\`{e}}r \emph{et~al.}, ``On adaptive attacks to adversarial example
  defenses,'' in \emph{Proc. {NeurIPS}, December 6-12, 2020, Virtual Event},
  2020.

\bibitem{DBLP:conf/iclr/MetzenGFB17}
J.~H. Metzen \emph{et~al.}, ``On detecting adversarial perturbations,'' in
  \emph{Proc. {ICLR}, Toulon, France, April 24-26, 2017}, 2017.

\end{thebibliography}

%

%
%
%




\end{document}